\documentclass[useAMS,usenatbib]{mn2e}

\usepackage[latin1]{inputenc}
\usepackage{graphicx}
\usepackage{epstopdf}

\usepackage{subfigure}
\usepackage{url}
\usepackage{rotate}
\usepackage{amsmath, float}
\usepackage{txfonts}
\usepackage{fleqn}
\usepackage{color}
\usepackage{comment}

\usepackage{mathrsfs}  % added packages
\usepackage{booktabs}
\usepackage{colortbl} 
\usepackage{xcolor}

\newcolumntype{G}{>{\columncolor[gray]{0.8}}l} % Gray column in tabular

\newcommand{\be}{\begin{equation}}
\newcommand{\ee}{\end{equation}}
\newcommand{\bdm}{\begin{displaymath}}
\newcommand{\edm}{\end{displaymath}}
\newcommand{\bea}{\begin{multline}}
\newcommand{\eea}{\end{multline}}
\newcommand{\ba}{\begin{align}}
\newcommand{\ea}{\end{align}}

\newcommand{\araa}{ARA\&A}		% Annual Review of Astron and Astrophys
\newcommand{\apj}{ApJ}			% Astrophysical Journal
\newcommand{\apjl}{ApJLett}		% Astrophysical Journal, Letters
\newcommand{\apjs}{ApJS}		% Astrophysical Journal, Supplement
\newcommand{\aap}{A\&A}			% Astronomy and Astrophysics
\newcommand{\mnras}{MNRAS}		% Monthly Notices of the RAS
\newcommand{\pra}{Phys.~Rev.~A}		% Physical Review A: General Physics
\newcommand{\prc}{Phys.~Rev.~C}		% Physical Review C
\newcommand{\prd}{Phys.~Rev.~D}		% Physical Review D
\newcommand{\pasj}{PASJ}		% Publications of the ASJ
\newcommand{\ssr}{Space~Sci.~Rev.}	% Space Science Reviews
\newcommand{\nat}{Nature}		% Nature
\newcommand{\aplett}{Astrophys.~Lett.} 	% Astrophysics Letters
\newcommand{\physrep}{Phys.~Rep.}   % Physics Reports
\newcommand{\jcap}{JCAP}
\newcommand{\pasa}{PASA}

\def\simlt{\mathrel{\hbox{\rlap{\hbox{\lower4pt\hbox{$\sim$}}}\hbox{$<$}}}}
\def\simgt{\mathrel{\hbox{\rlap{\hbox{\lower4pt\hbox{$\sim$}}}\hbox{$>$}}}}

\defcitealias{Pili_Bucciantini+14a}{PBD14}	
\defcitealias{Frieben_Rezzolla12a}{FR12}	
\defcitealias{Bucciantini_Pili+15a}{BPD15}	

\title[General relativistic models for rotating magnetized NSs]
{ General relativistic models for rotating magnetized neutron stars in conformally flat spacetime}
\author[A.~G. Pili, N. Bucciantini, L. Del Zanna]{
A.~G. Pili$^{1,2,3}$\thanks{E-mail: pili@arcetri.astro.it}, N. Bucciantini$^{2,1,3}$, L. Del Zanna$^{1,2,3}$ \\
$^{1}$Dipartimento di Fisica e Astronomia, Universit\`a degli Studi di Firenze, Via G. Sansone 1, 
I-50019 Sesto F.~no  (Firenze), Italy\\
$^{2}$INAF - Osservatorio Astrofisico di Arcetri, Largo E. Fermi 5, I-50125 Firenze, Italy\\
$^{3}$INFN - Sezione di Firenze, Via G. Sansone 1, I-50019 Sesto F.~no  (Firenze), Italy}

\begin{document}
 
\date{Accepted / Received}

\maketitle

\label{firstpage}

\begin{abstract}

The extraordinary energetic  activity of magnetars is usually explained in terms
of  dissipation of a huge  internal magnetic field of the order of $10^{15-16}$~G. 
How such a strong magnetic field can originate during
the formation of a neutron star  is still subject of active research. An important role
can be played by fast rotation: if magnetars are born as millisecond rotators dynamo mechanisms 
may efficiently amplify the magnetic field inherited from the progenitor star during the collapse.
In this case, the combination of rapid rotation and strong magnetic field determine the right
physical condition not only for the development of a powerful jet driven explosion, 
manifesting as a gamma ray burst, but also for a copious gravitational waves emission.
Strong magnetic fields are indeed able to induce substantial quadrupolar deformations in the star.
In this paper we analyze the joint effect of rotation and magnetization on the structure of
a polytropic and axisymmetric neutron star, within the ideal magneto-hydrodynamic regime. 
We will consider either purely toroidal or purely poloidal magnetic field geometries.
Through the sampling of a large parameter space, we generalize previous results in literature, 
inferring new quantitative relations that allow for a parametrization of  the induced deformation, 
that takes into account also the effects due to the stellar compactness and the current distribution.  
Finally, in the case of purely poloidal field, we also discuss how different prescription
on the surface charge distribution (a gauge freedom) modify the properties of the
surrounding electrosphere and its physical implications.
\end{abstract}

\begin{keywords}
gravitation - magnetic field - MHD - stars:magnetars - stars:neutron.
\end{keywords}

\section{Introduction}
Although there is a general consensus concerning the
main aspects of the  formation of a Neutron Star (NS), we still lack
a precise understanding of many key details of the physical
processes that lead from the collapse of the core of a massive 
progenitor, to the typical compact objects that we observe.

The compact remnant left behind a successful supernova (SN)
explosion is an hot and neutron rich object, the so called proto-NS (PNS), 
whose complex evolution is characterized by large entropy gradients 
\citep{Pons_Reddy+99a}, instabilities \citep{Urpin10a}, 
dynamo \citep{Duncan_Thompson92a}, 
and intense neutrino-driven winds \citep{Pons_Reddy+99a}.
It is only after a typical Kelvin-Helmholz timescale, $\sim 100$~s after  
core bounce, that the PNS becomes transparent to neutrinos, it rapidly
cools, and reaches a state that is dynamically very close to its final
equilibrium \citep{Burrows_Lattimer86a}, even if its thermal and 
magnetic evolution can last much longer \citep{Vigano_Rea+13a}, and 
late phase transitions are still possible 
\citep{Staff_Ouyed+06a,Drago_Pagliara15a,Pili_Bucciantini+16a}.

It is during this brief lapse of time that the magnetic properties of
the resulting NS, such as the magnetic field strength, geometry and 
distribution, will be set. Once the crust begins to form, it will tend to 
``freeze'' it in position, and only dissipative effects might then affect its 
evolution \citep{Vigano_Rea+13a}. The dynamics of the core collapse
is so fast that to first order the core evolves as if it was detached from the outer layers of 
the surrounding star. Hence, it is reasonable to expect that the magnetic properties might 
be related to the conditions in the core of the progenitor, like its rotational profile 
\citep{Duncan_Thompson92a,Spruit09a} or the presence of a seed field
\citep{Woltjer60a,Ruderman72a}. However, given the complexity of the 
problem, it is difficult to establish a clear relation among them.

Among all NSs species, the origin of magnetars is a particularly puzzling
problem, since they exhibit the strongest magnetic field of all NSs,
with a typical strength of the order of $10^{14-15}$~G.
Many evolutionary scenarios have been proposed so far 
(see \citealt{Spruit09a}, \citealt{Ferrario_Melatos+15a} 
and \citealt{Popov15a} for reviews): the magnetar magnetic field can be 
either completely inherited from the  progenitor, as 
the fossil-field scenario suggests \citep{Woltjer60a,Ruderman72a}, or 
generated by dynamo mechanisms during the PNS phase as 
proposed in the original magnetar model \citep{Duncan_Thompson92a}.
In the first hypothesis the magnetic field  is the result of magnetic flux freezing
in the core collapse of a strongly magnetized ($\sim 10^{4}$~G) OB star
\citep{Ferrario_Wickramasinghe06a,Ferrario_Wickramasinghe08a}.
Nevertheless, as pointed out by \citet{Spruit08a}, the paucity of suitable magnetized 
progenitors may be inconsistent with the magnetar birth-rate.
On the other hand, if the nascent NS rotates at millisecond period, the inherited magnetic
field can be further increased by differential rotation \citep{Burrows_Dessart+07a} 
and possible dynamo effects \citep{Bonanno_Rezzolla+03a,Rheinhardt_Geppert05a}.
In principle after core bounce, or alternatively after a merging event, there is a large
amount of available free energy that a huge magnetic field, as high as   $10^{17}$~G,
could  even be reached. Moreover, the combination of millisecond rotation and magnetar
magnetic field can easily drive relativistic outflows  whose energetic, of the order 
of $\sim 10^{49-50} \mbox{erg s}^{-1}$, can even power a typical Gamma-ray Burst (GRB).
This makes proto-magnetars possible engine candidates for GRBs \citep{Usov92a}.

Unfortunately, at present, it is still not clear how the evolution of 
the rotation of the stellar core proceeds before and after the collapse \citep{Ott_Burrows+06a}. 
If magnetars progenitors are massive main sequence stars\footnote{
		The association of magnetar  CXO~J164710.2-455216 with the young cluster Westerlund~1 
		had suggested that  some magnetars may originate from  high mass $\gtrsim 40 \mbox{M}_\odot$ main 
		sequence stars \citep{Muno_Clark+06a}.  Recently, \citet{Clark_Ritchie+14a} have provided a number of arguments
         in favor of the possible origin of CXO J164710.2-455216 in a massive 
         binary system ($\sim 41 \mbox{M}_\odot +  35 \mbox{M}_\odot$), identifying the putative pre-SN companion 
         with $M=9 \mbox{M}_\odot$.},
the magneto-frictional coupling between the core and the outer envelop during 
red supergiant phase can substantially spin-down the core before collapse
\citep{Heger_Woosley+05a,Yoon15a}, unless the progenitor star evolves in a close
binary system, where mass accretion and/or tidal synchronization can indeed enhance 
the rotation \citep{Popov_Prokhorov06a,Popov15a}. After the core bounce, the minimum rotational period
attainable by the newly-born PNS depends on its physical conditions (such as its temperature and entropy gradient) 
and its EoS. While theoretical works predict a minimum rotational period $P\sim1-3$~ms
\citep{Goussard_Haensel+98a,Villain_Pons+04a,Camelio_Gualtieri+16a}, some GRBs
light curves, if interpreted within the proto-magnetar model \citep{Bucciantini_Quataert+09a,
Metzger_Giannios+11a,Bucciantini_Metzger+12a}, point at the possibility of very fast rotators with $P\gtrsim 0.6$~ms
\citep{Rowlinson_Gompertz+14a,Rea_Gullon+15a}, near the mass-shedding limit.
Interestingly, recent 3D numerical simulations of core collapse 
SNe have shown that fast rotation can help the onset of neutrino driven 
explosions \citep{Nakamura_Kuroda+14a,Takiwaki_Kotake+16a,Gilkis16a}, while
magneto-rotational instability can grow magnetar strength magnetic fields
with a strong toroidal component \citep{Mosta_Ott+15a}.
Efficient magnetic field amplification has been obtained also in NS merging 
numerical simulations  by \citet{Giacomazzo_Zrake+15a} and \citet{Kiuchi_Cerda-Duran+15a}.

Such numerical studies follow the  evolution of the system for, at most, few tens of 
milliseconds after the birth of the PNS. At this time, the magnetic field configuration is 
still away from the final one: the magnetic field continues to 
evolve until a stabilizing crust is formed. Since, in the case of magnetars,
the Alfv\'enic crossing  time ($\sim0.01-0.1$~s) is much smaller than the 
typical Kelvin-Helmholtz time-scale, the magnetic field can in principle
decay against instabilities unless a stable equilibrium is reached 
before a crust forms \citep{Spruit08a}.  
However, numerical simulations have shown that the initial field may  evolve
toward a roughly axisymmetric mixed configuration, dubbed as \textit{Twisted Torus} (TT),
able to persist for many Alfv\'en times \citep{Braithwaite_Spruit06a,Braithwaite09a,Mitchell_Braithwaite+15a} .

The analysis of the long-term effects induced by the magnetic field on the 
structure and the properties of the NS requires a different approach, that
is based on the modelization of equilibrium configurations of magnetized NSs, 
taking into account the largest possible set of magnetic field morphologies
and current distributions.
Several authors have dealt with this problem considering  purely poloidal or 
purely toroidal magnetic field in static and/or rotating NSs
\citep{Bocquet_Bonazzola+95a,Bonazzola_Gourgoulhon96a,Konno01a,Cardall_Prakash+01a,Kiuchi_Yoshida08a,
Frieben_Rezzolla12a,Pili_Bucciantini+14a,Bucciantini_Pili+15a,Franzon_Dexheimer+16a}
but also mixed field configuration in Newtonian gravity 
\citep{Yoshida_Eriguchi06a,Lander_Jones09a,Mastrano_Melatos+11a,Fujisawa_Yoshida+12a,Lasky_Melatos13a,
Glampedakis_Lander+2013a,Fujisawa_Kisaka14a,Armaza_Reisenegger+15a,Mastrano_Suvorov+15a,Fujisawa_Eriguchi15a}
or  within the framework of General Relativity (GR) \citep{Ioka_Sasaki04a,Ciolfi_Ferrari+09a,Ciolfi_Ferrari+10a,
Pili_Bucciantini+14a,Pili_Bucciantini+15a,Bucciantini_Pili+15a,Uryu_Gourgoulhon+14a}.
A common motivation at the base of these works is the characterization of the stellar quadrupole deformation induced
by the magnetic field. Fast rotating newly born magnetars, by virtues of such deformations,
are currently considered as possible sources of detectable Gravitational Waves (GWs), especially if
the magnetic field is energetically dominated by its toroidal component \citep{DallOsso_Shore+09a}.
In this case indeed, the magnetically induced  deformation is prolate and dissipative processes 
can lead to the orthogonalization of the spin and the magnetic axes, in order to minimize 
the total rotational energy \citep{Cutler02a}. 
At the same time, this  `spin-flip' mechanism  maximizes  the efficiency of GW emission so that,
in principle, the initial spin-down could be mostly due to GWs rather than magnetic braking.
Nevertheless  \citet{Lasky_Glampedakis16a}  and   \citet{Moriya_Tauris16a} have recently  searched for GW 
signatures in the light curve of, respectively,  short GRBs and super-luminous SNe. They have shown that, 
if interpreted within the proto-magnetar model, the shape of such light curves indicates that most of the 
rotational energy losses are compatible with the hypothesis that magnetic braking largely prevails over 
GWs emission. This either implies small stellar deformations, constraining the toroidal field to 
be $\lesssim10^{16}$~G, or an inefficient spin-flip  mechanism \citep{Lasky_Glampedakis16a}.

In continuity with  \citet{Pili_Bucciantini+14a} (hereafter \citetalias{Pili_Bucciantini+14a}),
\citet{Bucciantini_Pili+15a} (hereafter \citetalias{Bucciantini_Pili+15a}) and \citet{Pili_Bucciantini+15a} 
that were limited to static equilibria, in this work we perform a vast and detailed parameter study of 
rapidly rotating magnetized NSs within the framework of GR. This allows us to establish general trends 
and expectations regarding how different morphologies of the magnetic field
and in particular, its coupling with rotation, affects the structure of the star.
The analysis of a large set of equilibria provides us the opportunity to 
extend previous results presented in literature and to derive new quantitative relations
between the induced deformations,  the energy content of the system and  the structure of the
magnetic field. In this work we will consider only strictly stationary configurations.
Hence, although mixed field configurations, and in particular those dominated by the toroidal magnetic field,
are favored on the basis of stability arguments  \citep{Akgun_Reisenegger+13a},
we will consider only axisymmetric configurations with either purely toroidal or purely poloidal magnetic field. 
Rotating configurations with mixed field have indeed a non-vanishing Poynting vector,
that prevents to consider them as strictly stationary equilibria. 

The Paper is organized as follows. In  Section~\ref{sec:formalism}
we will introduce the general formalism and the governing equations.
In Section~\ref{sec:numerical}  we describe the numerical scheme and discuss
the new features introduced in XNS (see \url{http://www.arcetri.astro.it/science/ahead/XNS/} for an updated version
of the code). Our results are presented in Section~\ref{sec:results}, while conclusion are drawn in 
Section~\ref{sec:conclusions}.

\section{Formalism}
\label{sec:formalism}

Let us briefly introduce here the general formalism that we have used
to describe and compute the spacetime structure (the metric), the
electromagnetic field, and the matter distribution in our equilibrium
models for steady state, magnetized and rotating NSs.
In the following we choose the signature $(-,+,+,+)$ for the 
spacetime metric and, unless otherwise stated, we adopt 
geometrized units with $c=G={\rm M}_{\odot}=1$ and with the factor $\sqrt{4\upi}$ 
reabsorbed in the definition of the electromagnetic fields.

\subsection{Metric}
\label{sec:metric}

Our approach to the numerical solution of Einstein equations together with the
equilibrium conditions for the matter distribution and the
electromagnetic field  is based on the so
called {\it 3+1 splitting} of the space-time metric and fluid
quantities (see for example \citealt{Gourgoulhon07a}, \citealt{Gourgoulhon10a}, \citealt{Alcubierre08a},
\citealt{Baumgarte_Shapiro10a}, \citealt{Bucciantini_Del-Zanna11a}, to which the reader is 
referred for a detailed description). In such formalism the generic line element can be written in 
the ADM form \citep{Arnowitt_Deser+62a}
\begin{equation}
{\rm d}s^2=-\alpha^2 {\rm d}t^2 +\gamma_{ij}({\rm d}x^i +\beta^i {\rm d}t)({\rm d}x^j+\beta^j {\rm d}t),
\end{equation}
where $\alpha$ is known as {\it lapse function}, $\beta^i$ is a purely
spatial vector known as {\it shift vector}, $\gamma_{ij}$ is the
3-metric induced  on the space-like 3-surface of the foliation, and where
we have used adapted coordinates: $x^\mu=[t,x^i]$. 
The components of any 4-vector projected on the space-like 3-surface 
are referred as its {\it Eulerian} components. 
In the following greek indexes will be used for 4-dimensional quantities, 
while latin ones for their respective 3-dimensional projections.

In the case of a stationary and axisymmetric equilibrium, 
the spacetime is endowed with a stationary Killing vector $\xi^{\mu}$
and a azimuthal Killing vector $\chi^{\mu}$ that locally define a 
time-like 2-plane $\Pi$. Any vector $\varv^\mu$ is said to be {\it toroidal} if it belongs to this plane, 
{\it  poloidal} if it is perpendicular \citep{Carter70a,Carter73a}.
For those particular forms of the matter-energy distribution, such that the
energy-momentum tensor $T^{\mu\nu}$ satisfies the relations \citep{Kundt_Trumper66a,Carter69a}
\begin{equation}
\xi_\mu T^{\mu[\nu}\xi^\kappa \chi^{\lambda]}=0,\quad \chi_\mu T^{\mu[\nu}\xi^\kappa \chi^{\lambda]}=0,
\end{equation}
where square brackets indicate antisymmetrization with respect to
enclosed indexes, the metric is {\it circular} and, adopting  spherical-like coordinates $x^\mu=[t,r,\theta,\phi]$
(hence $\xi^{\mu}=(\upartial_t)^\mu$ and $\chi^\mu=(\upartial_\phi)^\mu$), the line element simplifies to
\begin{equation}
\rmn{d} s^2 = - \alpha^2 \rmn{d} t^2 + \psi^4 (\rmn{d}r^2 + r^2 \rmn{d}\theta^2  ) + R^2(\rmn{d} \phi - \omega \rmn{d} t)^2.
\label{eq:circular}
\end{equation}
Here $R=\sqrt{\gamma_{\phi\phi}}$ is knows as {\it
quasi-isotropic radius}, $\psi$ is a {\it conformal factor} and  $\omega=-\beta^\phi$.
The determinant of the 3-metric is then $\sqrt{\gamma}=Rr\psi^4$.

The energy-momentum tensor of a fluid at thermodynamic equilibrium equipped with an
electromagnetic field, in the absence of magnetization effects
\citep{Chatterjee_Elghozi+15a,Franzon_Dexheimer+16a},  can be written  as:
\begin{equation}
T^{\mu\nu}=\rho h u^\mu u^\nu +pg^{\mu\nu}+F^\mu_\lambda F^{\nu\lambda}-\frac{1}{4}(F^{\lambda\kappa}F_{\lambda\kappa})g^{\mu\nu},
\label{eq:stressenergy}
\end{equation}
where $\rho$ is the rest mass density, $h=(e+p)/\rho$  the specific
enthalpy, $e$ the energy density, $p=p(\rho, e)$ the pressure (provided by some
form of equation of state, EoS), $u^{\mu}$ is the fluid 4-velocity and $F^{\mu\nu}$ 
is the Faraday electromagnetic tensor. The electromagnetic tensor can be defined either in terms of 
the comoving magnetic and electric field $b^\mu$ and $e^\mu$ (whenever a flow velocity can be defined), or in
terms of the so called {\it Eulerian} electric and magnetic field
$E^\mu$ and $B^\mu$ (purely spatial vectors), following the same 3+1
splitting of the metric \citep{Del-Zanna_Zanotti+07a,Bucciantini_Del-Zanna11a}. 
In the region of space occupied by matter,  the ideal MHD condition is 
$e^\mu =u_\nu F^{\mu\nu}=0$, and only the comoving magnetic field does not vanish.
When this condition is relaxed, non-ideal effects such as dynamo or reconnection 
processes may take place in both the interior and the magnetosphere of magnetars. 
For applications to the numerical modeling of relativistic plasmas, see
\citet{Bucciantini_Del-Zanna13a} and \citet{Del-Zanna_Papini+16a}.

Given the ideal-plasma stress energy tensor in Eq.~\eqref{eq:stressenergy},
under the assumption of stationarity and axisymmetry, the circularity condition holds provided the
4-velocity is toroidal, i.e.  $u^r=u^\theta=0$, and the magnetic field is either purely toroidal or purely poloidal. 
However it can be shown \citep{Oron02a}, and it has been verified 
(\citealt{Shibata_Sekiguchi05a,Dimmelmeier_Stergioulas+06a,Ott_Dimmelmeier+07a,
Bucciantini_Del-Zanna11a}; \citetalias{Pili_Bucciantini+14a}) that, considering purely toroidal flow,
 the metric can  be safety simplified neglecting off-diagonal terms  
(with the exception of $\beta^\phi$), even for a mixed magnetic field.
In particular, even for highly deformed star up to the {\it mass shedding limit}  and/or for
magnetic field with strength up to $10^{19}$G, the difference between 
$R$ and $\psi^2r\sin{\theta}$  is at most of the order of $10^{-3}$, 
so one can assume to a high level of accuracy that the metric is
\emph{conformally flat} [CFC assumption \citep{Isenberg08a,Wilson_Mathews03a}]:
\begin{equation}
\rmn{d} s^2 = -\alpha^2 \rmn{d} t^2 +\psi^4 [\rmn{d} r^2 + r^2\rmn{d}
\theta^2 +  r^2\sin^2\theta (\rmn{d} \phi - \omega \rmn{d}
t)^2],
\label{eq:cfc}
\end{equation}
with the volume element of the 3-metric given by $\sqrt{\gamma} = \psi^6 r^2 \sin\theta$.

Einstein equations in the CFC approximation can be conveniently recast
into a set of elliptical partial diferential equations (PDEs) for the metric quantities $\alpha$, $\psi$
and $\omega$, where the source terms contain the information about the
energy-momentum distribution \citep{Dimmelmeier_Font+02a,Bucciantini_Del-Zanna11a}:
\begin{equation}
\Delta \psi =-[2\pi E+\frac{1}{8}K_{ij}K^{ij}] \psi^5,
\label{eq:CFC1}
\end{equation}
\begin{equation}
\Delta (\alpha\psi) =[2\pi (E+2S)+\frac{7}{8}K_{ij}K^{ij}]
\alpha\psi^5 ,
\label{eq:CFC2}
\end{equation}
\begin{equation}
\Delta \omega=-16\pi\alpha\psi^4 S^\phi-2\psi^{10}K^{\phi j}\nabla_j(\alpha \psi^{-6}).
\label{eq:CFC3}
\end{equation}
Here $\Delta$ and $\nabla$ are the standard {\it Laplacian} and {\it Gradient}
operator with respect to the 3-metric $\gamma_{ij}$, $K^{ij}$ is the
extrinsic curvature, and the source term are given by:
\begin{equation}
E=\rho h \Gamma^2-p+\frac{1}{2}(E^i E_i +B^i B_i),
\end{equation}
\begin{equation}
S=\rho h (\Gamma^2-1)+3p+\frac{1}{2}(E^i E_i +B^i B_i),
\end{equation}
\begin{equation}
S_\phi=\rho h \Gamma \varv_\phi+ \epsilon_{\phi jk}E^jB^k,
\end{equation}
where $\epsilon_{ijk}$ is the 3-dimensional  Levi-Civita alternating tensor
and the $\phi$-component of the 3-velocity $\varv^{\phi}$ is linked to $u^{\phi}$
through the Lorentz factor $\Gamma$ by $u^{\phi}=\Gamma(\varv^\phi-\beta^\phi/\alpha)$,
where the normalization $u^{\mu}u_{\mu}=-1$ implies $\Gamma=(1-\varv^2)^{-1/2}$.

\subsection{Electromagnetic Fields}
\label{sec:emfields}

The 3+1 splitting allows one to write the Maxwell equations:
\begin{equation}
\nabla_\mu F^{\mu\nu}=-j^\nu, \quad \nabla_\mu( \epsilon^{\mu\nu\lambda\kappa}F_{\lambda\kappa})=0
\end{equation}
where $j^{\,\mu}$ is the {\it 4-current density}, and
$\epsilon^{\mu\nu\lambda\kappa}$ 4-dimensional  Levi-Civita alternating tensor, in terms of
the Eulerian electric and magnetic field
$E^i=[E^r,E^\theta,E^\phi]$,
$B^i=[B^r,B^\theta,B^\phi]$. 
The stationary Maxwell equations in the $3+1$ formalism can be written
in a compact form using standard 3-dimensional vector quantities, and
the related divergence and curl operators defined with respect to the diagonal
3-metric $\gamma_{ij}$: 
\begin{equation}
\mathbf{\nabla} \cdot \mathbf{B} = 0,
\label{eq:m1}
\end{equation}
\begin{equation}
\mathbf{\nabla} \times (\alpha \mathbf{E} + \mathbf{\bbeta} \times\mathbf{B}) = 0,
\label{eq:m2}
\end{equation}
\begin{equation}
\mathbf{\nabla} \cdot \mathbf{E} = \rho_e,
\label{eq:m3}
\end{equation}
\begin{equation}
\mathbf{\nabla} \times (\alpha \mathbf{B} - \mathbf{\bbeta} \times\mathbf{E}) = 
\alpha\mathbf{J} - \rho_e \mathbf{\bbeta},
\label{eq:m4}
\end{equation}
where $\rho_e$ is the charge density, and corresponds to the time-like
component of the 4-current $j^{\,\mu}$, while $\mathbf{J}$ is its
3-dimensional space-like projection.

Eq.~\eqref{eq:m1} implies that the magnetic field can be written as the
curl of vector potential $\mathbf{A}$ such that $B^i=\varepsilon^{ijk}\upartial_j (A_k)$. This, together with the assumption of axisymmetry,
$\upartial_\phi =0$, implies that the poloidal ($r,\theta$) components of the magnetic field
can be expressed in terms of the gradient of a scalar function $\Psi(r,\theta)
\equiv A_\phi$, called
\emph{magnetic flux function}. Analogously the $\phi$-component of the
magnetic field can be written using another scalar function $\mathcal{I}$
($B_{\phi}=\alpha^{-1}\mathcal{I}$) known as {\it current function}.
By introducing the orthonormal triad the magnetic vector field can be expressed as
\begin{equation}
\mathbf{B} = \frac{\mathbf{\nabla}\Psi}{R} \times \mathbf{e}_{\hat{\phi}} + 
\frac{\mathcal{I}}{\alpha R}\mathbf{e}_{\hat{\phi}},
\label{eq:B} 
\end{equation}
where $\nabla\Psi$ is the vector field associated with the gradient of $\Psi$ with components
 $(\nabla\Psi)_{\hat{i}}=\upartial_i \Psi/\sqrt{\gamma_{ii}}$.
The isosurfaces $\Psi(r,\theta)=cost$, are known as {\it magnetic surfaces}, and any 
scalar function $f$ satisfying $\mathbf{B}\cdot \mathbf{\nabla} f =0$ will be constant
on them: $f=f(\Psi)$. 

It is possible to show, starting from Eq.~\eqref{eq:m4}, that the poloidal component of
the 3-current is related to the curl of the azimuthal magnetic field, 
$\alpha J^i=\varepsilon^{ij\phi}\upartial_j (\alpha B_\phi)$, 
such that:
\begin{equation}
\mathbf{J} = \frac{\mathbf{\nabla}\mathcal{I}}{\alpha R} \times \mathbf{e}_{\hat{\phi}} + 
J_{\hat{\phi}}\mathbf{e}_{\hat{\phi}},
\label{eq:J}
\end{equation}
where:
\begin{equation}
\frac{\alpha}{R}J_{\hat{\phi}} = -  \nabla\cdot \left( \frac{\alpha}{R^2} \nabla\Psi \right)
+ \mathbf{E} \cdot\nabla\omega.
\label{eq:Jphi}
\end{equation}
From the other sourceless Maxwell equation, Eq.~\eqref{eq:m2},  under the same
constraints of stationarity and axisymmetry, one finds
$\alpha E_\phi=0$. The same equation implies that the poloidal
components can be written as the gradient of a scalar function  $\Phi$ as:
\begin{equation}
\alpha \mathbf{E} + \mathbf{\bbeta} \times\mathbf{B} = \alpha \mathbf{E}  - \omega \mathbf{\nabla}\Psi = \mathbf{\nabla}\Phi.
\end{equation}

Assuming conformal flatness, Eq.~\eqref{eq:cfc},  the Maxwell-Gauss
equation, Eq.~\eqref{eq:m3}, can be written as an elliptical PDE for the
electromagnetic potentials $\Phi$ and $\Psi$, as a function of the
charge and current density.
\begin{eqnarray}
\Delta \Phi \negmedspace & = \negmedspace &  \psi^4 \left[  \alpha\rho_{ e}+ \omega\psi^4 r^2 \sin^2 \theta J^{\phi}\right]- 
	   \frac{\omega\psi^4 r^2 \sin^2\theta}{\alpha^2}\upartial\omega\upartial \Phi \nonumber\\
	& &  \left[1+\frac{\omega^2\psi^4 r^2 \sin^2\theta}{\alpha^2}\right] \upartial\omega\upartial \Psi 
	    - \upartial\ln(\alpha^{-1}\psi^2) \left[ \upartial \Phi + 2\omega\upartial \Psi \right] -  \nonumber\\
	& & \frac{2\omega}{r}\left[\upartial_r \Psi + \frac{1}{r\tan \theta}\upartial_\theta \Psi \right].
	\label{eq:MGpot}
\end{eqnarray}
The same can be done for the Maxwell-Amp\`ere equation Eq.~\eqref{eq:m4}:
\begin{equation}
 \tilde{\Delta}_3\tilde{\Psi} = 
 \psi^4 r \sin \theta \left( \frac{\upartial \omega \upartial \Phi + \omega \upartial \omega \upartial \Psi }{\alpha^2} 
 -\psi^4 J^{\phi} \right) +  \upartial \ln{(\alpha^{-1}\psi^2)} \upartial  \Psi, 
 \label{eq:MApot}
\end{equation}
where, for convenience, we have introduced  the quantity
$\tilde{\Psi}=\Psi/(r \sin \theta )$, and the new operators are
defined as in \citetalias{Pili_Bucciantini+14a}:
\begin{equation}
\partial f\partial g = \partial_r f\partial_r g
+\frac{1}{r^2}\partial_\theta f\partial_\theta g,
\end{equation}
\begin{equation}
\tilde{\Delta}_3=\Delta-\frac{1}{r^2\sin^2{\theta}}.
\end{equation}
These equations completely define the electromagnetic field in the
entire space, once the charge and current distribution are known,
independently of the fluid properties.

If there is an observer that measures a vanishing electric field in his 
reference frame
(for example if one assumes the NS to be perfectly conducting,  the 
comoving electric field inside vanishes), then the condition
$\mathbf{E}\cdot\mathbf{B}=0$ holds.  
This establishes a relation between the two electromagnetic  potentials: 
$\mathbf{B}\cdot \mathbf{\nabla}\Phi =0 \Rightarrow \Phi=\Phi(\Psi)$.
In such a situation one can introduce a velocity $\mathbf{v}$
so that
\begin{equation}
\mathbf{E}=-\mathbf{v} \times \mathbf{B}
\label{eq:Ohmlaw}
\end{equation}
where, defining $\Omega=\Omega(\Psi) = -\rmn{d} \Phi/\rmn{d} \Psi$, we have
\begin{equation}
\mathbf{v}=\varv \, \mathbf{e}_{\hat{\phi}} \quad \mbox{with} \quad
 \varv=\frac{R(\Omega-\omega)}{\alpha} \mathbf{e}_{\hat{\phi}}.
\end{equation}
Notice that here $\Omega$ represents the rotational rate of the 
magnetic field lines with respect to infinity.
Is then possible to express the electric field and the charge density in terms of the magnetic flux function:
\begin{equation}
\mathbf{E} = - \frac{\Omega-\omega}{\alpha}\mathbf{\nabla}\Psi,
\label{eq:E}
\end{equation}
\begin{equation}
\rho_e = - \mathbf{\nabla} \cdot \left(\frac{\Omega-\omega}{\alpha}\mathbf{\nabla}\Psi \right).
\end{equation}

As long as the magnetic field is confined inside the NS, so is also
the electric field. However this condition can naturally be  enforced only for
toroidal magnetic field, where the electric field vanishes by
construction. In the case of a poloidal magnetic field this is expected
to extend outside the NS surface
( \citealt{Lander_Jones09a,Ciolfi_Ferrari+09a,Ciolfi_Ferrari+10a,Ciolfi_Rezzolla13a};
\citetalias{Pili_Bucciantini+14a}; \citetalias{Bucciantini_Pili+15a}), unless an 
ad hoc singular toroidal current is added at the surface itself \citep{Fujisawa_Eriguchi15a}. One then
needs to define the condition holding outside the NS surface, in order
to decide how to extend the field from the interior, where the
typically high conductivity of degenerate matter can enforce ideal MHD.

\subsubsection{Outer Vacuum}
\label{sec:vacuum}

The general assumption that is made in the literature \citep{Bocquet_Bonazzola+95a,Franzon_Dexheimer+16a},
and the one that for consistency is adopted in this work,
is that outside the NS is surrounded by a vacuum.  In this case one
cannot define any meaningful reference frame in the exterior, such
that it is not possible to enforce any relation between $\Psi$ and $\Phi$. The
equations of the potentials must be solved separately assuming
$\rho_e=J^\phi=0$, only subject to the requirement of continuity at
the stellar surface. In this case it is well known \citep{Goldreich_Julian68a,Michel_Li99a} that
one will have regions with $\mathbf{E}\cdot\mathbf{B}\neq 0$, which are
known as {\it vacuum gaps}. Moreover as we will discuss in
Sect.~\ref{sec:nummax} the solution is not unique.

\subsubsection{Force-Free Limits}
\label{sec:ffe}

On the other hand  if one assume the existence of a low density plasma, that, without
affecting the dynamics can provide the required changes and
currents, the condition $\mathbf{E}\cdot\mathbf{B}=0$ can be extended outside
the NS surface. This is the base of the so called {\it degenerate electro-dynamics},
and it is the prescription generally adopted in magnetospheric models
that focus just on the exterior
\citep{Michel73a,Contopoulos_Kazanas99a,Timokhin06a,Spitkovsky06a,Tchekhovskoy_Spitkovsky+13a,Petri16a}, 
and that has been recently extended to global models \citep{Etienne_Paschalidis+15a}. In
this case, the Lorentz force per unit volume acting on the plasma is: 
\begin{equation}
\mathbf{L}= \rho_e \mathbf{E} + \mathbf{J} \times \mathbf{B} =
\left(\mathbf{J} - \rho_e \mathbf{v} \right)\times \mathbf{B},
\end{equation}
which, using Eqs.~\ref{eq:B}, \ref{eq:J}, becomes:
\begin{equation}
\label{eq:lorentz}
\mathbf{L}=\left( \! \frac{J_{\hat{\phi}}}{R} - \rho_e \frac{\varv}{R}  \right) \! \mathbf{\nabla} \Psi 
 -  \frac{\mathcal{I}\,\mathbf{\nabla}\mathcal{I} }{\alpha^2 R^2} +
\frac{\mathbf{\nabla}\mathcal{I} \! \times \! \mathbf{\nabla}\Psi\cdot \mathbf{e}_{\hat{\phi}}}{\alpha R^2}
  \mathbf{e}_{\hat{\phi}}.
\end{equation}
Given the negligible dynamical effects of the plasma the Lorentz force
must vanish: $\mathbf{L} = 0$. This case is referred as
\emph{force-free electrodynamics} (FFE). One may notice immediately
that  the azimuthal  component
of the Lorentz force vanishes if, and only if,
$\mathcal{I}=\mathcal{I}(\Psi)$. Hence:
\begin{equation}
\mathbf{L} = \left( \! \frac{J_{\hat{\phi}}}{R} -
\frac{\mathcal{I}}{\alpha^2 R^2}\frac{\rmn{d} \mathcal{I}}{\rmn{d} \Psi}
- \rho_e \frac{\Omega-\omega}{\alpha} \! \right) \! \mathbf{\nabla} \Psi =0.
\label{eq:lorentz2}
\end{equation}
By making use of Eq.~\eqref{eq:Jphi}, this can be written as a single
equation for  $\Psi$, known as the relativistic {\it Grad-Shafranov equation}.
After some algebra one recovers the {\it pulsar equation}:
\begin{equation}
\nabla \! \cdot \! \left[ \frac{\alpha}{R^2} \left( 1 \! - \! \varv^2 \right) \! \nabla\Psi \right]
+ \frac{\varv}{R} \frac{\rmn{d} \Omega}{\rmn{d} \Psi}| \nabla\Psi |^2 + 
\frac{\mathcal{I}}{\alpha R^2}\frac{\rmn{d} \mathcal{I}}{\rmn{d} \Psi} = 0.
\label{eq:ffegs}
\end{equation}
Notice that in the force-free regime there is nothing to prevent
$\varv>1$, being this just the
drift velocity associated to the motion of immaterial fieldlines. This
happens at a surface defined by $R=R_L=\alpha (\Omega - \omega)^{-1}$
and called \emph{Light Cylinder}. Regularity of the solution of the
Grad-Shafranov equation at such surface introduces 
 further  constraints on $\mathcal{I}(\Psi)$. Note that contary to solutions in
 vacuum, where there is no net energy flow associated to the outer
 electromagnetic field, in the FFE case there is a net energy flow
 along those field-lines that cross the Light Cylinder. This implies that 
 the FFE assumption is not fully consistent with our requirement
 of a strict stationary system.

\subsection{Matter}
\label{sec:matter}

Let us discuss here the equilibrium conditions for the matter
distribution, under the simultaneous action of gravity and an
electromagnetic field.
Here we will consider for simplicity a toroidal flow 
\begin{equation}
u^\mu = u^t (1, 0, 0, \Omega), \quad \Omega\equiv u^\phi/u^t.
\end{equation}
where now $\Omega$ is the fluid angular velocity as measured by
an observer at rest at spatial infinity. Notice that within the ideal MHD regime,
the angular velocity  appearing in the Ohm's law (that is the same as in \eqref{eq:Ohmlaw})
actually coincides with the fluid rotational rate.

The requirement of stationarity implies that the rotational rate
$\Omega$ should be constant on magnetic surfaces.
However very ad hoc forms for the angular momentum distribution and 
the current distribution are required in order to satisfy such requirement. 
For example the simplest prescription of {\it constant specific angular momentum} 
leads to a rotation largely stratified on cylinders, known as {\it von Zeipel
cylinders} \citep{von-Zeipel24a}, while the simplest prescriptions for the current
distribution (\citealt{Lander_Jones09a,Ciolfi_Ferrari+09a}; \citetalias{Pili_Bucciantini+14a}) lead to 
dipole-like magnetic surfaces. The other possibility is to consider solid body rotation 
with $\Omega=\mbox{const}$ \citep{Oron02a}, as we do in this work.
In this case the ideal MHD condition $\nabla \Phi= -\Omega \nabla \Psi $ can 
be easily integrated in 
\begin{equation}
\label{eq:MHDcond}
\Phi=-\Omega\Psi+C
\end{equation}
where $C$ is a integration constant that define the arbitrary monopolar 
charge of the star.
 
The dynamics of matter is describend by the relativistic {\it Euler
  equation}, which in the presence of a generic external force (per unit volume) $f_\mu$ is
\begin{equation}
\rho h a_\mu + \upartial_\mu p + u_\mu u^\nu\upartial_\nu p = f_\mu,
\end{equation}
where $a^\mu$ is the 4-acceleration.
Recalling that $u^i=0, \, (i=r,\theta)$ and $\upartial_t=\upartial_\phi=0$, so that
$u^\nu \upartial_\nu = 0$, one has:
\begin{equation}
a_\mu = u^\nu (\upartial_\nu u_\mu - \Gamma^\lambda_{\mu\nu}u_\lambda)=
-\tfrac{1}{2}u^\nu u^\lambda\upartial_\mu g_{\nu\lambda},
\end{equation}
and its spatial projection in the $3+1$ formulation is:
\begin{equation}
a_i = \frac{\Gamma^2}{2\alpha^2}[ \upartial_i (\alpha^2 - R^2 \omega^2)
+ 2\Omega \upartial_i (R^2\omega) - \Omega^2 \upartial_i R^2 ].
\end{equation}
Recalling the definition of $\varv$, and given the relation 
$\varv^2\Gamma^2\upartial_i\ln \varv = \upartial_i\ln\Gamma$,
one finally gets:
\begin{equation}
\frac{\upartial_i p}{\rho h} + \upartial_i \ln\alpha -
\upartial_i\ln\Gamma = \frac{L_i}{\rho h}.
%\frac{\upartial_i p}{\rho h} + \upartial_i \ln\alpha - \upartial_i\ln\Gamma + 
%\frac{\Gamma^2\varv^2}{\Omega-\omega}\upartial_i\Omega = \frac{f_i}{\rho h},
\label{eq:Euler}
\end{equation}
where we have also specialized the external force  to the Lorentz force $L_i$. 
Notice that axisymmetry implies necessarily $f_{\phi}=0$, whatever the nature of the force.

In order to cast this equation into an integrable form, suitable for
numerical solutions, two assumptions are required:
\begin{itemize}
\item a barotropic EoS $p=p(\rho)$,  as in the case of a polytropic law:
\begin{equation}
p=K\rho^{1+1/n} \Rightarrow h = 1 + (n+1) K \rho^{1/n},
\end{equation}
where $n$ is the {\it polytropic index}, such that $\upartial_i p /(\rho h)=\upartial_i \ln{h}$.  
\item an external conservative force with potential $\mathcal{M}$:
\begin{equation}
\mathbf{L} = \rho h \mathbf{\nabla} \mathcal{M}.
\label{eq:M}
\end{equation}
\end{itemize}
The first one is usually justified by the fact that matter in neutron stars
can be considered fully degenerate (zero temperature). The
second one, on the other hand,  restricts the possible choices of the current
distribution (see e.g. \citealt{Akgun_Reisenegger+13a} for examples where this constrain is
relaxed), but is the only one that permits to compute equilibria in
the fully non perturbative regime. 

Under those two assumptions one can integrate Euler's equation to
derive the so called  {\it Bernoulli integral}:
\begin{equation}
\ln \frac{h}{h_c} + \ln\frac{\alpha}{\alpha_c} - \ln\frac{\Gamma}{\Gamma_c} 
= \mathcal{M}-\mathcal{M}_c,
%\ln \frac{h}{h_c} + \ln\frac{\alpha}{\alpha_c} - \ln\frac{\Gamma}{\Gamma_c} - 
%\frac{A^2}{2}(\Omega_c-\Omega)^2 = \mathcal{M}-\mathcal{M}_c,
\end{equation}
where we have indicated with the label $c$ a reference position, for
instance the center of a rotating neutron star.

Interestingly the MHD condition in Eq.~\eqref{eq:MHDcond}, 
together with the requirement of integrability, can be easily translated 
into a condition on the charge and current distribution:
\begin{equation}
\label{eq:rhomhd}
\rho_{ e}=\frac{R^2(\Omega-\omega)}{\alpha} \rho h \Gamma^2 \frac{\rmn{d} \mathcal{M}}{\rmn{d}  \Psi}
            -\frac{\Omega-\omega}{\alpha\psi^4 \varv^2}\upartial \ln \Gamma^2 \upartial \Psi
            -\frac{1}{\psi^4}\upartial\omega\upartial \Psi,
\end{equation}
\begin{equation}
\label{eq:Jmhd}
J^{\phi}= \rho h \Gamma^2 \frac{\rmn{d}  \mathcal{M}}{\rmn{d}  \Psi}
		-\frac{1}{\psi^8 r^2 \sin^2 \theta} \upartial \ln \Gamma^2 \upartial \Psi
		+\frac{\omega-\Omega}{\alpha^2\psi^4}\upartial \omega \upartial \Psi.
\end{equation}
Note however that it is not sufficient to impose these forms for the
source terms into Eq.~\eqref{eq:MGpot} in order to ensure ideal MHD inside the
NS, because Eq.~\eqref{eq:MGpot} defines the electromagnetic field minus an arbitrary
harmonic function. This harmonic function (which guarantees ideal MHD
inside) corresponds to a singular source term (a surface charge), that
have been neglected in deriving the integrability conditions, where we
only considered distributed forces. 

\subsection{Currents}
\label{sec:currents}

The morphology of the magnetic field is entirely controlled by the analytic form of 
the free functions $\mathcal{M}$ and $\mathcal{I}$. As discussed in the previous section 
 the magnetization  function  $\mathcal{M}$ is associated with the Lorentz force 
term appearing  in the Euler equation, Eq.~\eqref{eq:Euler}. The current function  $\mathcal{I}$, instead, is
strictly related only to the toroidal component of the magnetic field.

If the magnetic field has a poloidal component then $\Psi\neq 0$ and  $\mathcal{M}$
can be expressed as a function of the magnetic potential $\Psi$ alone
because of the orthogonality relation  $\mathbf{L}\cdot \mathbf{B}=0$. 
A common choice is to express $\mathcal{M}$ as a linear function of  $\Psi$ 
\citep{Ciolfi_Ferrari+09a,Lander_Jones09a} even if more general analytic forms,
including also a non-linear dependence,  have been recently investigated 
(\citealt{Fujisawa_Yoshida+12a,Ciolfi_Rezzolla13a,Bera_Bhattacharya14a};
\citetalias{Bucciantini_Pili+15a}).
In particular, as in \citetalias{Bucciantini_Pili+15a}, we adopt
\begin{equation}
\mathcal{M}(\Psi)=k_{\rm pol}\Psi\left( 1+ \frac{\xi}{|\nu+1|}\Psi^\nu \right),
\label{eq:Mpol}
\end{equation}
where  $k_{\rm pol}$ is the so-called \textit{ poloidal magnetization constant}, $\nu$ is the
\textit{poloidal magnetization index} of the non linear term. Given
that the effect of non-liner terms have been already discussed in \citepalias{Bucciantini_Pili+15a},
in the present work we will focus on configurations with $\xi=0$, in order 
to limit the parameter space. 

Notice that, in the definition of toroidal currents distribution
$J^\phi$ and of the charge density $\rho_e$,
Eqs~\ref{eq:Jmhd}-\ref{eq:rhomhd}, only  the magnetization functions $\mathcal{M}$
enters.  This because in purely poloidal configurations
$\mathcal{I}=0$. $\mathcal{I}$  in fact encodes information about the
poloidal currents distribution, and is different from zero only in the
presence of a toroidal magnetic field. However, in the case of a
purely toroidal magnetic field, $\Psi=0$ and $\mathcal{M}$ can  now be
directly connected to $\mathcal{I}$. Indeed, using
Eq.~\eqref{eq:lorentz} with $\Psi=\mbox{const}$, the integrability condition in
Eq.~\eqref{eq:M} reads:
\begin{equation}
\nabla \mathcal{M}=-\frac{\mathcal{I} \, \nabla\mathcal{I}}{\rho h \alpha^2 R^2}.
\end{equation}
Defining the new quantity $G=\rho h \alpha^2 R^2$
 the previous equation can be easily integrated if we assume a barotropic-like
 dependency for the current function $\mathcal{I}=\mathcal{I}(G)$, namely
 \begin{equation}
 \mathcal{I}=K_m G^m
 \end{equation}
 where $K_m$ is the \textit{toroidal magnetization constant} and $m \geq 1$
 is the\textit{ toroidal magnetization index} (see also \citealt{Kiuchi_Yoshida08a},
 \citealt{Lander_Jones09a}, \citetalias{Frieben_Rezzolla12a}  and \citealt{Fujisawa15a}).
 With this assumption the magnetization function $\mathcal{M}$ is given by
 \begin{equation}
\mathcal{M}=-\frac{m K_m^2}{2m-1}G^{2m-1},
\label{eq:Mtor}
\end{equation}
and the magnetic field is related to the enthalpy per unit volume through
\begin{equation}
B_\phi=\alpha^{-1} K_m G^m=\alpha^{-1} K_m (\,\rho h\, \alpha^2 R^2)^m.
\end{equation}
Here the magnetization constant $K_m$ regulates the strength
of the magnetic field (more specifically the magnetic flux trough the
meridional plane), while the magnetization index $m$ is related to the distribution 
of the magnetic field inside the star. 

In this work we will consider exclusively purely poloidal or purely toroidal magnetic fields
with all the currents confined inside the star with, at most the addiction of surface terms.
In general, rotation with mixed morphologies leads to configurations where the Poynting
vector does not vanish and, hence again, the system can not be considered strictly stationary.

\section{Numerical Scheme}
\label{sec:numerical}

The numerical equilibrium models presented in this work are obtained through the 
\texttt{XNS} code. This code has been already used to compute equilibrium solutions
for static configurations with different topologies of the
magnetic field (\citetalias{Pili_Bucciantini+14a}; \citetalias{Bucciantini_Pili+15a}; \citealt{Pili_Bucciantini+15a}), 
or for rotating unmagnetized star with realistic EoSs \citep{Pili_Bucciantini+16a}.
It has also been extensively validated in the Newtonian-limit 
\citep{Das_Mukhopadhyay15a,Bera_Bhattacharya16a,Mukhopadhyay15a}.
In particular \citet{Subramanian_Mukhopadhyay15a} have verified that XNS
provides solutions that are indistinguishable, within the numerical accuracy, 
from those obtained with other numerical codes.
In this work the XNS code has been modified to handle rotating configurations
endowed with a poloidal magnetic field. For the reader convenience we briefly
summarize here its main features (see  \citealt{Bucciantini_Del-Zanna11a} and \citetalias{Pili_Bucciantini+14a}
for a complete and detailed description), while in the next subsection we will
discuss in more detail our modifications.

The XNS code solves self-consistently the Einstein-Maxwell equations system
in the case of an axisymmetric and stationary space-time under the hypothesis
of conformal flatness and maximal slicing.
The metric solver operates in the so called  
{\it eXtended Conformally Flat Condition} (XCFC) \citep{Cordero-Carrion_Cerda-Duran+09a},
which extends the CFC Eqs.~\eqref{eq:CFC1}-\eqref{eq:CFC3} in a numerical
stable form that can be solved using standard and accurate numerical techniques,
with a typical relative error $10^{-4}$ with respect to the exact GR solution 
(\citealt{Bucciantini_Del-Zanna11a}; \citetalias{Pili_Bucciantini+14a}).
The main idea at the base of XCFC is to express  $K^{ij}$ in terms of an auxiliary 
vector  $W^i$ which is related to the longitudinal part of $K^{ij}$ itself.
The transverse-traceless component of the extrinsic curvature $K^{ij}$ is instead
neglected, being  typically smaller that the non-conformal content of the 
spatial metric.

Einstein equations are turned into a system of two scalar Poisson-like Partial 
Differential Equations (PDEs), one for the conformal factor $\psi$ and 
one for the lapse function $\alpha$, and two vector Poisson-like equations, 
one for the shift vector $\beta^i$ and the other for $W^i$. Such equations
are fully decoupled so that they can be solved hierarchically  (first the one 
for  $W^i$, then $\psi$, $\alpha$ and finally $\beta^i$). Moreover they can be cast into a
form that guarantees the {\it local uniqueness}. 

In the case of rotating stars with no meridional flow (i.e. $\varv^r=\varv^\theta=0$)
only $W^\phi$, and $\beta^\phi$ are different from zero. The PDEs
associated to these two quantities have the general form:
\begin{equation}
\Delta X^{\hat{\phi}} = H^{\hat{\phi}},
\label{eq:vectorPoisson}
\end{equation}
where the $\textbf{X}=\textbf{X}(r, \theta)$ is the generic unknown vector field
 and $\textbf{H}$ is the associated source term. For $W^\phi$, the source
 term depends on $S^\phi$. For  $\omega=-\beta^\phi$ (solved at the end), 
 $\textbf{H}$ depends also on previously determined quantities.
 Analogously the PDEs for $\alpha$ and $\psi$ can be generally written as: 
 \begin{equation}
\Delta q = s q^p
\label{eq:scalarpoisson}
\end{equation}
where, again, $q=q(r,\theta)$ is the generic unknown function while $s$
is the scalar source term which depends on the stress-energy content
of the space-time and on previously computed metric terms.

Solutions to these equations are computed with a semi-spectral method.
The scalar functions, such as  $\psi$, $\alpha$ and $\Phi$, are
expanded into a linear combination of spherical harmonics $Y_\ell(\theta)$ 
\begin{equation}
q(r,\theta)=\sum_{\ell=0}^{\infty} C_\ell(r) Y_\ell(\theta),
\end{equation}
while vector quantities $W^\phi$, $\beta^\phi$ and $\Psi$ are expanded as
\begin{equation}
X^{\phi}(r,\theta)=\sum_{\ell=0}^{\infty} K_\ell(r) Y_\ell '(\theta),
\end{equation}
where $'$ stands for the derivative with respect to $\theta$ (notice
that axisymmetry excludes the harmonic degrees $m\neq 0$ ). 
Adopting a second order radial discretization, the harmonic expansion reduces each PDE to a set 
of radial ordinary  differential equations, one for each coefficient $C_\ell(r)$ (or analogously $K_\ell$),
that is solved via direct inversion of  tridiagonal matrices. The harmonic decomposition
ensures also the correct behavior of the solution on the symmetry axis, at the centre and at
the outer boundary of the domain which, in our case, is set at a finite distance from the stellar surface.
In particular at the center of the star the harmonic coefficients  $C_\ell$ and  $K_\ell$
go to zero with parity $(-1)^\ell$ and $(-1)^{\ell+1}$ respectively,  while at the outer 
radius they scale as $r^{-(\ell+1)}$. 

Remarkably Maxwell equations~\eqref{eq:MGpot}-\eqref{eq:MApot}
share the same mathematical  structure of XCFC equations:
the scalar equation for $\Phi$ is equivalent to Eq.~\eqref{eq:scalarpoisson};
the equation for  $\Psi$ (the $\phi$-component
of the magnetic potential) is a vector Poisson-like
equation analogous to Eq.~\eqref{eq:vectorPoisson} where
$H=H(\Psi,\upartial \Psi)$.

\subsection{Numerical resolution of Maxwell equations}
\label{sec:nummax}

In the case of rotating NSs with a purely poloidal field, we solve separately Eqs.~\eqref{eq:MApot} and~\eqref{eq:MGpot},
for the potential $\Psi$ and $\Phi$ respectively. Equations are  iteratively solved with the source terms given by 
Eqs.~\eqref{eq:rhomhd} and~\eqref{eq:Jmhd}. Accordingly with the vacuum assumption, the current density 
$J^{\phi}$ and the charge density $\rho_e$ are set to zero outside the star.

Since  Eqs.~\eqref{eq:MApot} and~\eqref{eq:MGpot} are solved at once on the numerical grid
both $\Psi$ and $\Phi$ extends smoothly outside the star. This is not consistent
with the fact that a magnetized rotating perfect conductor naturally acquires 
a surface charge density, which in turn manifests in the derivatives of $\Phi$.
Indeed, as anticipated in Sec.~\ref{sec:matter}, the newly obtained potential $\Phi$ does not satisfy the  perfect 
conducting relation \eqref{eq:MHDcond} inside the star but differs from the MHD solution
$\Phi_{\rm MHD}=-\Omega \Psi + C$ by an harmonic function $\Phi_a$
\begin{equation}
\Phi=\Phi_{\rm MHD}+\Phi_a \quad \mbox{with} \quad  \Delta \Phi_a=0 . 
\end{equation}
This harmonic function $\Phi_a$, can be set by requiring Eq.~\eqref{eq:MHDcond} to
hold just at the stellar surface $\mathcal{S}_{\rm NS}$:
\begin{equation}
\Phi_a |_{\mathcal{S}_{\rm NS}}= (\Phi - \Phi_{MHD}) |_{\mathcal{S}_{\rm NS}}.
\label{eq:surfacecondition}
\end{equation}
Being an harmonic function, it can be expanded in
spherical harmonics as 
\begin{equation}
\Phi_a=\sum_{l=0}^{N_\ell}  Y_\ell(\theta) \times  \begin{cases}
  a_\ell r^\ell \quad \quad\mbox{ inside the star},\\
b_\ell r^{-(\ell+1)}\quad \mbox{ outside the star}.
\end{cases}
\end{equation}
The coefficients $a_\ell$ and $b_\ell$ are found  by solving 
the  $N_\ell+1$ equations that derive from the evaluation of 
Eq.~\eqref{eq:surfacecondition} on $N_\ell+1$ collocation points located 
along the surface $\mathcal{S}_{\rm NS}$. These points
are not evenly distributed: in the case of highly oblate NSs,
a proper redistribution of the collocation points is chosen in order 
to improve the convergence and avoid aliasing effects. In particular
while for the interior solution the collocation points are mainly distributed
in the vicinity of the stellar equator (at larger radii), in the other case 
they are clustered near the pole. In addition,  points do not coincide with any of the
grid-point locating $\mathcal{S}_{\rm NS}$ but they are chosen on top of the 
super-ellipsoid that best fits the discretized  stellar surface $\mathcal{S}_{\rm NS}$. 
We have indeed verified that for largely deformed stars choosing collocation points on
the discretized surface can excite high frequency numerical noise, 
that can compromise  the accuracy with which high-$\ell$
coefficients are computed. Given the iterative nature of the algorithm,
this might compromise the overall convergence.
The equation of  the super-ellipsoid in polar coordinates is given by  
\begin{equation}
r=\left[\left(\frac{\cos \theta}{r_{\rm p} }\right)^{n_{\rm s}} +
  \left(\frac{\sin \theta}{r_{\rm eq}
    }\right)^{n_{\rm s}}\right]^{-\frac{1}{n_{\rm s}}},
\label{eq:supellip}
\end{equation}
where $r_{\rm p}$ is the polar radius, $r_{\rm eq}$ is the equatorial radius
and $n_{\rm s}$ regulates the shape of the ellipse. Interestingly in almost all cases
$\mathcal{S}_{\rm NS}$ can be approximated by a super-ellipsoid with $ 1.5 \lesssim n_{\rm s} \lesssim 3$
with an error of at most one grid point ($\lesssim 1$\%).
Once $\Phi_{\rm a}$ has been obtained, the potential $\Phi$ can be corrected in 
$\Phi_{\rm new}=\Phi-\Phi_{\rm a}$ to satisfy ideal MHD inside the star. As expected
$\Phi_{\rm new}$ is now continuous but not differentiable across the surface.

Note that the ideal MHD condition \eqref{eq:MHDcond}, does not
completely set the scalar potential $\Phi_{\rm new}$, which is still defined
minus an arbitrary constant $C$. As pointed out by \citet{Bocquet_Bonazzola+95a}, the 
constant $C$ corresponds to an arbitrary charge, that can be added to the star. 
The value of $C$ cannot be determined purely based on symmetric or equilibrium
considerations, but only based on physical arguments. 
A common choice, often done in the literature, is to require that the NS is globally
neutral, based on the physical argument that any charged body 
in space can attract charges from the surrounding to neutralize itself
\citep{Goldreich_Julian68a,Bocquet_Bonazzola+95a,Franzon_Schramm15a}. 
This argument is however justified only for non-rotating systems.
In the rotating case there is a clear ambiguity about which reference
frame to consider. It is indeed well known that a neutrally charged 
rotating NS can easily pull charges from its surface, thus creating a 
magnetosphere and charging itself \citep{Goldreich_Julian68a}.

Other different choices have been presented even if less often. One
can assume a net charge, in order to minimize the electric field,
responsible for the extraction of charges from the surface
\citep{Michel74a}, or in order to minimize the electromagnetic energy in the
space outside the star \citep{Ruffini_Treves73a}. 
All of these different choices lead to the same internal  electromagnetic field giving
the same  deformation of the matter distribution inside the NS. 
They only  differ in the structure of the external electric field, 
the surface charge, and the associated surface Lorentz force, as we will show.

In this work we consider both neutrally charged NS, minimizing 
iteratively the monopolar content of $\Phi_{\rm new}$ with the same
approach discussed in \citep{Bocquet_Bonazzola+95a}, and
electromagnetic configurations that minimize the polar Lorentz force.
The latter case is realized by simply minimizing the discontinuity of the electric
field at the pole.

Finally, it is evident from the discussion above, that  also the
magnetic flux $\Psi$ is defined minus an harmonic term. In general
this degree of freedom has been neglected, leading to a smooth
magnetic field. In principle, once one relaxes the requirement
for a smooth electric field, covariance demands that a similar
requirement  should be relaxed for the magnetic field too:
there is an arbitrary surface current that can only be fixed 
on physical arguments.

\subsection{Algorithm and Setup}

 The following points summarize the computational work-flow of the iterative procedure used in XNS:
\begin{enumerate}
\item  we start with an initial guess [the TOV solution at the  beginning \citep{Tolman39a}];
\item the  XCFC Einstein equations are solved for the metric functions;
\item  the three-velocity $\varv^{\phi}$ and the Lorentz factor $\Gamma$ are 
		obtained on top of the new metric;
\item  depending on the morphology for the magnetic field
          the Lorentz force contribution to the Bernoulli integral  $\mathcal{M}$
          is evaluated through different approaches:
        \begin{itemize}
         \item if the magnetic field configuration is purely toroidal the potential $\mathcal{M}$ is given 
          by Eq.~\eqref{eq:Mtor};
         \item  in the case of a static configuration endowed with a poloidal magnetic field, 
                  the magnetic potential $\Psi$ is obtained from Eq.~\eqref{eq:MApot} and 
                  Eq.~\eqref{eq:Jmhd}, finally determining $\mathcal{M}$;
         \item  if the magnetic field is purely poloidal but the star is rotating we iteratively solve together both the
                  Maxwell-Gauss equation Eq.~\eqref{eq:MGpot} and the Amp\`ere equation Eq.~\eqref{eq:MApot}
                  (respectively for the potentials   $\Phi$ and  $\Psi$), initializing the source terms
                  using  Eq.~\eqref{eq:Mpol} and assuming vacuum outside the star;
        \end{itemize}
\item the Bernoulli integral is finally solved via a Newton method and the fluid quantities 
         are updated in order to start a new iterative cycle; 
\item steps from (ii) to (v) are repeated until convergence to a desired tolerance is achieved.
\end{enumerate}

Numerical equilibrium configurations are obtained using 30 spherical harmonics on top of a uniform 
numerical grid in spherical coordinates covering the ranges $r=[0,30]$ (in geometrized units) 
and $\theta=[0,\upi]$, with  600 points in the radial direction and 300 points in the azimuthal one. 
The radial domain has been chosen such that its outer boundary is far enough from  the stellar surface  so
that the asymptotic behavior of the different metric terms can be imposed properly.
In some cases of strongly magnetized star with purely toroidal field, the outer boundary of the grid 
has been set to $r=50$, in order to model also highly inflated configurations. 
The convergence tolerance is fixed to $\sim10^{-8}$ (convergence is checked looking at the maximum deviation between 
successive solutions). Globally, we have verified that, because of the 
discretization errors, the overall accuracy of the final solutions is of the order of $\sim 10^{-3}$.

We want to remark here that the solutions to both  Einstein and Maxwell equations
are searched over the entire numerical grid without a matching conditions at 
the stellar surface, and this  necessitates correcting the $\Phi$ potential in the case 
of rotating poloidal magnetized NSs. We found that the correction procedure
described in the previous subsection can be realized with $N_{\ell}\sim 20$
in order to enforce Eq.~\eqref{eq:surfacecondition} to discretization accuracy.

%We want to remark here that both Einstein equations and Maxwell
%equations are solved over the entire numerical domain including the interior of the star, 
%where the source terms are fully confined, and the external vacuum space-time.
%In our scheme no matching procedure between the internal and the external vacuum solutions 
%is needed at the surface, which is defined as the locus of points where the baryonic density 
%drop below a reference value, and  whose shape is free to vary according to the solution of 
%the Bernoilli integral. For numerical reason, the baryon density outside the star 
%has been set to a  small fiducial value, i.e. $10^{-6}\rho_{\rm c}$. 

\section{Numerical Results}
\label{sec:results}

In this section we present a detailed investigation of the parameter space for the equilibrium configurations
of rotating magnetized neutron endowed either with a purely toroidal or
a purely poloidal magnetic field. 
Given that our focus is on the role of rotation and magnetic field in shaping the structure of
the NS we considered just a simple polytropic EoS $p=K \rho^{1+1/n}$, 
with an adiabatic index $n=1$ and a polytropic constant 
$K=1.601\times 10^5 \mbox{cm}^5 \mbox{g}^{-1} \mbox{s}^{-2}$
(corresponding to $K_a=110$ in geometrized units), in line with our
previous work \citepalias{Pili_Bucciantini+14a}. All stellar quantities (such as the gravitational mass $M$, 
the absolute value of the gravitational binding energy $W$, the surface ellitpicity $e_{\rm s}$,
the mean deformation $\bar{e}$ and so on) are defined as in \citet{Kiuchi_Yoshida08a}, \citet{Frieben_Rezzolla12a}
(hereafter FR12) and \citetalias{Pili_Bucciantini+14a}.

\subsection{Toroidal Magnetic Field}

In continuity with what has been done in \citetalias{Pili_Bucciantini+14a} 
we have computed a large set of equilibria
varying  the central density, the magnetization, the magnetic field profile, and the
rotation frequency. The latter is however maintained 
below $\Omega=5.08 \times 10^3 \mbox{s}^{-1}$  (corresponding to $\Omega=2.5\times10^{-2}$ in 
non-dimensional units). 
At this frequency the  space of solutions is quite narrow due to mass shedding,
mostly limited to very compact configurations, while less compact stars are 
more prone to magnetic and rotational deformation.

\begin{figure*}
	\centering
	\includegraphics[width=.99\textwidth]{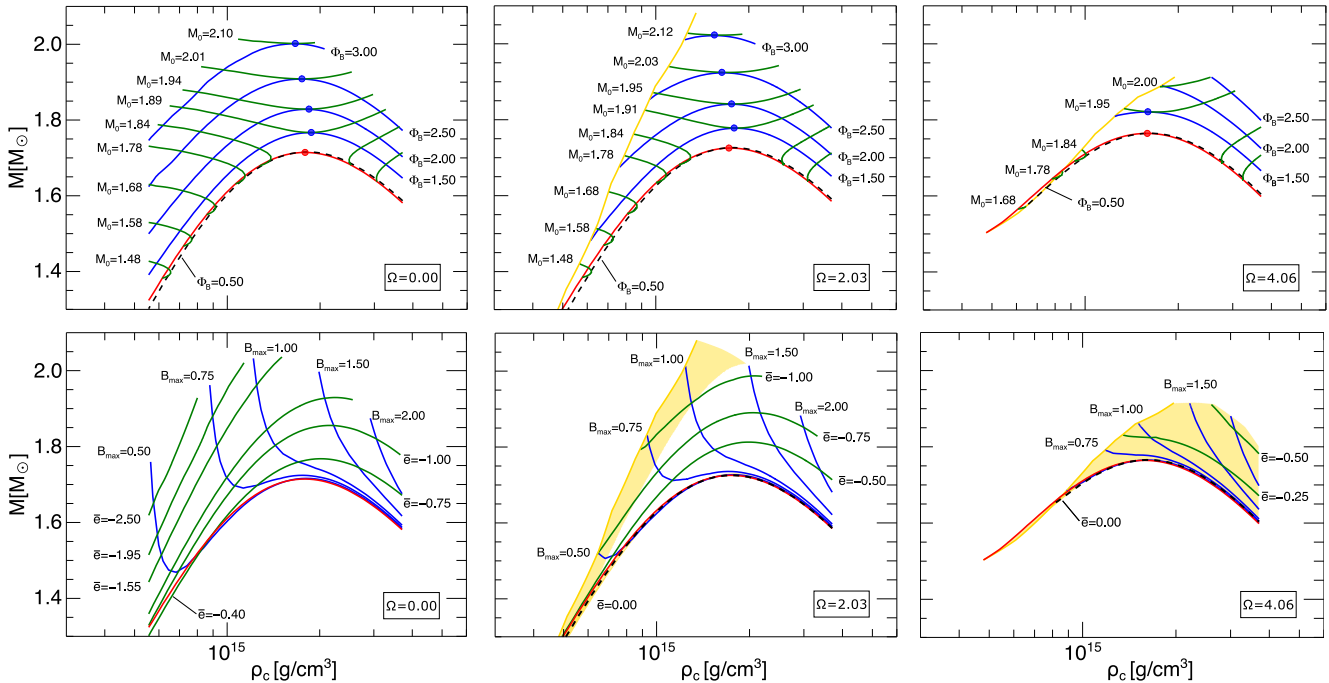}
	\caption{ Space of physical solutions for magnetized NSs with magnetic index $m=1$.
	Top row: equilibrium sequences with fixed baryonic mass $M_0$ (green lines) and fixed 
	magnetic flux $\Phi_{\rm B}$   (blue lines) for different values of the rotational rate $\Omega$. 
	The black dashed lines represent configurations with low value of the magnetic flux.  
	Bottom row: equilibrium sequences at fixed maximum magnetic field strength
	$B_{\rm max}$  (blue lines) and fixed deformation rate $\bar{e}$. Here the black dashed lines represent
	 magnetized	configurations with vanishing $\bar{e}$, while the yellow shaded regions correspond to 
	 those configurations having positive surface ellipticity. In all cases the red lines represent the 
	 unmagnetized sequences.	The baryonic mass $M_0$ is expressed in unity of $\mbox{M}_\odot$,  $\Phi_{\rm B}$ in 	
	unity of $10^{30}\mbox{G\,cm}^2$, $B_{\rm max}$ in unity of $10^{18}$~G and the rotational rate is
	expressed in $10^{3}\mbox{s}^{-1}$.
     }
	\label{fig:parsm1}
\end{figure*}
The parameter space is shown in Fig.~\ref{fig:parsm1}, for $m=1$ and different values of the
rotational rate $\Omega$. Models are given in terms of the central 
density $\rho_c$ and the gravitational mass $M$. For each value of  $\Omega$ we plot
sequences of constant baryonic mass $M_0$ and of constant magnetic flux $\Phi_{\rm B}$, and 
sequences of constant  deformation ratio $\bar{e}$ and fixed maximum 
magnetic field strength $B_{\rm max}$. 
Notice that at low central density the space of solutions is limited by two physical boundaries: 
the unmagnetized limit (red line) and the mass shedding limit (yellow line). 

\begin{figure*}
	\centering
	\includegraphics[width=.33\textwidth]{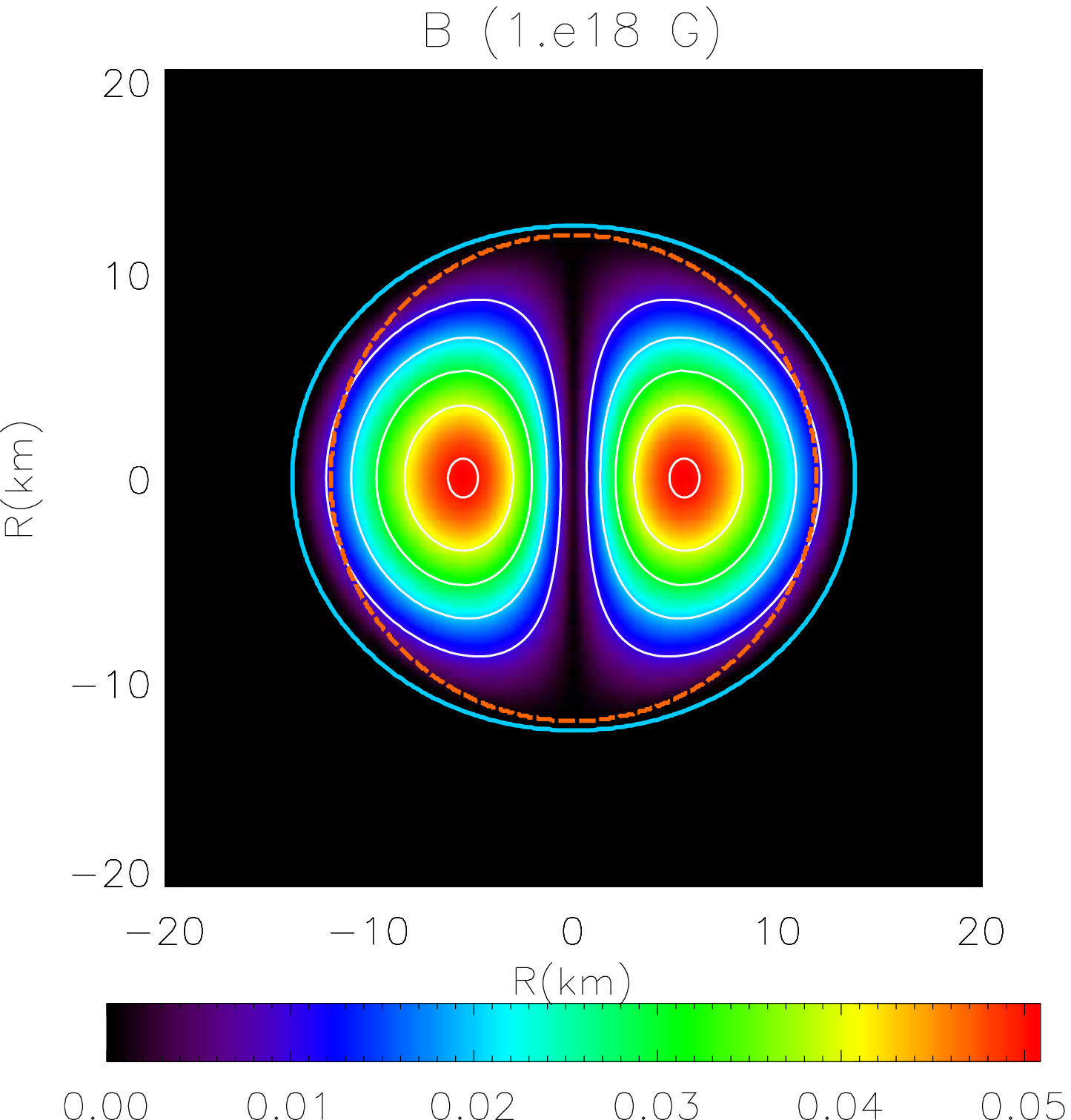}
	\includegraphics[width=.33\textwidth]{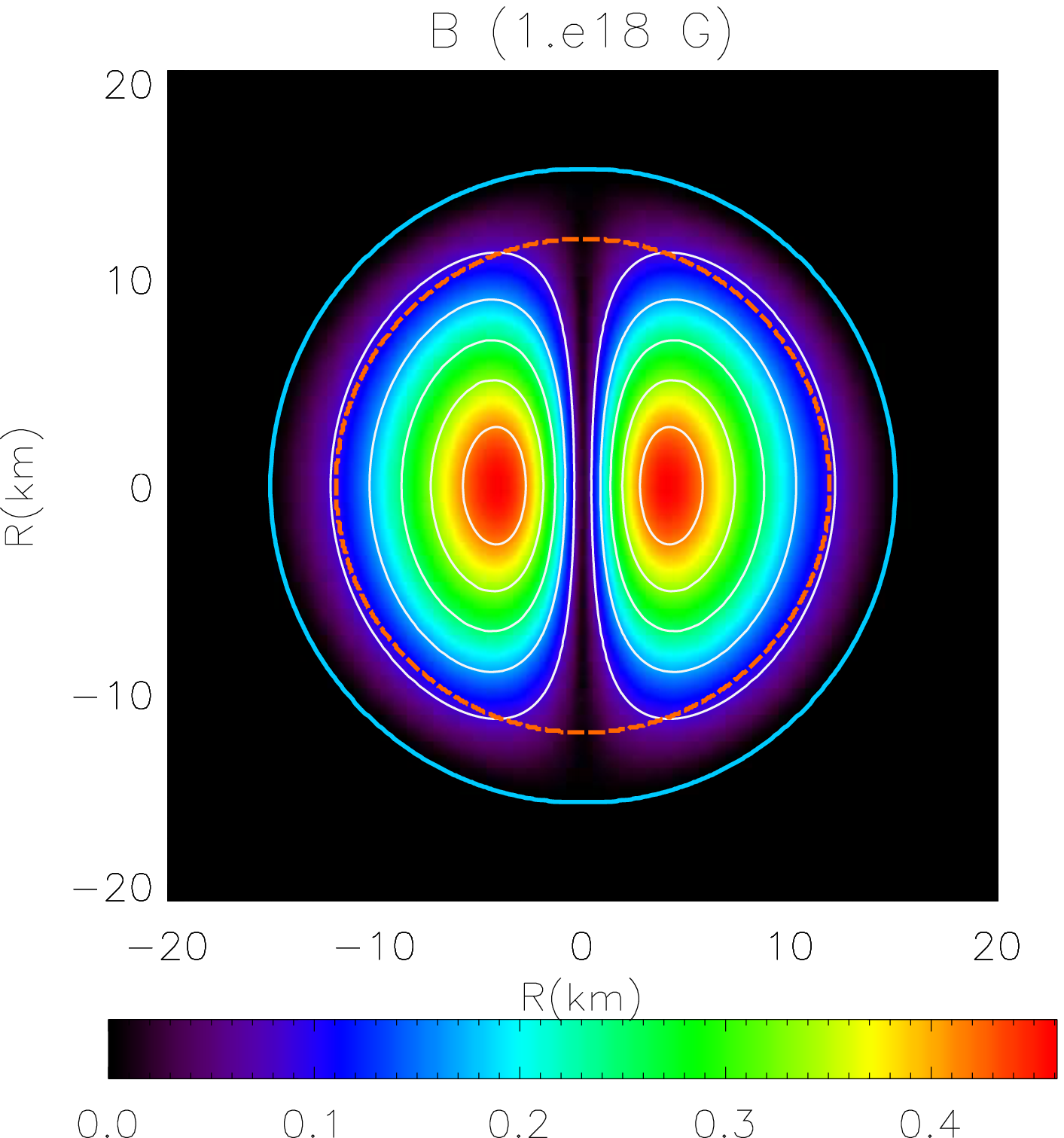}
	\includegraphics[width=.33\textwidth]{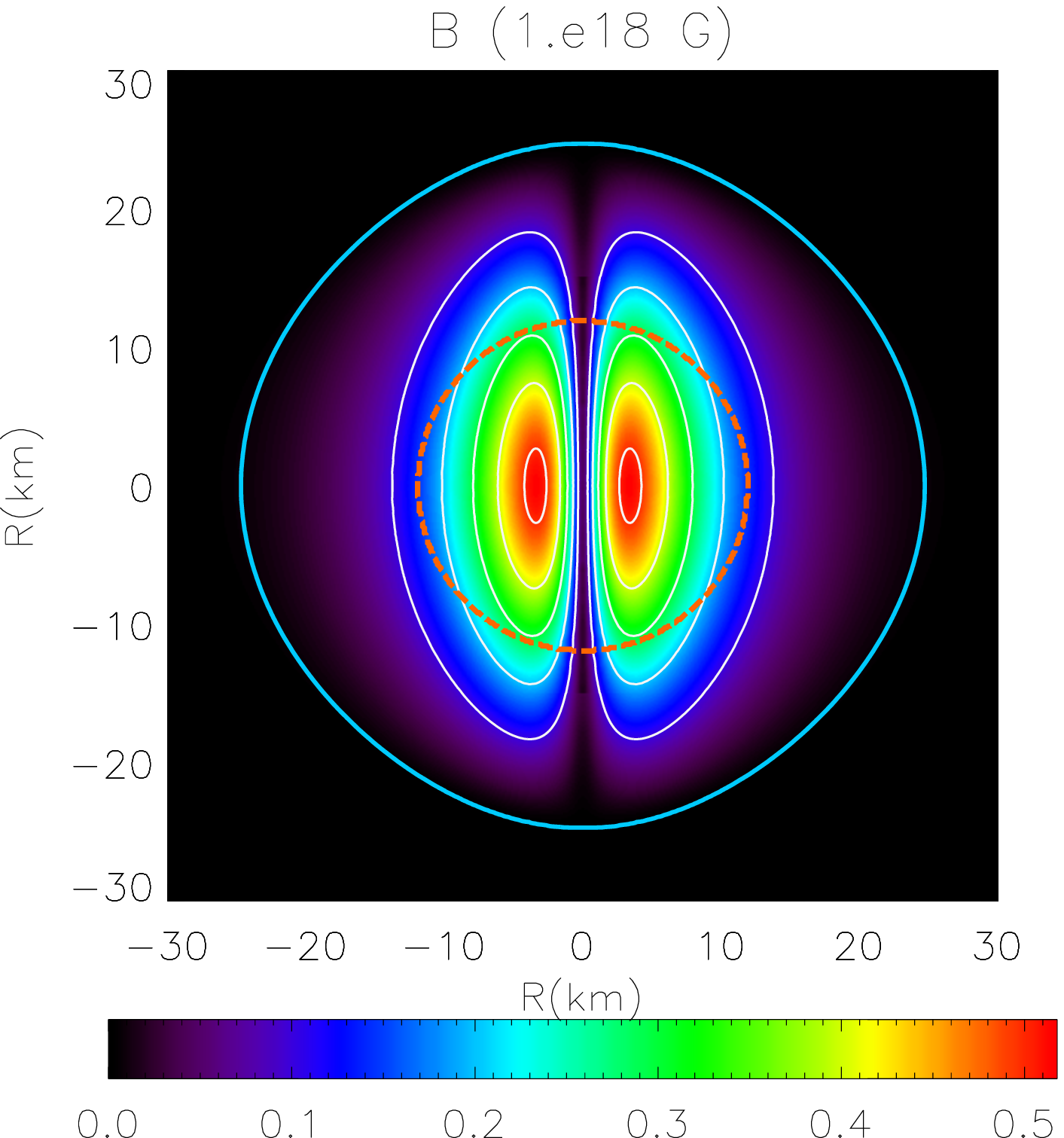}\\
    \includegraphics[width=.33\textwidth]{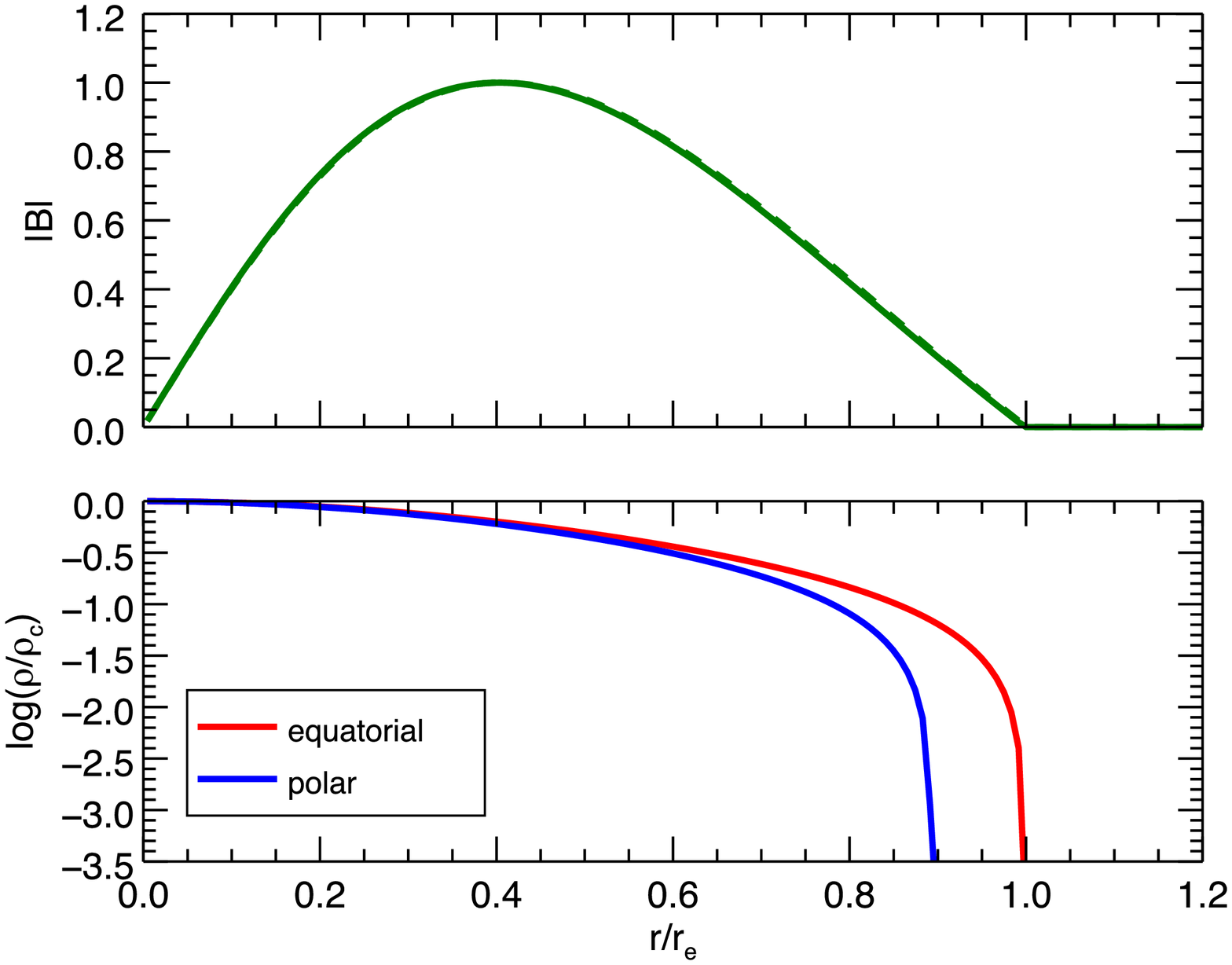}
	\includegraphics[width=.33\textwidth]{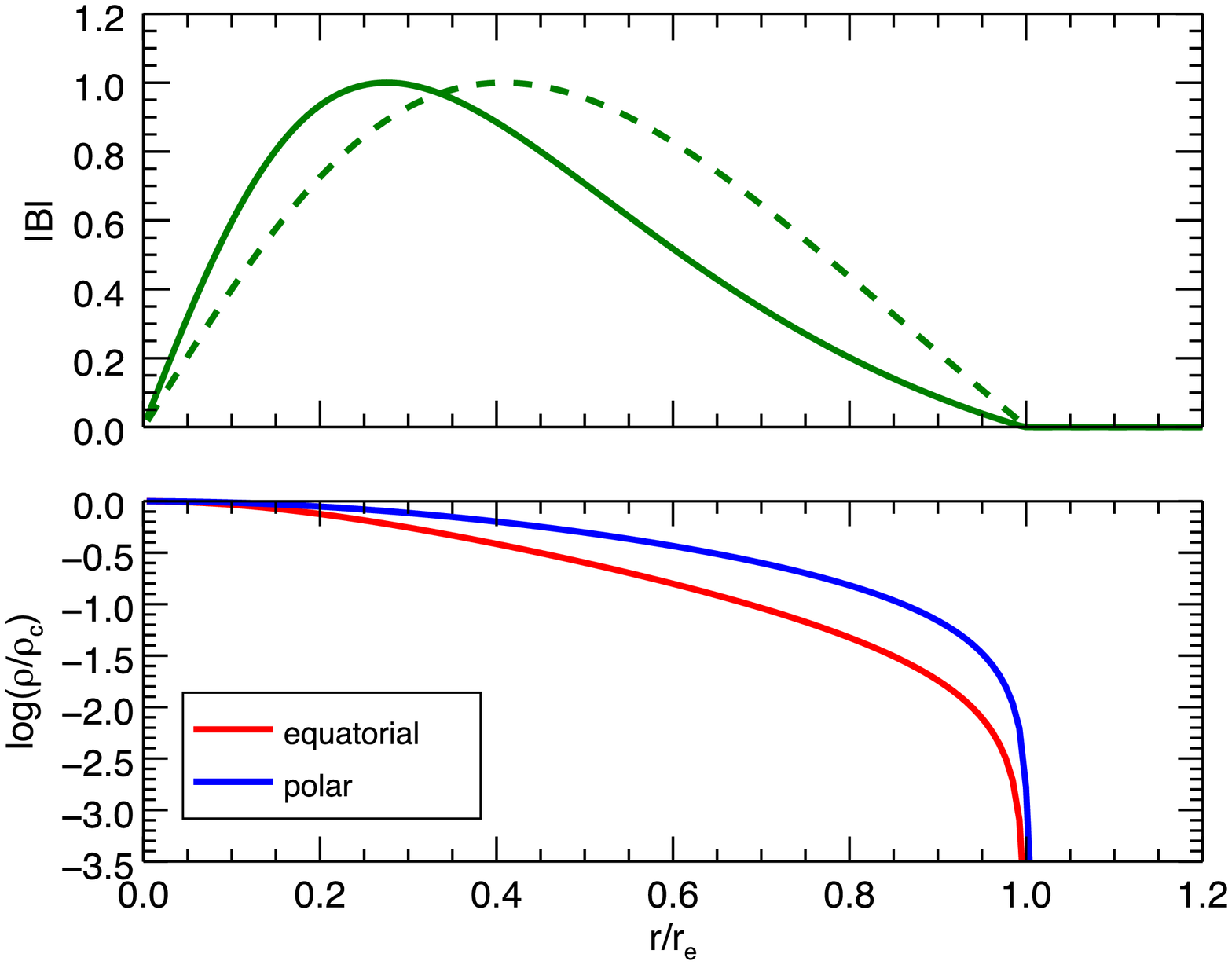}
	\includegraphics[width=.33\textwidth]{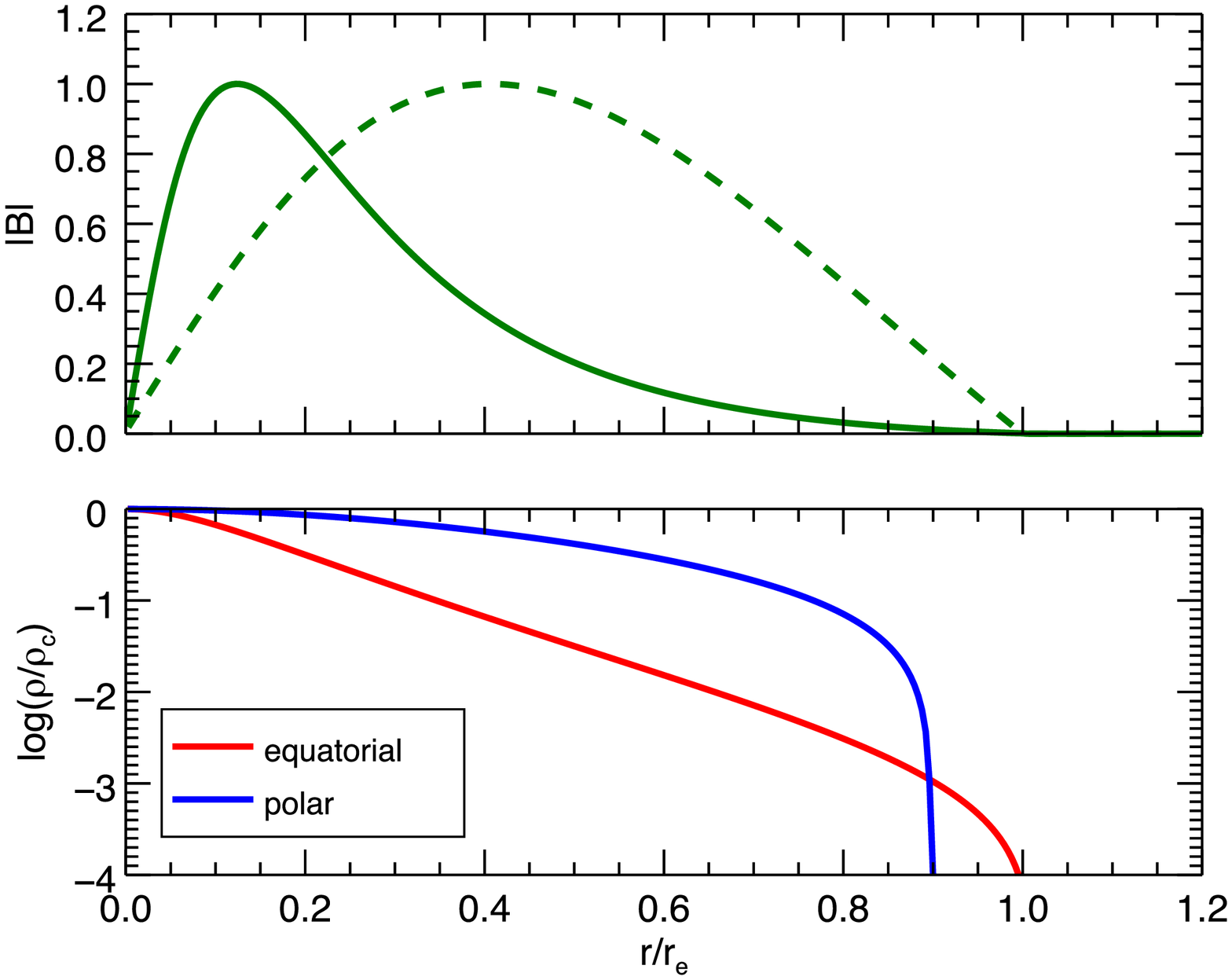}
	\caption{ Top panels: distribution and isocontours of the magnetic field strength $B=\sqrt{B^{\phi}B_{\phi}}$ 
	for equilibrium  configurations with the same gravitational mass	 $M=1.55 \mbox{M}_\odot$ but
    different rotational rate: $\Omega=3.05\times10^{3}\mbox{s}^{-1}$ (left panel),  
    $\Omega=2.03\times10^{3}\mbox{s}^{-1}$ (central and right panel). 
    The blue lines represent the stellar surface while the dashed lines represent the surface
    of the non rotating and unmagnetized model with the same gravitational mass.
	Bottom panels: profiles of the magnetic field strength and of the baryon density normalized to the peak value along 
	the equatorial direction $\theta = \upi/2 $. The green dashed lines represent the magnetic field distribution in the 
	weak magnetization limit ($B_{\rm max} \lesssim 10^{16}$~G). Radii are normalized to the equatorial radius $r_{\rm e}$. 
	Global physical quantities for these equilibrium  configurations are listed in Tab.~\ref{tab:tableconfm1}.
              }
	\label{fig:conf155m1}
\end{figure*}
\begin{table*}
\centering
\caption{ \label{tab:tableconfm1}
Global quantities for the configuration shown in Fig.~\ref{fig:conf155m1} with gravitational mass $M=1.55 \mbox{M}_\odot$.
}
\begin{tabular}{l*{11}c}
\toprule
\toprule
$B_{\rm max}$ & $\Omega$ & $\rho_{\rm c}$ & $M_0$ & $R_{\rm circ}$ & $r_{\rm p}/r_{\rm e}$ & $\bar{e}$ & $n_{\rm s}$ & $H/W$ & $T/W$ & $H/M$ & $T/M$  \\ 
$10^{17}$~G & $10^3\mbox{s}^{-1}$ & $10^{14} \mbox{g}/\mbox{cm}^3$ & $\mbox{M}_\odot$ & km &  & &  & $10^{-1}$ & $10^{-1}$ & $10^{-2}$ & $10^{-2}$  \\
\midrule
0.51 & 3.05 & 7.31 & 1.67 & 15.6 & 0.90  &  0.09 & 2.00 & 0.01 & 0.25 & 0.02 & 3.81 \\ 
4.63 & 2.03 & 8.24 & 1.66 & 16.9 & 1.00  & -0.23 & 2.00 & 1.03 & 0.12 & 1.32 & 0.16  \\ 
5.20 & 2.03 & 6.63  & 1.62 & 28.9 & 0.90 & -0.79 & 1.76 & 2.30 & 0.19 & 2.44 & 0.21  \\ 
\bottomrule
\end{tabular} 
\end{table*}
Let us begin by discussing the general interplay between rotation and magnetic field. 
As already pointed out by \citetalias{Frieben_Rezzolla12a}, depending on the strength of
the magnetic field versus the rotation rate, we can obtain three main typologies of
deformation, as shown in Fig.~\ref{fig:conf155m1} and Tab.~\ref{tab:tableconfm1}:
if the rotation is dominant the resulting configuration is purely oblate with both positive 
surface ellipticity $e_{\rm s}$ and mean deformation rate $\bar{e}$ 
(left panel of Fig.~\ref{fig:conf155m1}); conversely, if the effects due to the magnetic field  prevail,
the star is purely prolate with $\bar{e}<0$ and $e_{\rm s}<0$, or at most $e_{\rm s}=0$
(central panel of Fig.~\ref{fig:conf155m1}); for intermediate configurations, when
magnetization and rotation can counterbalance, the morphology of the star is only apparently oblate,
with $e_{\rm s}>0$, but globally prolate, with $\bar{e}<0$ (right panel of Fig.~\ref{fig:conf155m1}). 

While  rigid rotation flattens the star toward the equatorial plane
(increases its {\it oblateness}), the Lorentz force squeezes the star in the 
direction of the magnetic axis inflating also the outer stellar layers
(increases its {\it prolateness}). Moreover while the rotation acts mainly in the 
outer region, where the specific rotational energy is larger, the magnetic field 
affects mainly the inner regions where its strength peaks. As a result, close to mass
shedding, the star shows a peculiar diamond-like shape. In general the
surface  can be approximated by standard ellipsoids  with $n_{\rm s}$ in the
range $1.9-2.1$. It is only for these peculiar diamond-shaped surfaces that the
super-ellipsoid index lowers to $\sim 1.7$.

As already discussed in \citetalias{Pili_Bucciantini+14a} the gravitational mass of an equilibrium configuration
generally grows with the magnetic flux $\Phi_{\rm B}$. This trend is reversed only in a small region of the 
parameter space characterized by weak magnetization ($\Phi_{\rm B} \lesssim 5\times 10^{29} \rm{\,G \, cm}^2$)
and low central density ($\rho \lesssim 10^{15}  \rm{\, g/cm}^3$), that is present also in the case 
of fast rotators. Moreover along sequences with fixed $\Phi_{\rm B}$,  the  configurations with the higher 
gravitational mass are very close to those having the higher value for the baryonic mass $M_0$ 
(they are coincident within the approximation of our code as can be seen in Fig.~\ref{fig:parsm1}).
These trends remain qualitatively unchanged also in the case of rotation.
However, comparing the same sequences for different values of the  rotational frequency $\Omega$, 
it is clear that also the rotation contributes to increase both the gravitational and the baryonic mass.  
For example the maximum mass on  the unmagnetized sequences $\Phi_{\rm B}=0$, 
changes from $M=1.710 \mbox{M}_\odot$ in the non rotating case to  $M=1.805 \mbox{M}_\odot$ for 
$\Omega = 5.3 \times 10^3\mbox{s}^{-1}$, while the related central density drops by $\sim 30\%$ 
(and from $M=1.770 \mbox{M}_\odot$ in the non rotating case to  
$M=1.820 \mbox{M}_\odot$ for $\Omega = 4.2\times 10^{3}\mbox{s}^{-1}$ if 
$\Phi_{\rm B}=2.0\times 10^{30}\mbox{G\,cm}^2$). The increase of
the gravitational and the baryonic mass, at a given central density, is a simple volume effect and
it is mainly linked to the growth of the stellar radius caused by the rotation and the magnetic
field. The magnetic energy  and the rotational energy indeed contribute together to
the value of the gravitational mass for at most few percents.

The locus of points with $\bar{e}=0$ is shown in the bottom panels of Fig.~\ref{fig:parsm1}
as a black dashed line. Configurations that are found below this line, characterized by weaker 
magnetization, have $\bar{e}>0$. In the same  figure we also show the region of the parameter 
space where the equatorial radius of the star is larger than the polar one: $e_{\rm s}>0$. 
This yellow shaded region includes not only the purely oblate configurations with $\bar{e}>0$
but also strongly magnetized  equilibria  located in proximity of the mass shedding limit.
Notice that for rapid rotators with  $\Omega \gtrsim 3 \times 10^3 \mbox{s}^{-1}$ almost all
the obtained equilibria appear as oblate ellipsoids. Finally, at higher magnetization the mass 
shedding limit occurs at higher densities with respect to the non-magnetized case. 
This happens because the toroidal magnetic field  significantly expands and rarefies the outer 
layers of the star making them  volatile to centrifugal effects.
\begin{figure*}
	\centering
	\includegraphics[width=.33\textwidth]{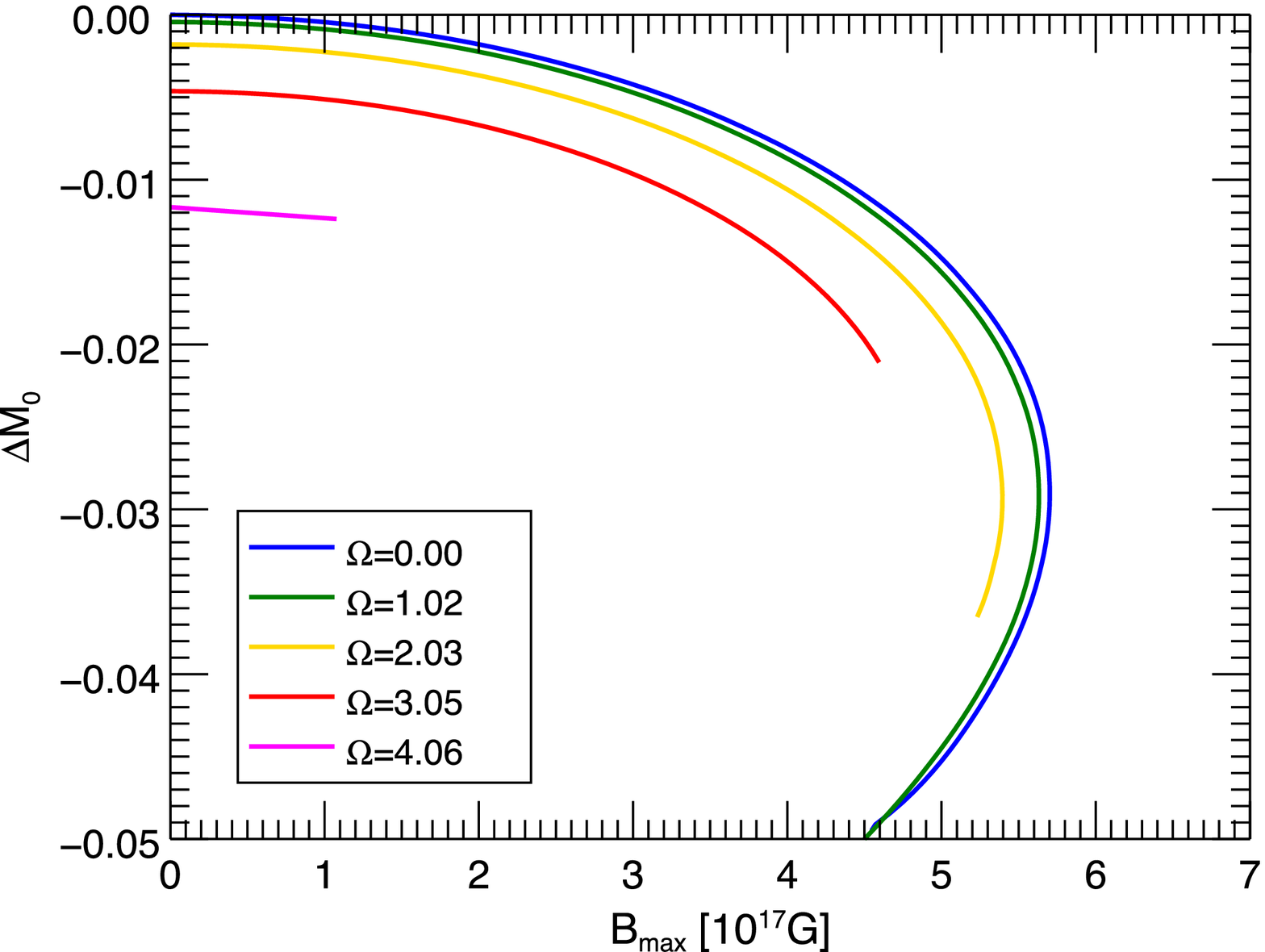}
	\includegraphics[width=.33\textwidth]{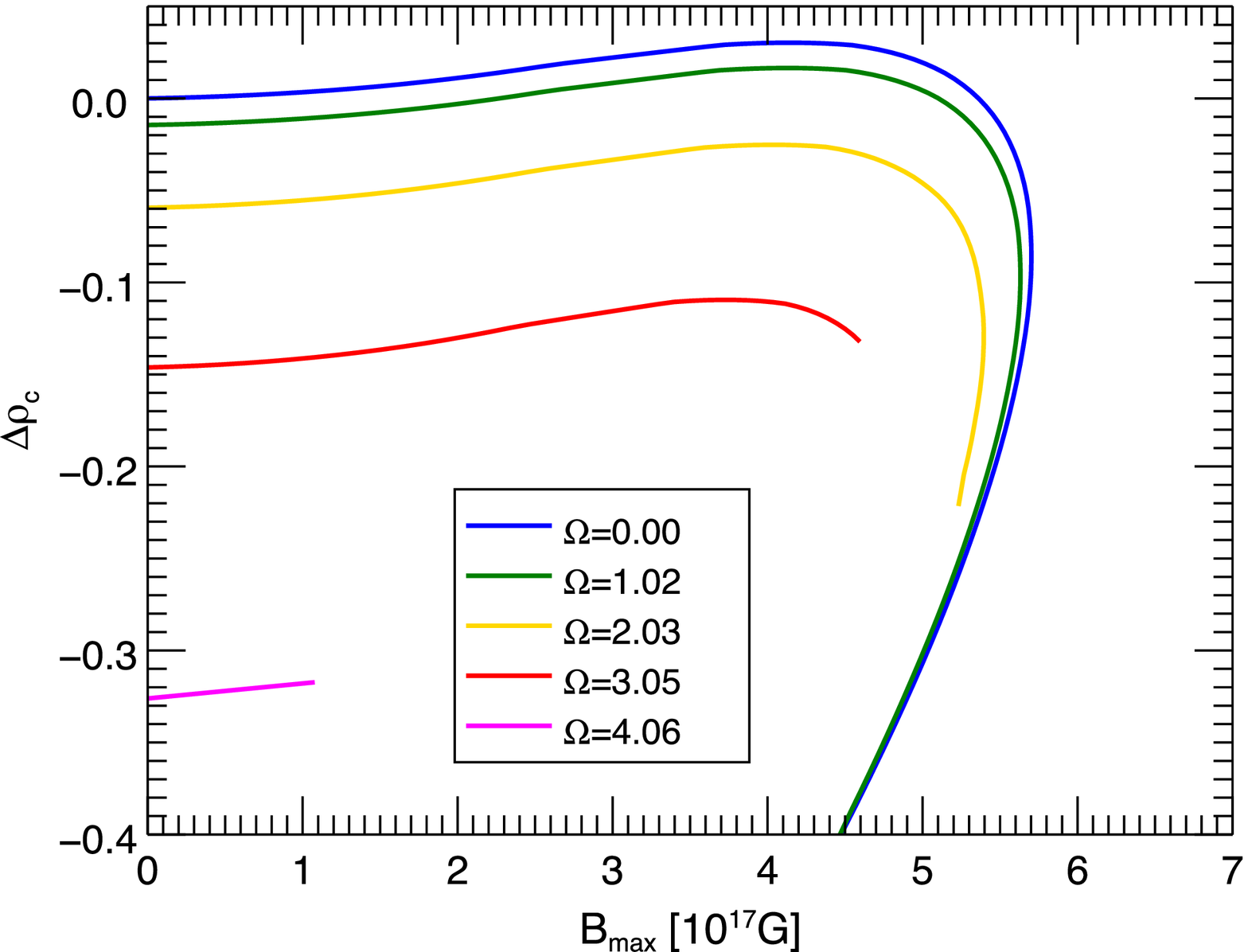}
	\includegraphics[width=.33\textwidth]{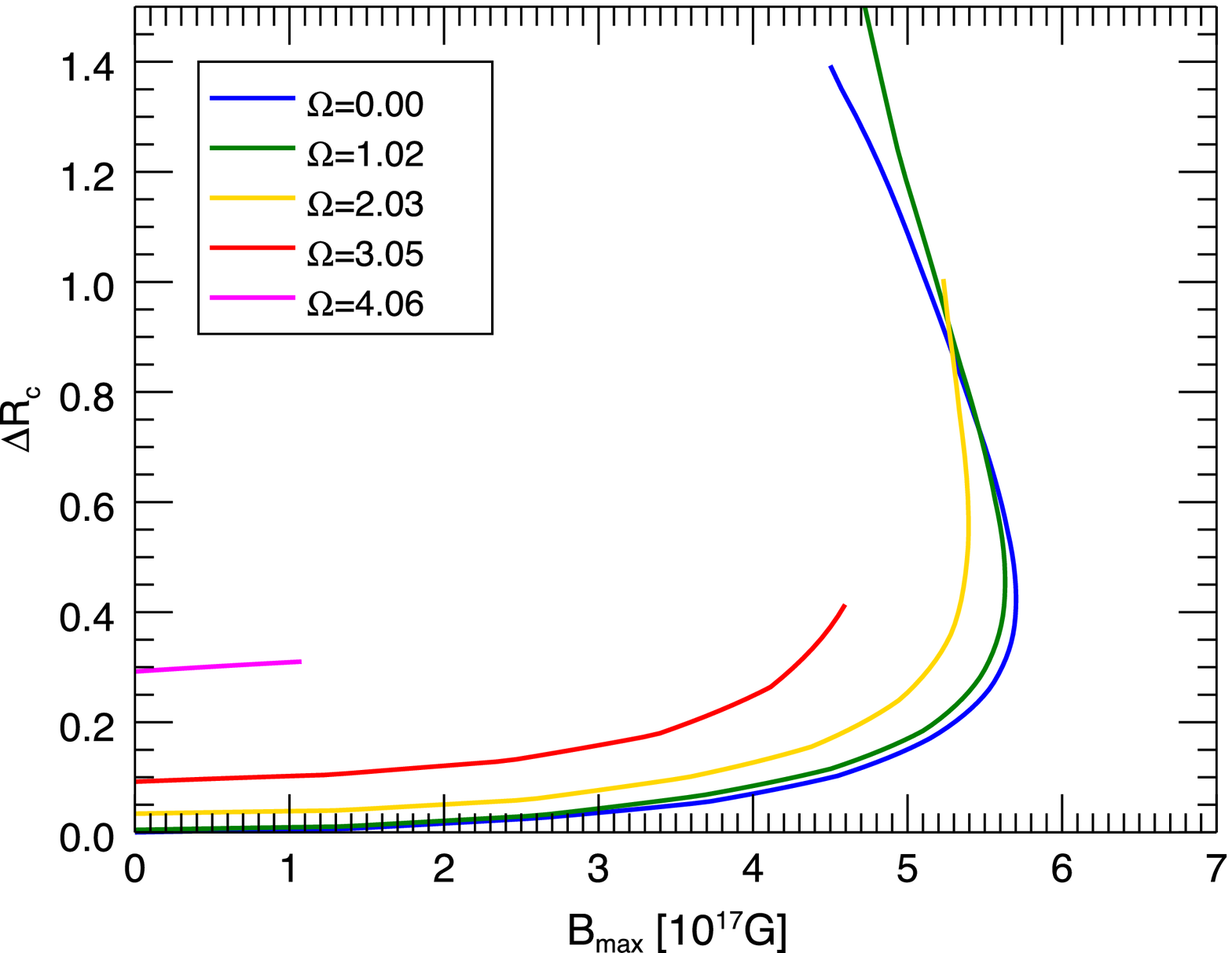}
	\caption{ Variation with respect to the unmagnetized and non-rotating reference model of the 
	baryon central density $\rho_{\rm c}$, of the baryon mass $M_0$ and of the circumferential 
	radius $R_{\rm circ}$ along the equilibrium sequences with fixed gravitational mass 
	$M=1.55 {\rm M}_\odot$, fixed magnetization index $m=1$ but with different $\Omega$ expressed in unity of $10^{3}\,\mbox{s}^{-1}$.
              }
	\label{fig:quantities155m1}
\end{figure*}

\subsubsection{Results at fixed gravitational mass and  magnetic
  polytropic index}

We present in this subsection a detailed analysis of models at fixed
gravitational mass, and fixed magnetic polytropic index $m$. For
simplicity, and for consistency with previously published results
(\citetalias{Pili_Bucciantini+14a}; \citetalias{Frieben_Rezzolla12a}), we will focus on
models with $M=1.55 \mbox{M}_{\odot}$ and $m=1$. We want to stress
however that the results we found, from a qualitative point of view, apply also to
cases with different $m$. Quantitative differences with respect to the distribution of
magnetic field will be discussed in the next subsection.

In analogy to what has been done in \citetalias{Pili_Bucciantini+14a}, to which we refer for a
discussion on how deviations of the various quantities are defined, in 
Fig.~\ref{fig:quantities155m1} we show the variation of the central density $\rho_c$,
of the baryonic mass $M_0$ and of the circumferential radius $R_{\rm c}$ as a function 
of  $B_{\rm max}$ and  for different values of $\Omega$. As expected, 
as the rotational rate increases, the central density together with the baryonic mass drops,
while the equatorial circumferential radius $R_{\rm c}$ expands. 

Interestingly,  the effect of the magnetic field along these equilibrium sequences is qualitatively 
independent from the specific value of $\Omega$, tracing the same behaviors of  static 
equilibria discussed in in \citetalias{Pili_Bucciantini+14a}: for small values of the magnetization parameter $K_{m}$, 
corresponding to $B_{\rm max}\lesssim 4\times 10^{17} \mbox{G}$,
the magnetic tension compresses the core of the star causing a growth in 
the central density $\rho_{\rm c}$; at higher values of  the magnetization the 
effect of the magnetic pressure becomes dominant and the outer layers 
of the star expand, while the central density starts to drop and the 
field strength reaches a maximum. The rotation acts in two ways: it produces an offset, which can be
safely computed for unmagnetized models, and it increases  the effectiveness of the magnetic
field (at higher rotations a lower value of $B_{\rm max}$ is required to achieve the
same deviation). Indeed we verified that the curves can be superimposed 
adding an offset corresponding to the unmagnetized rotators ($B_{\rm max}=0$), 
and slightly rescaling  the magnetic field with $\Omega$. At higher rotation rate, when the unmagnetized 
star is already close to the {\it mass shedding limit}, the magnetic field  can be increased only marginally, 
and the magnetic non-linear regime of highly inflated stars is never reached. 
We found that while at $\Omega=2\times 10^3 \mbox{s}^{-1}$ 
the star reaches the mass shedding when $B_{\rm max}=5.1\times 10^{17}$~G, and
$H/M = 0.0245$, at  $\Omega=4 \times 10^3 \mbox{s}^{-1}$ this happens for 
$B_{\rm max}=1.1\times 10^{17}$~G, and $H/M = 0.001$.

Another interesting parameter, that describes the joint effect of magnetic field and rotation
is the apparent ellipticity. We find that as the rotation rate increases, this shows a peculiar trend.
At low magnetization, the surface shape is always oblate, as expected for an unmagnetized
rotator.  As the magnetic field increases,  the oblateness diminishes 
and the shape can become prolate (this happens only for $\Omega<2.5 \times
10^3 \mbox{s}^{-1}$). Then as the magnetic field begins to inflate the
outer layer of the star, the local centrifugal support is enhanced,
and the star becomes oblate again (we observe this already at
$\Omega=10^3 \mbox{s}^{-1}$).  It was suggested by \citetalias{Frieben_Rezzolla12a}, that
at the mass shedding all models show apparent oblateness: $R_{\rm eq} > R_{\rm pol}$.
This is consistent with our findings. For example models with
$\Omega=10^3 \mbox{s}^{-1}$ have $e_s=-0.02$ at mass shedding. At lower rotation rates
the accuracy with which we sample the surface, does not allow us to
draw any conclusion. However this second transition toward apparent
oblateness, only takes place in a range of magnetic energy extremely
close to the threshold for mass shedding, when the circumferential
radius almost doubles its size.  

\begin{figure*}
	\centering
	\includegraphics[width=.33\textwidth]{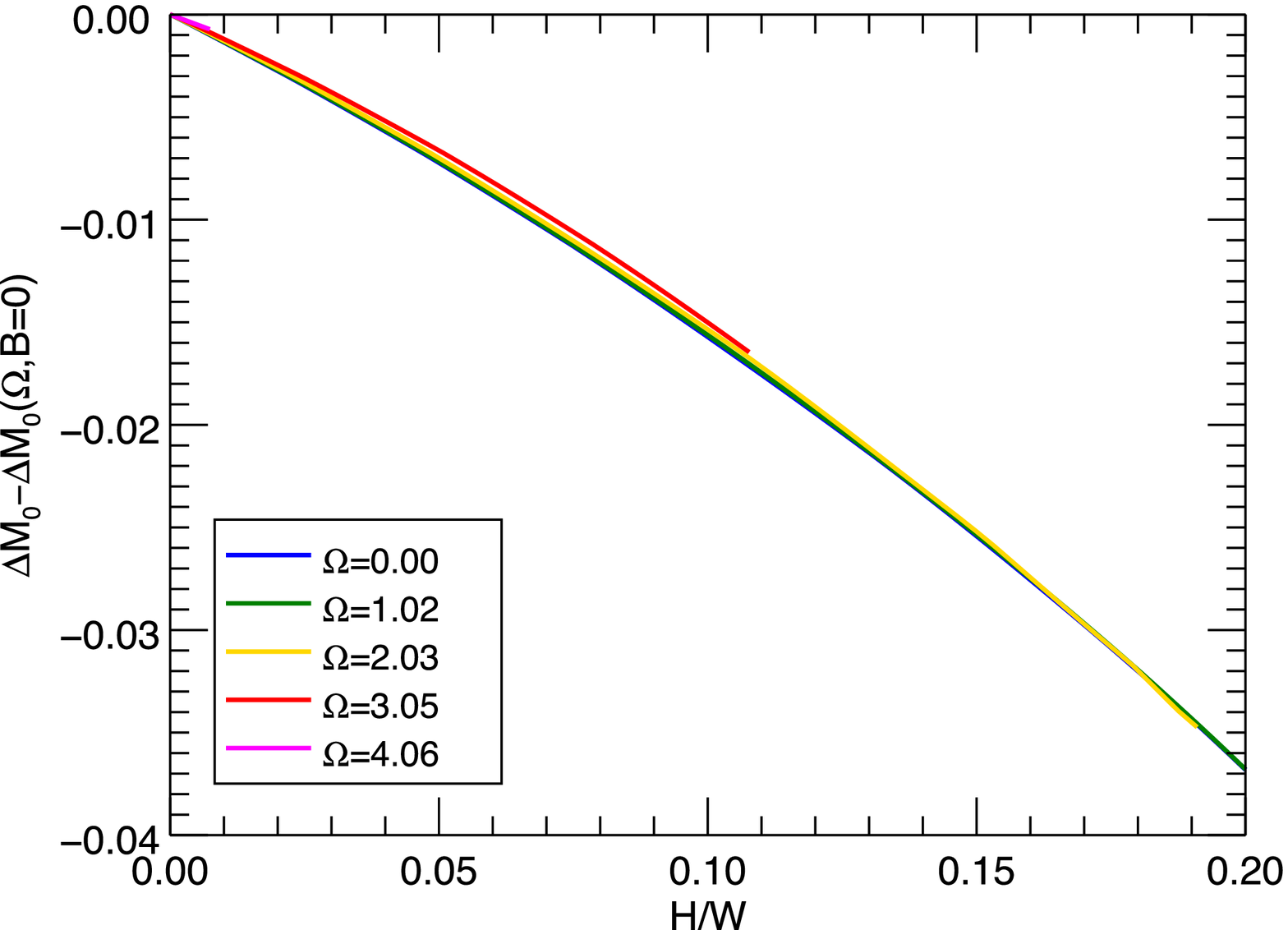}
	\includegraphics[width=.33\textwidth]{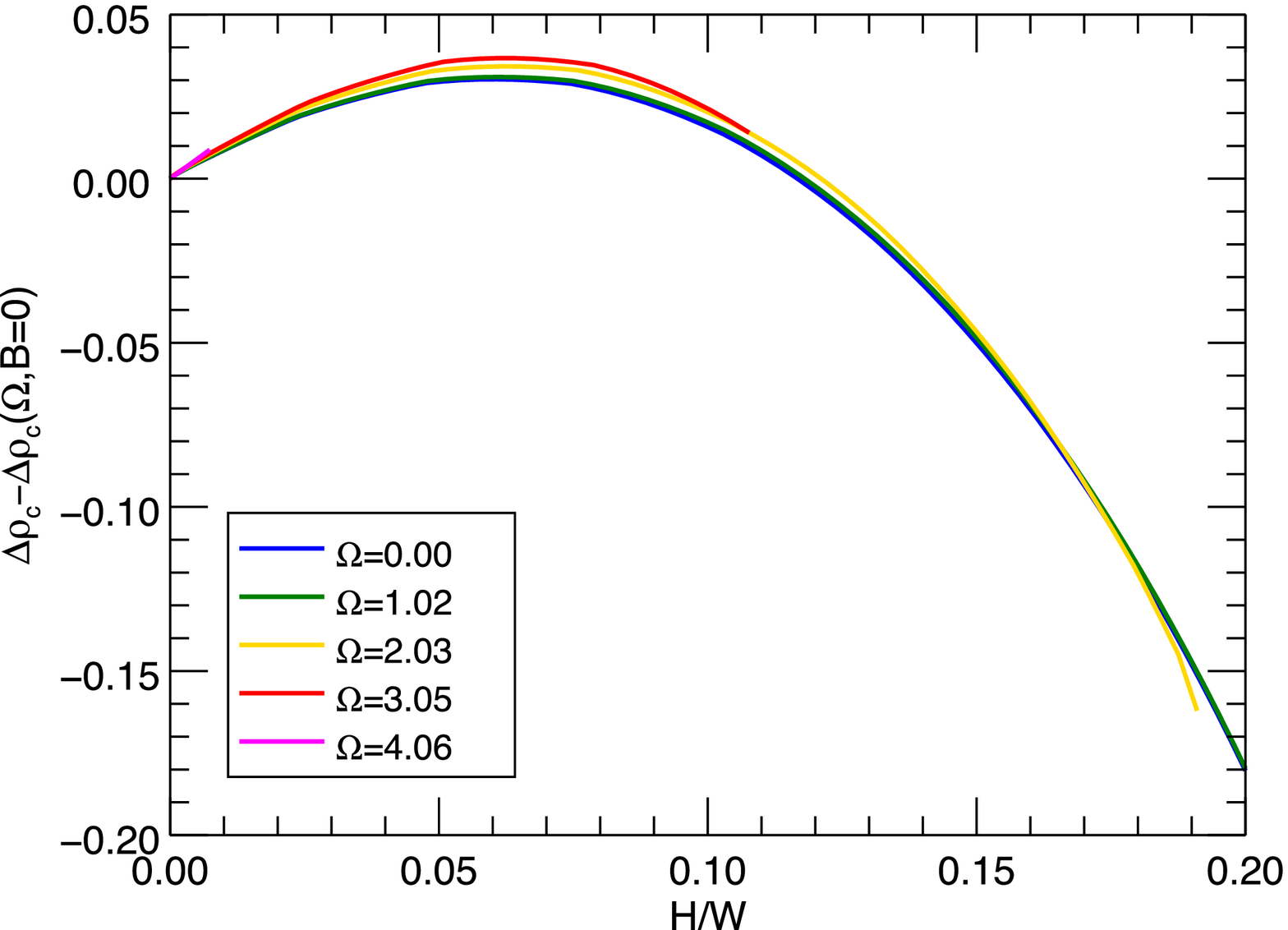}
	\includegraphics[width=.33\textwidth]{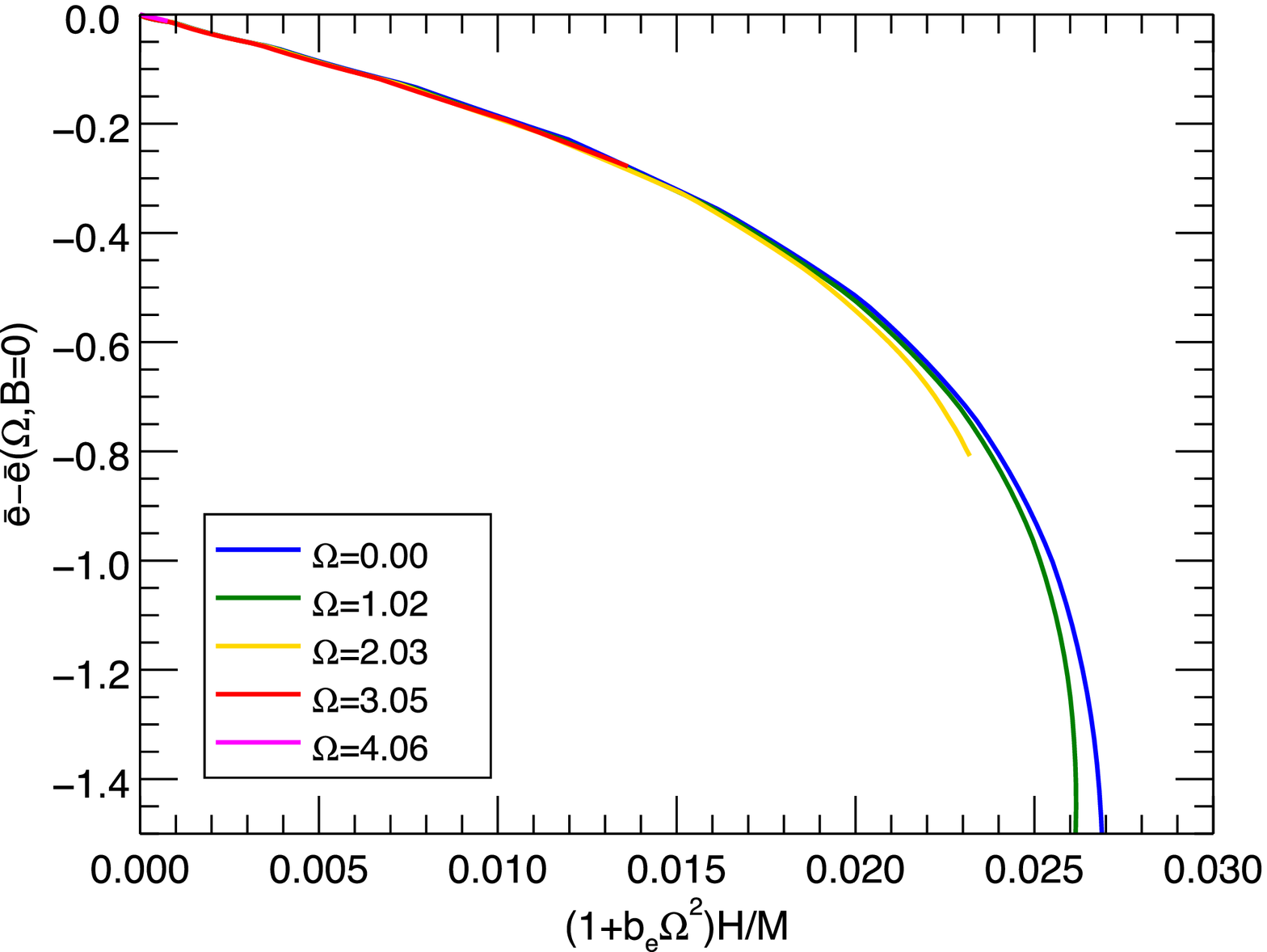}
	\caption{Variation of the baryonic mass, of the central density and of the deformation rate $\bar{e}$ with
	respect to the unmagnetized rotating reference model, as a function of the
	magnetic energy to binding energy ratio $H/W$ or to gravitational mass $H/M$.
    The equilibrium sequences with different $\Omega$ are computed holding constant the gravitational 
    mass $M=1.55 {\rm M}_\odot$.
    }
	\label{fig:def155m1}
\end{figure*}

The trends shown in Fig.~\ref{fig:def155m1} suggest that it should be
possible to find how global quantities change with similarity variable,
such as the energy ratios $H/M$ and $H/W$. In the Newtonian case, 
in the limit $\Omega,B_{\rm max}\rightarrow 0$, it can be shown that 
deviations should scale bilinearly in $H/W$ and $T/W$ \citep{Cutler02a}. 
These ratios represent, in fact, the relative energies of terms leading to 
deformations (magnetic field and rotation) with respect to the gravitational 
binding energy that tends to sphericize the star.  We find for example that 
quantities like the baryon mass variation $\Delta M_0$,  the change in central density $\Delta \rho_{\rm c}$
or in  circumferential radius $\Delta R_{\rm circ}$, and the deformation ratio
$\bar{e}$ can be fitted at fixed $M$ and $m$ for all values of $\Omega$ and 
$B_{\rm max}$ as
\begin{equation}
\Delta q = \mathcal{G}_q(\Omega, H=0)+\mathcal{F}_q\left([1+a_q\Omega^2_{\rm ms}]H/W\right),
\label{eq:ss1}
\end{equation}
where the subscript $q$ stands for a generic stellar quantities,
as shown in Fig.~\ref{fig:def155m1}. Here $\Omega_{\rm ms}$ is the rotation rate in units of 
the frequency of a millisecond rotator. The function $\mathcal{F}$ is linear in $H/W$ in the 
limit $H\rightarrow 0$, as expected, while  $a_q$ represents the non-linear coupling between magnetic
field and rotation. This non-linear coupling term is small because rotation and magnetization cannot 
be increased independently in an arbitrary way, due to mass shedding. In the case of $\bar{e}$,
we find $a_{\bar{e}}=-0.96$. %$a_{\bar{e}}=-0.1$.
Within this parametrization the role of rotation is completely factored out
in $\mathcal{G}$ (even if the kinetic energy enters also in the
definition of $W$), which is linear in $\Omega^2$ in the limit
$\Omega\rightarrow 0$. It is remarkable that this self-similar scaling holds for
highly deformed star in the full GR regime.

Interestingly, the deformation $\bar{e}$ can be fitted equivalently in terms of $H/M$ as:
\begin{equation}
\bar{e} = \mathcal{G}_{\bar{e}}(\Omega,
H=0)+\mathcal{B}_{\bar{e}}\left([1+b_{\bar{e}}\Omega^2_{\rm ms}]H/M\right),
\label{eq:ss2}
\end{equation}
with $b_{\bar{e}}=-0.48$, 
as shown in Fig.~\ref{fig:def155m1}. 
This reflects the fact that the non-linear coupling between
rotation and magnetic field is negligible.

The bilinear approximation for the deformation is found to hold,
with an error $<10\%$ in the range $H/M \lesssim 0.01$,
$T/M\lesssim 0.006$ (equivalently
$H/W\lesssim 0.07$, $T/W\lesssim 0.04$ or $B_{\rm
max}\lesssim 4\times 10^{17}$~G, $\Omega \lesssim 10^{3} \mbox{s}^{-1}$), where one can write:
\begin{equation}
\bar{e}\simeq - d_B \, B_{17}^2 + d_\Omega \, \Omega_{\rm ms}^2,
\label{eq:param1}
\end{equation}
with $d_B\simeq 9\times 10^{-3}$ and $d_\Omega \simeq 0.3$
($B_{17}$ is the maximum magnetic field strength in units $10^{17}$~G).
In this linear regime the effects induced by rotation and magnetic field
on the global deformation of the star $\bar{e}$ cancel if 
$B_{17} \simeq 6 \, \Omega_{\rm ms}$, corresponding to  $H/T\simeq 1.2$.

In the same limit $\Omega, B_{\rm max}\rightarrow 0$, also the
apparent ellipticity of the surface can be fitted with a similar bilinear
dependence:
\begin{equation}
e_S\simeq - s_B \,B_{17}^2+ s_\Omega\, \Omega_{\rm ms}^2,
\label{eq:parames1}
\end{equation}
with $s_B\simeq 2.5\times 10^{-3}$ and $s_\Omega \simeq 0.4$.
Notice that, concerning the apparent ellipticity,  the effects induced by rotation and magnetic field
cancel at   $B_{17} \simeq 13 \Omega_{\rm ms}$, corresponding to a ratio of magnetic to 
kinetic energy $H/T\simeq 6.8$. This is about a factor 5 higher than for $\bar{e}$,
indicating that apparently oblate stars can have a matter distribution with a net prolate quadrupole.

It is evident that in terms of quadrupolar deformation, the contribution of the
magnetic energy is analogous (even if acting in the opposite way) to
that of the rotational energy, and the two tend to compensate each
other close to equipartition, while rotational energy  is slightly more
efficient in determining the shape of the surface. This because
magnetic field tends to act in the interior, while rotation mostly
affects the outer layers. Note that while formally a parametrization in
terms of $\Omega^2$ and $B_{\rm max}^2$ is equivalent to one in $T$
and $H$, the latter holds with the same accuracy for a $\sim 50\%$ larger 
range of magnetic field strengths and rotation rates.

\subsubsection{The role of magnetic field distribution at fixed
  gravitational mass}

Let us now discuss the effects of different magnetic field
distributions parametrized by the magnetization index $m$. 
A comparison in the  non rotating case was already presented in
\citetalias{Pili_Bucciantini+14a}. Here some of the results are reviewed in view of the trends
found in the previous sub-section. We recall that the self
similarity scalings found previously apply also to other values of $m$.

\begin{figure*}
	\centering
	\includegraphics[width=.33\textwidth]{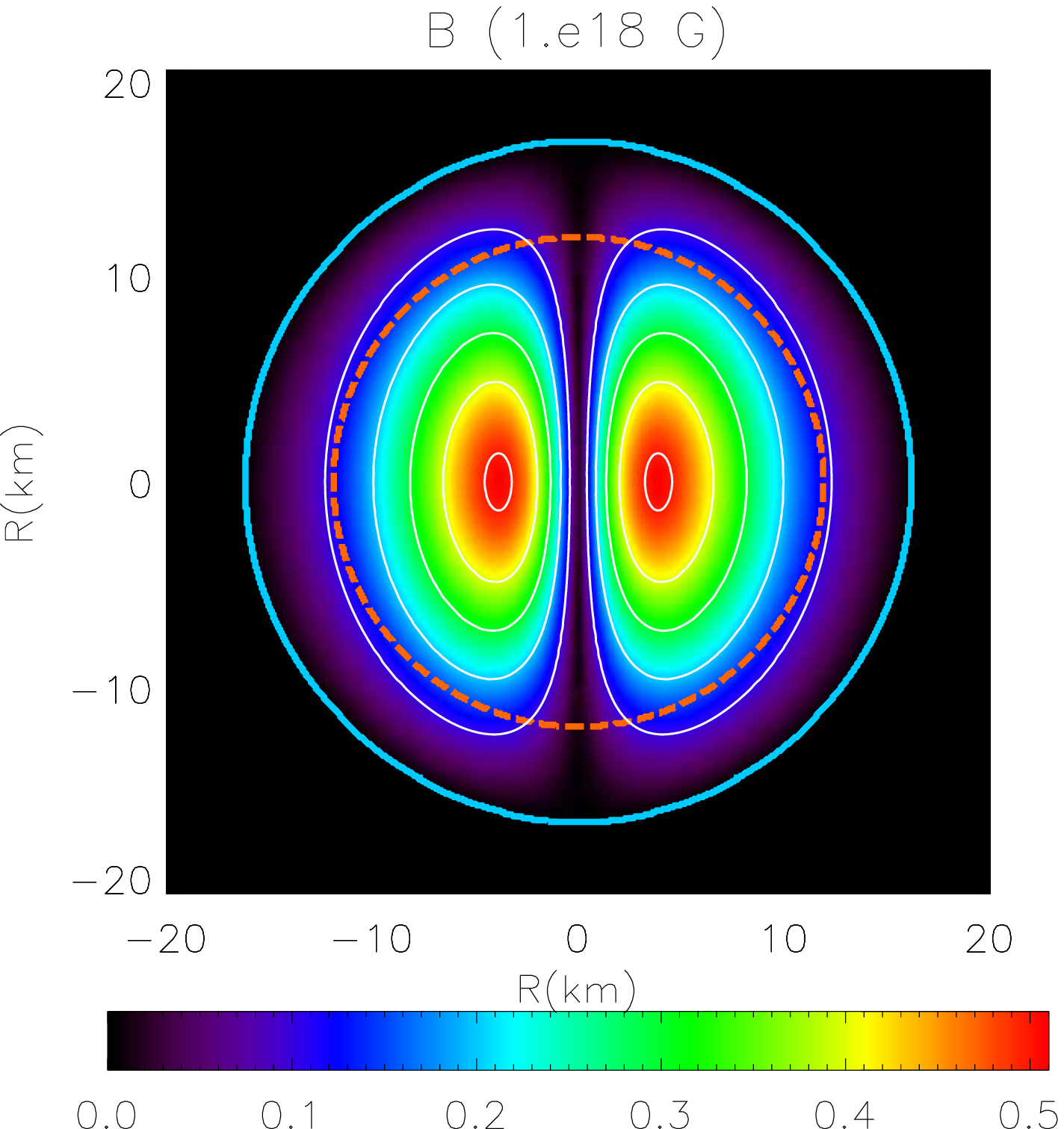}
	\includegraphics[width=.33\textwidth]{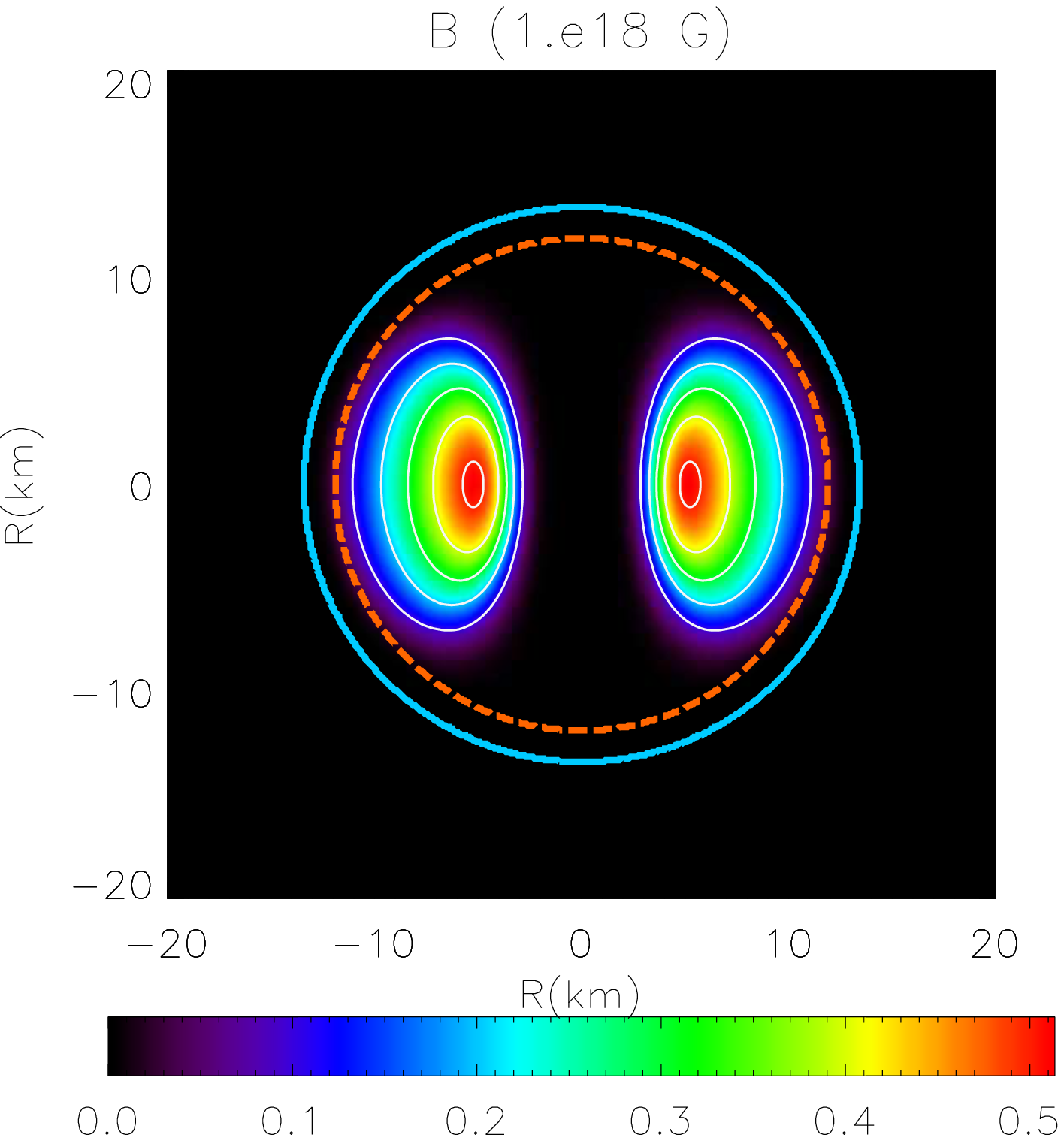}
	\includegraphics[width=.33\textwidth]{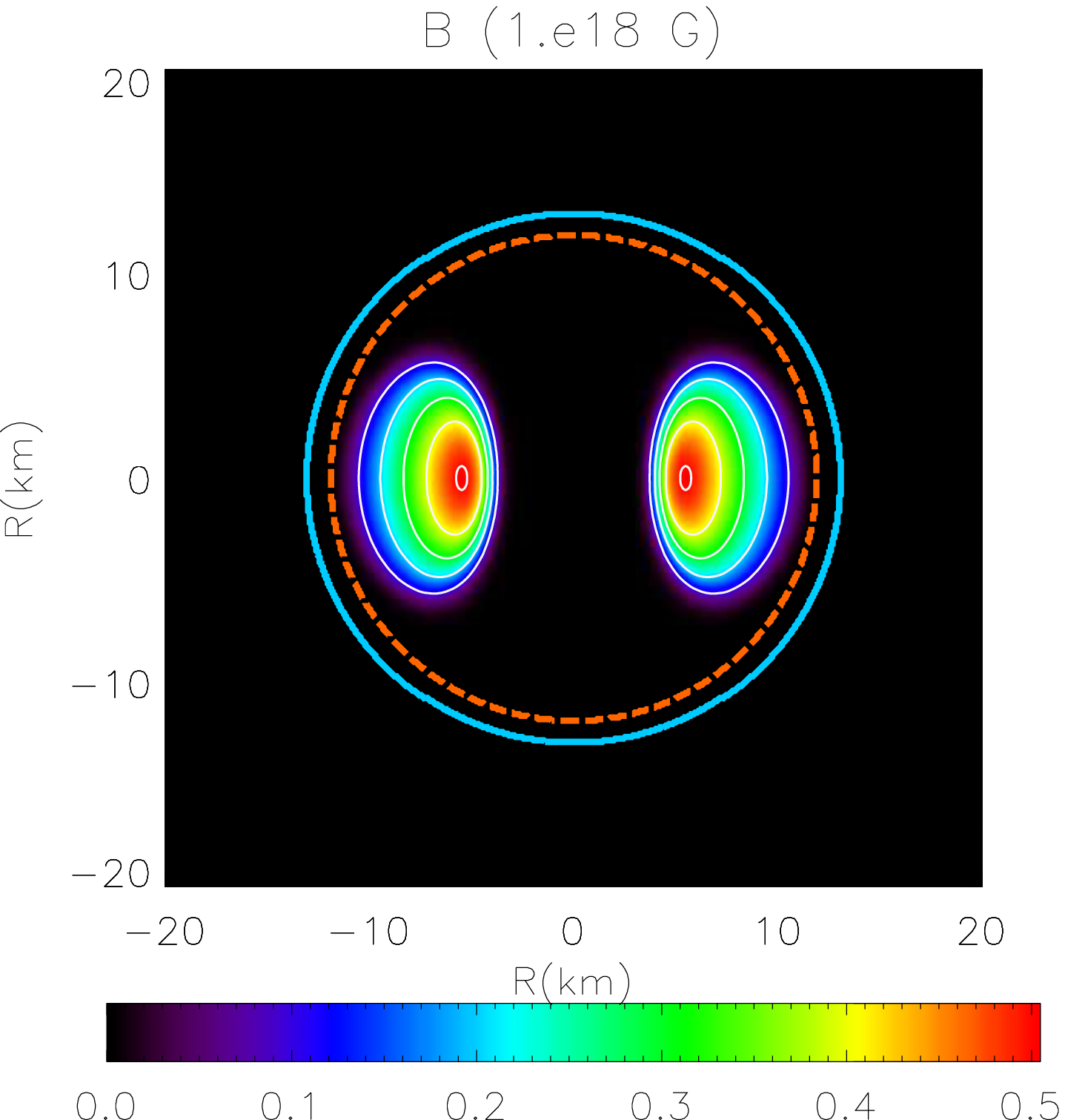}\\
    \includegraphics[width=.33\textwidth]{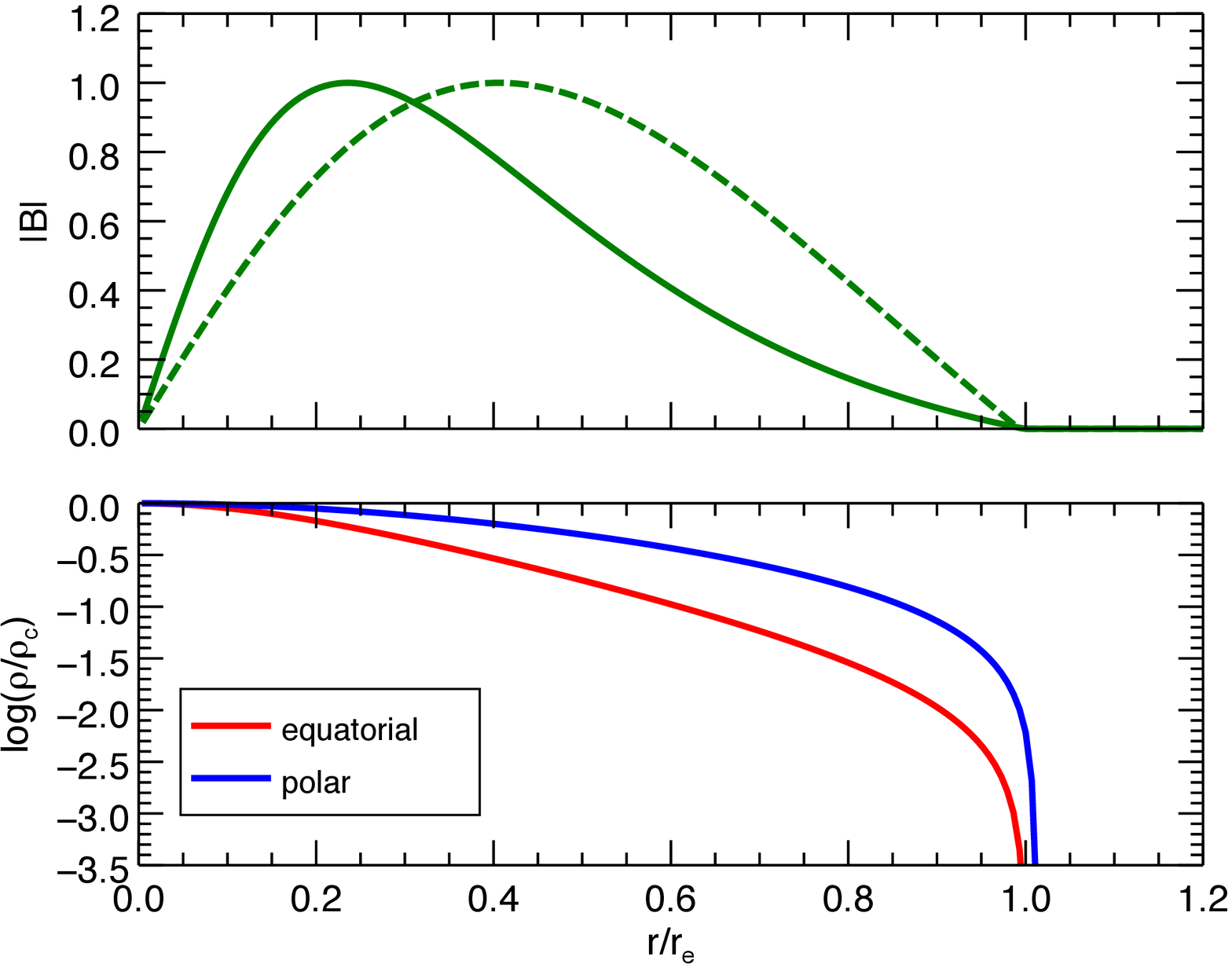}
	\includegraphics[width=.33\textwidth]{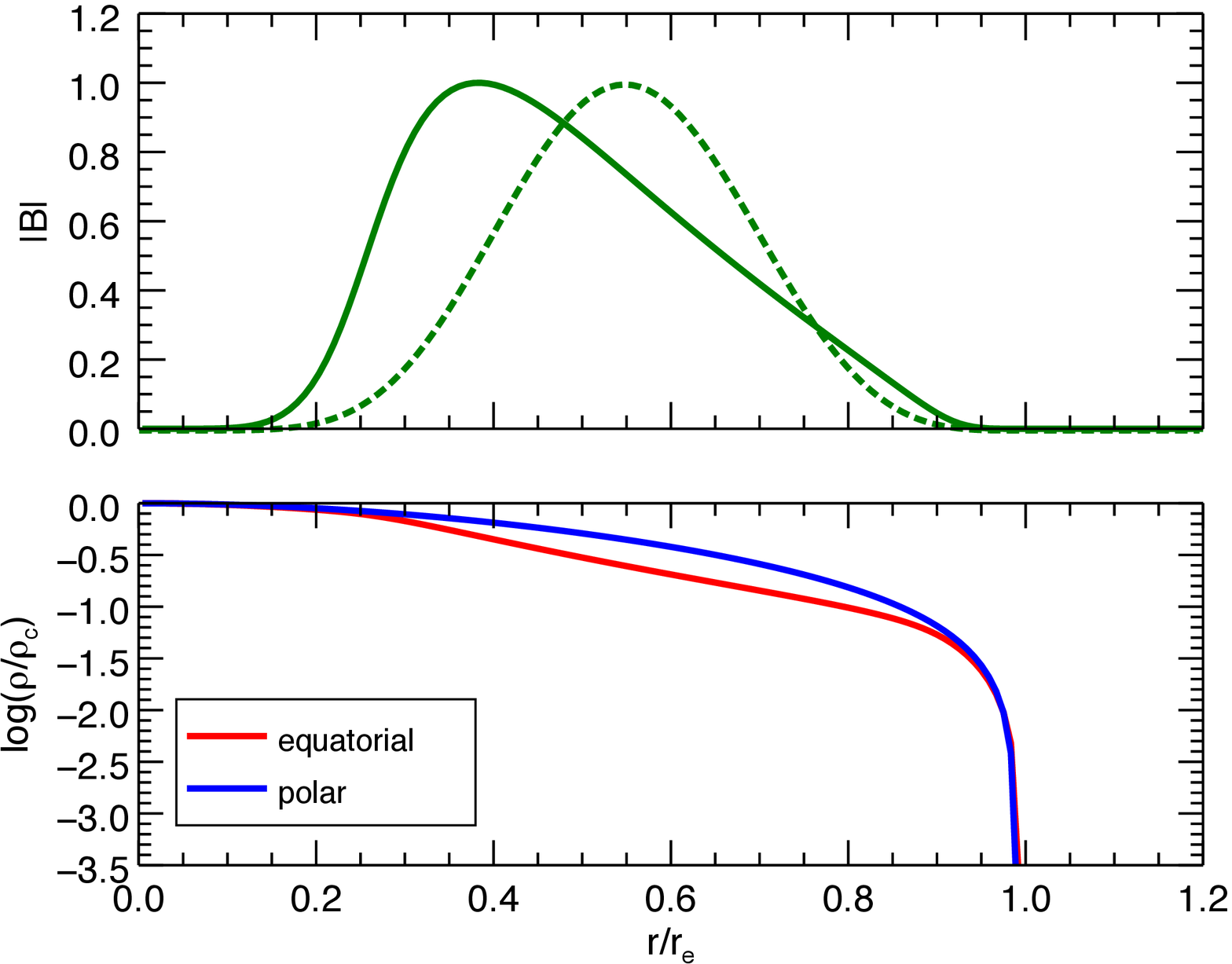}
	\includegraphics[width=.33\textwidth]{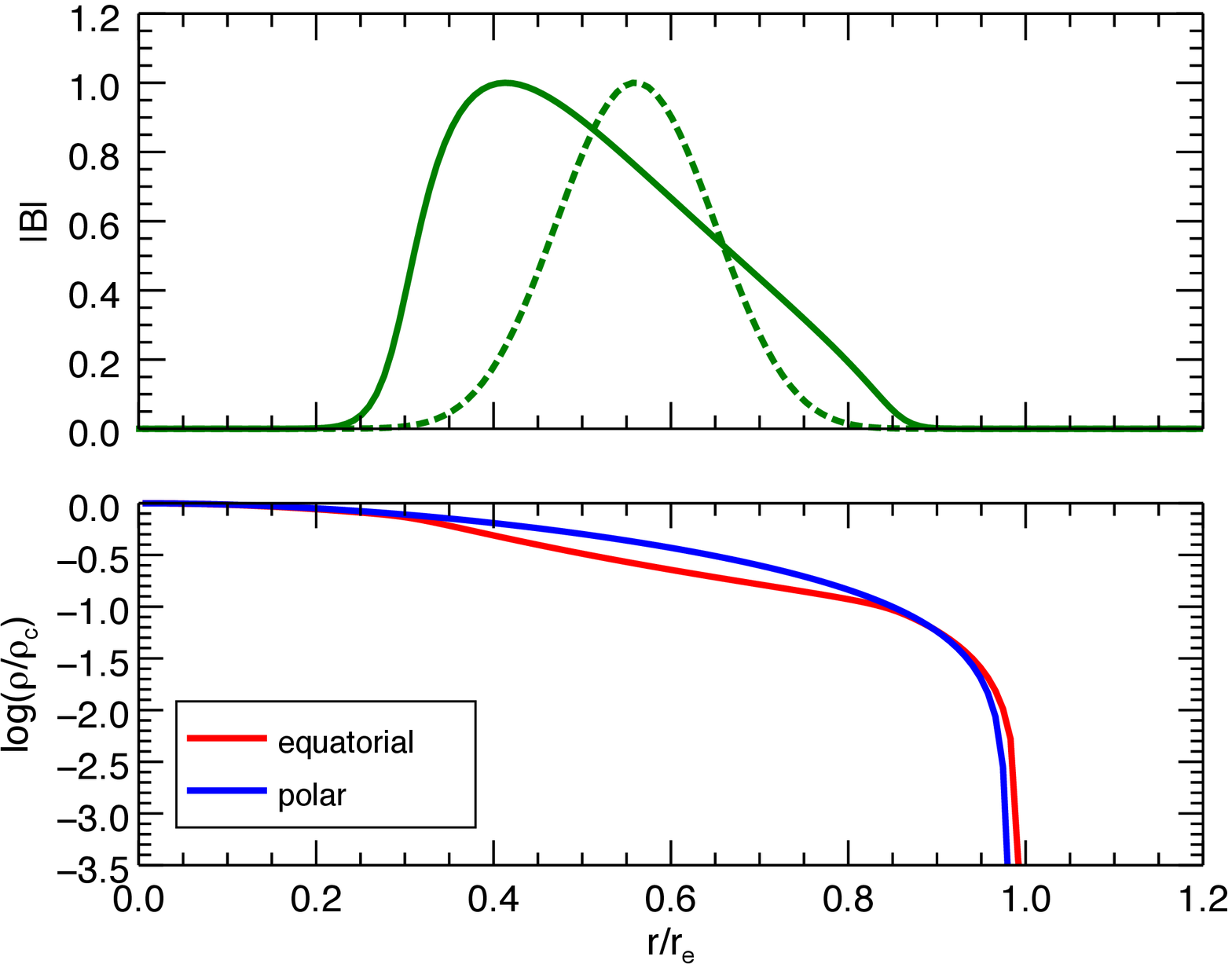}
	\caption{ Top panels: distribution and isocontours of the magnetic field strength $B=\sqrt{B^{\phi}B_{\phi}}$
	 for equilibrium configurations with the same gravitational mass	 $M=1.55 \mbox{M}_\odot$ and
	 rotational rate $\Omega=1.015 \times 10^3\mbox{s}^{-1}$ but 
	 with different values of the magnetization index: $m=1$ (left), $m=4$ (middle) and $m=10$ (right).  
	 The blue line  is the stellar surface.	Bottom panels: radial profiles of the magnetic field strength and of the
	  baryon density normalized to peak values along the equatorial direction 
	 $\theta = \upi/2 $ and the polar direction  $\theta = 0 $. The green dashed lines represent the magnetic
	 field radial profile in the weak magnetization limit ($B_{\rm max} \lesssim 10^{16}$~G). 
	 Radii are normalized to the equatorial value $r_{\rm e}$.
	 Global physical quantities for these equilibrium  configurations are listed in Tab.~\ref{tab:tableconfmh}.
              }
	\label{fig:conf155mh}
\end{figure*}

In Fig.~\ref{fig:conf155mh} we show a comparison among equilibria 
with the same gravitational mass and the same value of the
maximum magnetic field strength but with different values of $m$. A quantitative
characterization of  these configurations is given in Tab.~\ref{tab:tableconfmh}.
\begin{table*}
\centering
\caption{ \label{tab:tableconfmh}
Global quantities for the configuration shown in Fig.~\ref{fig:conf155mh} with 
gravitational mass $M=1.55 \mbox{M}_\odot$ and rotation frequency $\Omega=2.03\times10^{3}\mbox{s}^{-1}$.
}
\begin{tabular}{l*{12}c}
\toprule
\toprule
 & $B_{\rm max}$ & $\Omega$ & $\rho_{\rm c}$ & $M_0$ & $R_{\rm circ}$ & $r_{\rm p}/r_{\rm e}$ & $\bar{e}$ & $n_{\rm s}$ & $H/W$ & $T/W$ & $H/M$ & $T/M$  \\ 
 & $10^{17}$~G & $10^3\mbox{s}^{-1}$ & $10^{14} \mbox{g}/\mbox{cm}^3$ & ${\rm M}_\odot$ & km &  & &  & $10^{-2}$ & $10^{-2}$ & $10^{-2}$ & $10^{-3}$  \\
\midrule
m=1 & 5.12  & 2.03 & 8.07 & 1.65 & 18.3 & 1.01 & -0.35 & 2.00 & 12.3 & 1.16 & 1.76 & 1.58 \\ 
m=2 & 5.15  & 2.03 & 7.93 & 1.66 & 16.2 & 1.01 & -0.20 & 2.00 & 9.00 & 1.10 & 1.29 & 1.50 \\ 
m=4 & 5.15  & 2.03 & 7.95 & 1.66 & 15.6 & 1.03 & -0.12 & 2.00 & 6.65 & 1.05 & 0.98 & 1.55 \\ 
m=10 & 5.07& 2.03 & 7.98 & 1.67 & 15.3 & 1.03 & -0.07 & 2.00 & 4.81 & 1.03 & 0.72 & 1.55  \\ 
\bottomrule
\end{tabular} 
\end{table*}

As the value of $m$ increases the magnetic field distribution concentrates
toward the surface of the star, with a larger fraction of the star that behaves 
as if it was essentially unmagnetized. As a consequence the magnetic energy 
(at fixed $B_{\rm max}$) is smaller. This effect however is progressively less pronounced 
at higher magnetization, as shown by the comparison with the perturbative regime 
in Fig.~\ref{fig:conf155mh}, where we plot the magnetic field distribution 
in the radial direction at the equator for both the strong and weak field 
regime\footnote{Here and in the remainder
     of this paper we refer to \textit{weak field regime} as the regime where the effects induced by 
      the magnetic field on the NS can be safely computed adopting a perturbative approach.
      In the explored parameter space, this weak magnetization limit applies as long as  $B_{\rm max} \lesssim 10^{16}$~G
      (see also \citealt{Pili_Bucciantini+15a} or \citetalias{Bucciantini_Pili+15a} for a discussion).}.

\begin{figure*}
	\centering
	\includegraphics[width=.99\textwidth]{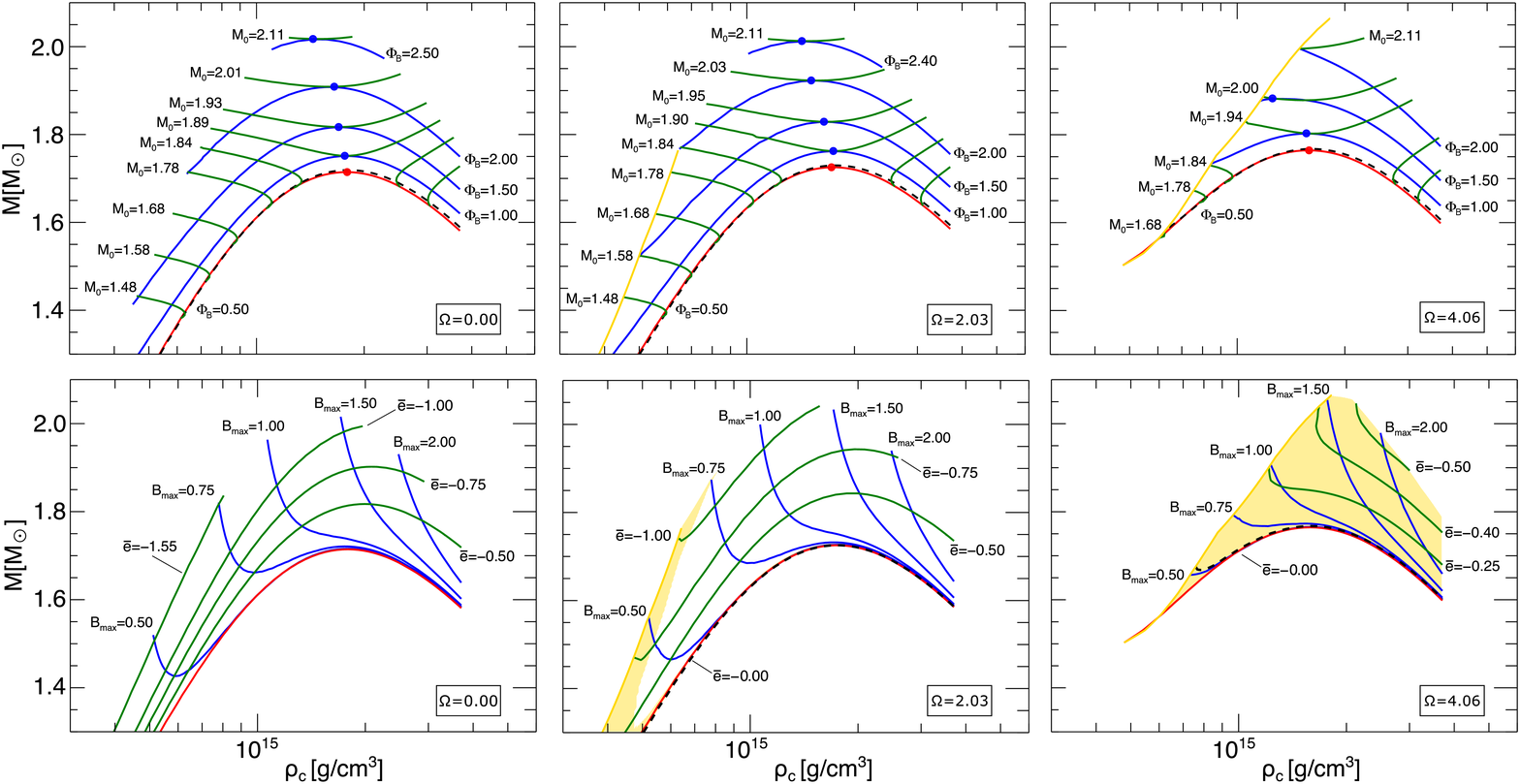}
	\caption{ Space of physical solutions for magnetized NSs with magnetic index $m=2$ and rigid rotation.
	Top row: equilibrium sequences with fixed baryonic mass $M_0$ (green lines) and fixed magnetic
	flux $\Phi_{\rm B}$ (blue lines) for different values of the rotational rate $\Omega$. 
	The black dashed lines represent configurations with a low value of the magnetic flux.  
	Bottom row: equilibrium sequences at fixed maximum magnetic field strength
	$B_{\rm max}$  (blue lines) and fixed deformation rate $\bar{e}$. Here the black dashed line 
	represents magnetized configurations with $\bar{e}=0$. The yellow shaded regions indicate 
	those configurations having  $e_{\rm s}>0$.	The red lines represent the unmagnetized sequences.
	The baryonic mass $M_0$ is expressed in unity of $\mbox{M}_\odot$,  $\Phi_{\rm B}$ in 	
	unity of $10^{30}\mbox{G\,cm}^2$ and $B_{\rm max}$ in unity of $10^{18}$~G.
              }
	\label{fig:quantities155m2}
\end{figure*}

\begin{figure*}
	\centering
    \includegraphics[width=.99\textwidth]{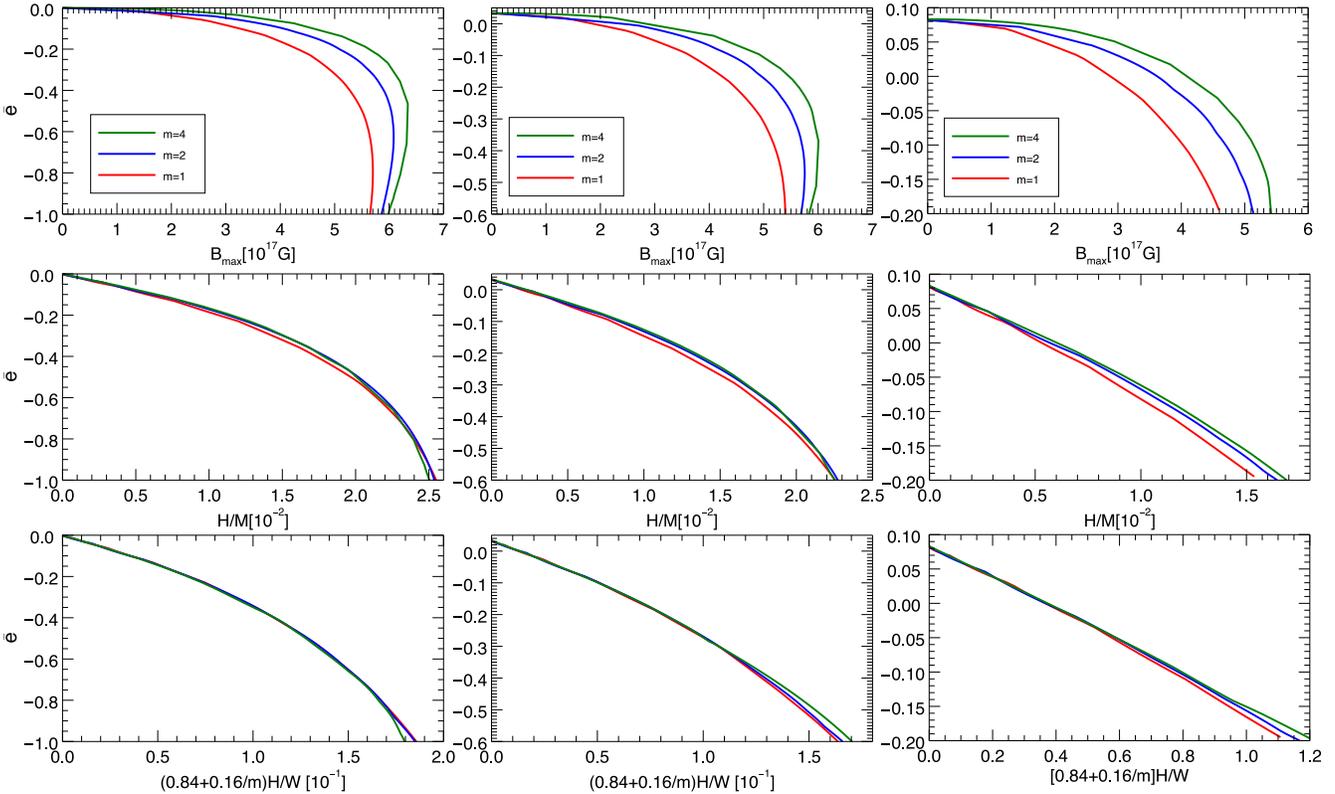}
	\caption{ Mean deformation rate $\bar{e}$ as a function of the maximum magnetic field 
	strength (top), of the magnetic energy to gravitational mass ratio (middle) and of the
	the effective magnetic energy ratio $[H/W]_{\rm eff}$ for different values of the magnetization index $m\in\lbrace1,2,4 \rbrace$  
	along sequences of fixed gravitational mass $M=1.55 {\rm M}_{\odot}$ in the 
	static case (left), with $\Omega=2.03 \times 10^{3} \mbox{s}^{-1}$ (centre) and  $\Omega=3.05 \times 10^{3} \mbox{s}^{-1}$(right). 
     }
	\label{fig:Defmh.eps}
\end{figure*}

Hence, if parametrized in terms of the magnetic field strength, the effects of the magnetic
field against rotation is strongly reduced at higher  $m$, leading to smaller deformations and 
ellipticities. This is also evident from Fig.~\ref{fig:quantities155m2} where we plot the space 
of physical solutions in the case $m=2$. Indeed, with respect to the $m=1$ case (see Fig.~\ref{fig:parsm1}),
the mass-shedding line moves to lower densities and the region characterized by $e_{\rm s}>0$ 
shrinks toward models with lower compactness. In good part this is due to the fact that the
inflationary effect on the outer layers of the star is also suppressed.
In Fig.~\ref{fig:Defmh.eps} we compare  the trends of $\bar{e}$, along our usual equilibrium sequences 
with $M=1.551 \mbox{M}_\odot$, considering cases with $m= 1, 2, 4$. 
It is evident that  in terms of $B_{\rm max}$, the various cases show clearly distinct trends.
However once parametrized in terms of $H/W$,  the dependence on $m$
for the baryonic mass $\Delta M_0$ becomes negligible (less than at most $5\%$ for the fastest rotating
configurations). In the case of $\bar{e}$ the parametrization in  Eq.~\eqref{eq:ss1} leaves a residual dependence
on $m$ which is at most $15\%$. This can be however reabsorbed defining an effective energy ratio:
\begin{equation}
\bar{e} = \mathcal{G}_{\bar{e}}(\Omega, H=0)+\mathcal{F}_{\bar{e}}\left([1+a_{\bar{e}}\Omega^2_{\rm ms}]\left[0.84+\frac{0.16}{m}\right]\frac{H}{W}\right),
\label{eq:ss3}
\end{equation}
which now generalizes Eq.~\eqref{eq:ss1} for different magnetic field distributions. 
The same holds for the apparent ellipticity, with major differences arising only close to the 
saturation fields, $B_{\rm max} \gtrsim 4\times 10^{17}$~G.
Interestingly we found that, once parametrized in terms of $H/M$ as in Eq.~\eqref{eq:ss2},
both $\bar{e}$ and $e_S$ show only a weak dependency (within 5\%) on the parameter $m$
even in the non-linear phase (see Fig.~\ref{fig:Defmh.eps}). Moreover from an observational
point of view, one may prefer a parametrization in terms of the gravitational
mass which is indeed a measurable quantity.

In the bilinear regime, the parametrization given by Eqs.~\ref{eq:param1}-\ref{eq:parames1}, 
of the mean deformation $\bar{e}$ and of the surface ellipticity $e_s$, can be generalized as:
\begin{eqnarray}
\bar{e}\simeq - \frac{d_B}{m} \, B_{17}^2 + d_\Omega \, \Omega_{\rm ms}^2,\label{eq:param1m}\\
e_{\rm s}\simeq - \frac{s_B}{m} \, B_{17}^2+ s_\Omega \, \Omega_{\rm ms}^2.
\label{eq:paramsm}
\end{eqnarray}
Now, since in the bilinear regime 
$\bar{e}\propto 17 H/M$, Eq.~\eqref{eq:param1m} also implies that 
$B_{17}=42 \sqrt{mH/M}$.

Given the small residual effect due to $m$, it is reasonable to conclude
that the quadrupolar and surface deformation  give only a direct
indication of the magnetic energy content rather than of
the current distribution, at least in the case of purely toroidal
magnetic fields. 

\subsubsection{Trends at different mass}

Let us finish our discussion of models with toroidal magnetic field,
by analyzing how results change at different masses. In general we
found that the trends and scalings found in the previous sections,
still hold at different masses in stable branches of the mass density relation. 
Obviously at lower gravitational mass (corresponding also to a lower
compactness) the effect of rotation and magnetic field are
enhanced.

\begin{figure}
	\centering
	\includegraphics[width=.45\textwidth]{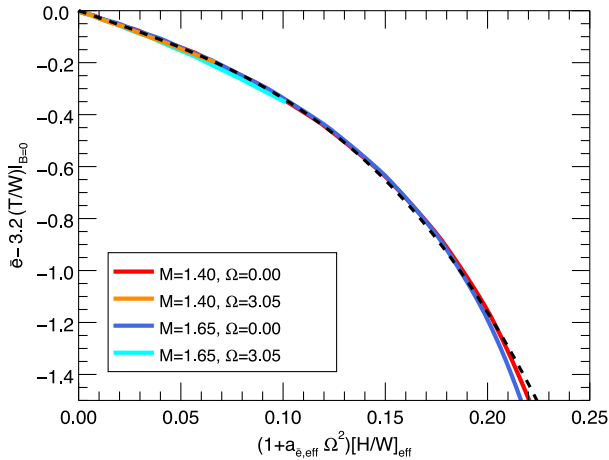}
	\caption{ Deformation ratio $\bar{e}$ with respect to the unmagnetized model
	              in terms of the effective magnetic energy ratio $\left[ H/W \right]_{\rm eff}$
	              for configurations having gravitational mass $1.40-1.65\, {\rm M}_\odot$ and
	              rotational frequency $\Omega=0.0-3.05\times10^3\,\mbox{s}^{-1}$.
	              The black dashed line represents Eq.~\eqref{eq:trendM}.
              }
	\label{fig:compareMass}
\end{figure}

For the deformation ratio $\bar{e}$, in the case of unmagnetized rotators,
it is possible to reabsorb the mass differences in the term due to rotation by 
using $T/W$, instead of $\Omega^2$ or $T/M$. We also found that
the trend is linear in $T/W$ almost all the way up to the fastest rotators.
Concerning the effects of the magnetic field on $\bar{e}$ and $e_S$, we find that 
they can be rescaled  defining an effective normalized (with respect to our fiducial
model with $M=1.55 \mbox{M}_{\odot}$) magnetic energy ratio and rotational coupling term,
such that:
\begin{equation}
\left[ \frac{H}{W} \right]_{\rm eff} = \left[0.84+\frac{0.16}{m}\right]
\frac{1.55\mbox{M}_\odot}{M}\frac{H}{W},
\label{eq:heff}
\end{equation}
\begin{equation}
a_{\bar{e}\; {\rm eff}}=- \left( 3.94-2.98\frac{M}{1.55\mbox{M}_\odot}\right).
\label{eq:aeff}
\end{equation}
This behaviour was already found to hold in the linear regime at fixed mass
and for different EoS by \citetalias{Frieben_Rezzolla12a}.  
Here we show that it can be also generalized for different masses and magnetic 
field distributions. Notice that, while the effects due to rotation scales as $T/W$,
the magnetization effects go as $H/WM$. This might be related to the way 
mass stratification couples with rotation and magnetic field: for a rigid rotator,
the rotational stratification is independent of density and mass; on the contrary, assuming
the magnetic barotropic law Eq.~\eqref{eq:Mtor}, as mass and stratification change so does 
the magnetic field distribution. We find that the following functional form:
\begin{eqnarray}
\bar{e}\simeq 3.2 \frac{T}{W} \bigg|_{B=0}
+\mathcal{F}\left(\left[1+a_{\bar{e}, {\rm eff}} \Omega^2_{\rm ms}\right]\left[\frac{H}{W}\right]_{\rm eff}
\right)\nonumber \\
{\rm with}\quad \mathcal{F}(x)=-2.71 x - 0.068(10x)^{3.2}
\label{eq:trendM}
\end{eqnarray}
fits the deformation ratio for all values of $\Omega$, $H$, $M$ and $m$ up to $\bar{e}\simeq
1$, with an error less than $5\%$ as shown in Fig.~\ref{fig:compareMass}.

Focusing on the bilinear regime, it is natural to expects that the 
coefficients in Eq.~\eqref{eq:param1m} are functions of mass.
As shown in Tab.~\ref{tab:tableparam1} both $d_\Omega$ and $d_{\rm B}$ 
are decreasing functions of $M$, while their ratio $d_\Omega/d_{\rm B}$ grows with $M$.
\begin{table}
\centering
\caption{ \label{tab:tableparam1}
Mass dependency for the $\bar{e}$ expansion coefficients in Eq.~\eqref{eq:param1m}.
}
\begin{tabular}{l*{6}c}
\toprule
\toprule
 $M$ & $W_0$ &  $d_{\Omega} $ & $ d_{\rm B} $ & $s_{\Omega} $ & $ s_{\rm B} $ \\ 
 $\mbox{M}_\odot$ &  $\mbox{M}_\odot$  & $10^{-1}$  & $10^{-3}$ & $10^{-1}$ & $10^{-3}$        \\
\midrule
1.40 & 0.18  & 4.5    & 15   &  6.0  &  4.8   \\
1.45 & 0.20  & 4.0    & 13   &  5.1  &  4.1   \\ 
1.50 & 0.23  & 3.6   &  11   &  4.6  &  3.6   \\ 
1.55 & 0.25 & 3.1   &  9.5  &  3.8  &  2.6   \\ 
1.60 & 0.28  & 2.8   &  7.0  &  3.5  &  2.2   \\
1.65 & 0.32  & 2.3   &  5.5  &  3.1  &  1.5   \\
\bottomrule
\end{tabular} 
\end{table}
In this regime we can also simply Eq.~\eqref{eq:trendM} evaluating its limit as 
$H,T\rightarrow 0$, for which $W\rightarrow W_0= {\rm const}$, so that 
\begin{equation}
\bar{e}\simeq \frac{C_{\bar{e}}}{W_0}\left[T-1.3  \frac{H}{M/\mbox{M}_\odot}\right],
\label{eq:selfsime}
\end{equation}
where the coefficient $C_{\bar e}$ depends on the specific EoS and the mass
of the reference unmagnetized stationary configuration:
for our fiducial model with $M=1.551\mbox{M}_{\sun}$ we have  $C_{\bar{e}}=3.2$ and $W_0=0.25 \mbox{M}_\odot$. 
We recall that even the coefficient $1.3$ in general has a residual dependence on $m$,
but it can be taken as constant with an accuracy $\sim 10\%$.
We can conclude that, in the perturbative regime, the deformation ratio
is linear in the quantity $T-1.3H/(M/\mbox{M}_\odot)$, which can be
considered as a self-similarity variable, with all the information
about the EoS, factored out in a single proportionality
coefficient.

Repeating the same analysis for the surface deformation in the bilinear regime,
we found that the coefficients $s_B$ and $s_\Omega$ in Eq.~\eqref{eq:paramsm}
show analogous trends with those found for the deformation ratio.
Coefficients are given in Tab.~\ref{tab:tableparam1}.
Obviously a parametrization in terms of $\Omega^2$ and $B_{\rm max}^2$ is not
optimal, once different masses are taken into consideration (at fixed
mass $\Omega^2 \propto T$ and $B_{\rm max}^2 \propto H$, so it is
equivalent to a parametrization in terms of kinetic and magnetic
energies). We found that using the effective magnetic energy defined by
Eq.~\eqref{eq:heff}, and $T/W$ for the rotational contribution, the mass
dependence can be reabsorbed analogously to Eq.~\eqref{eq:selfsime}, 
with an accuracy $\lesssim 10\%$
\begin{equation}
e_{\rm S}\simeq  \frac{C_{e_{\rm s}}}{W_0}\left[T-0.23\frac{H}{M/\mbox{M}_\odot}\right],
\end{equation}
with $C_{e_{\rm s}}\sim4.2$. Note that given the accuracy of our
solution, the surface deformation of our models cannot be determined  better than $0.01$, and models with $e_{\rm S} \gtrsim 0.05$ are
already in the non-linear regime.

In the case $M=1.40 \mbox{M}_{\odot}$ we can compare our results to those
obtained by \citet{Cutler02a} and \citetalias{Frieben_Rezzolla12a}.
In the Newtonian regime they both provide the distortion
coefficients for the parametrization $\bar{e}= a_{\rm \Omega} T/W + a_{\rm B} H/W$:
\citetalias{Frieben_Rezzolla12a} obtain $a_{\rm \Omega}=3.8$ and $a_{\rm B}=3.5$,
\citet{Cutler02a} instead has $a_{\rm \Omega}=a_{\rm B}=3.75$. 
Differences between distortion coefficients depends on the choices for the EoS: both
\citet{Cutler02a} and  \citetalias{Frieben_Rezzolla12a} adopt  a NS having stellar radius $R=10$~km 
as reference model but while the former adopt an incompressible fluid, the latter a polytropic with $n=1$.
In our case we use $n=1$ as well, but with a different polytropic constant leading to a NS with $R_{\rm circ}=15$~km
and  we obtain $a_{\rm \Omega}=3.2$ and $a_{\rm B}=2.97$ (see Eq.~\eqref{eq:selfsime}).
Leading to less compact configurations, our choice should  in principle imply
higher value for $a_{\rm \Omega}$ and $a_{\rm B}$. However our calculations are made within
GR and as pointed out by \citetalias{Frieben_Rezzolla12a}, a Newtonian treatment overestimates
the correct distortion coefficients. Interestingly for $M=1.4 \mbox{M}_\odot$, we find a ratio 
$a_{\Omega}/a_{\rm B}\sim1.078 $, similar to the value $1.085$ found by \citetalias{Frieben_Rezzolla12a}
in the Newtonian limit. This may suggest that the relative contributions of  the magnetic field 
and of the rotation are the same independently from the compactness of the star, that may only 
affect their absolute values.

A comparison in full GR is only possible with \citetalias{Frieben_Rezzolla12a} (see also \citetalias{Pili_Bucciantini+14a}),
where however they adopt different parametrisations with respect to ours. In particular 
they express both the surface ellipticity $e_{\rm s}$ and the the quadrupole deformation $e_{\rm q}$
using the  average magnetic field strength  $<\!B^2\!>$ instead of its maximum value $B^2_{\rm max}$:
\begin{equation}
e_s=-b_{\rm B}<\!B^2_{\rm 15}\!> + b_{\rm \Omega}\,\Omega^2, \quad e_q=-c_{\rm B}<\! B^2_{\rm 15}\!> + c_{\rm \Omega}\,\Omega^2
\end{equation}
where the magnetic field is normalized to $10^{15}$~G and $\Omega$  is expressed in $\mbox{s}^{-1}$
(to be compared with our Eqs.~\ref{eq:paramsm}~and~\ref{eq:param1m}).
Considering again their reference model at  $M=1.40\mbox{M}_{\odot} $ and $m=1$, we find that 
$<\!B^2\!> \sim 0.25 B^2_{\rm max}$ while $e_{\rm q}\simeq 0.5 \bar{e}$ (see also \citealt{Pili_Bucciantini+15a}). 
Hence the coefficients given in  Tab.~\ref{tab:tableparam1} translate into 
$b_{\rm B}=1.9\times 10^{-6}$, $b_{\rm \Omega}=1.5\times 10^{-8}$
and $c_{\rm B}=2.9\times 10^{-6}$, $c_{\rm \Omega}=5.7\times 10^{-8}$
in good agreement with \citetalias{Frieben_Rezzolla12a} considering  
$R_{\rm circ}\simeq15\mbox{km}$ (see Fig.13 of their paper).
\begin{figure}
	\centering
	\includegraphics[width=.49\textwidth]{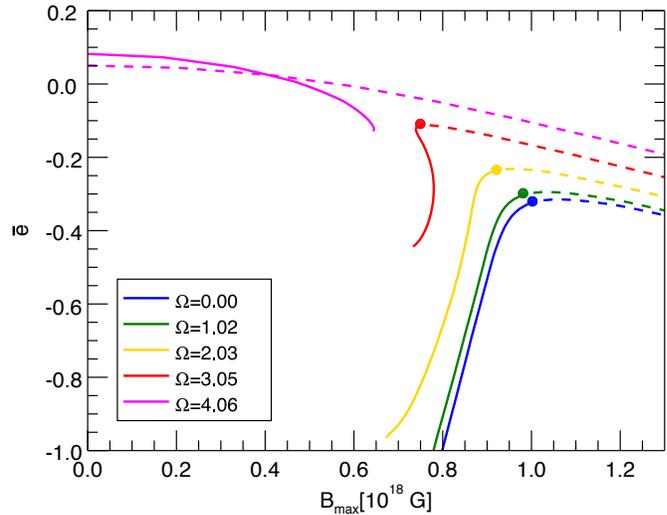}
	\caption{ Variation of the mean deformation ratio $\bar{e}$  along equilibrium sequences characterized by
				 fixed gravitational mass $M=1.75 \mbox{M}_\odot$ and different value of $\Omega$ with $m=1$.
				 Filled circles represent the configurations with highest $M_0$ and they split each sequence
				 into a low $\rho_{\rm c}$ (solid lines) and high $\rho_{\rm c}$ (dashed lines) branch.
              }
	\label{fig:quantities175m1}
\end{figure}

In this section we consider just equilibrium sequences with gravitational mass
below the maximum mass of the static and unmagnetized sequences (i.e. $1.72 \mbox{M}_\odot$ for
our choice of EoS) and central density 
$\rho_c<1.8 \times 10^{15} \mbox{g cm}^{-3}$. This corresponds to selecting sequences that are 
connected to the stable unmagnetized static branch and for which a comparison with a static 
unmagnetized reference model is meaningful.

Trends become more complex if one considers supramassive sequences
(see \citealt{Kiuchi_Yoshida08a} for discussion). An example is shown in
Fig.~\ref{fig:quantities175m1}  where we plot $\bar{e}$ as a function of 
$\Omega$ and $B_{\rm max}$ considering $M=1.75\mbox{M}_\odot$.
Notice that with such gravitational mass, no static configuration exists
if $\Phi_{\rm B} < 1.31\times 10^{30} \mbox{G cm}^2$, neither 
unmagnetized configurations with $\Omega \lesssim 4\times 10^{3} \mbox{s}^{-1}$.
The behavior of $\bar{e}$ in this case can be understood by looking 
also at Fig.~\ref{fig:parsm1}. The configurations with the minimal deformation
are very close to the ones that, at fixed gravitational mass, have the highest $M_0$.
Moving away from this configuration, either toward lower or
higher central densities, the absolute value of the deformation rises. 
 At lower densities, despite the fact that the magnetic field also diminishes, the
deformation rises because the star becomes less compact. At higher
density, despite the star becoming more compact, the magnetic field
rises and its effect on the deformation becomes stronger. As one can
see the interplay between the magnetic field and the compactness can
be quite different depending on the branch in the mass density sequence.
For $\Omega \gtrsim 4\times 10^{3} \mbox{s}^{-1}$, two unmagnetized configurations
become possible (one stable one unstable) and  the $\bar{e}$
sequence splits in two separated branches: the one connected with the unmagnetized configuration 
with lower $\rho_{\rm c}$, the other connected with the unmagnetized
configuration with higher $\rho_{\rm c}$. Being this latter more compact, 
both the effect of rotation and of  magnetic field are less pronounced.

\subsection{Poloidal Magnetic Field}

In this section we will characterize the effects of a poloidal magnetic  field, focusing
on the simplest choice with $\xi=0$ (see Eq.~\eqref{eq:Mpol}) and with a vanishing net charge. Our investigation is
again limited to $\Omega\lesssim 5.1\times 10^{3} \mbox{ s}^{-1}$
since at this frequency and for our choice of the EoS  the parameter
space is substantially reduced due to mass shedding.
\begin{figure*}
\centering
	\includegraphics[width=.95\textwidth]{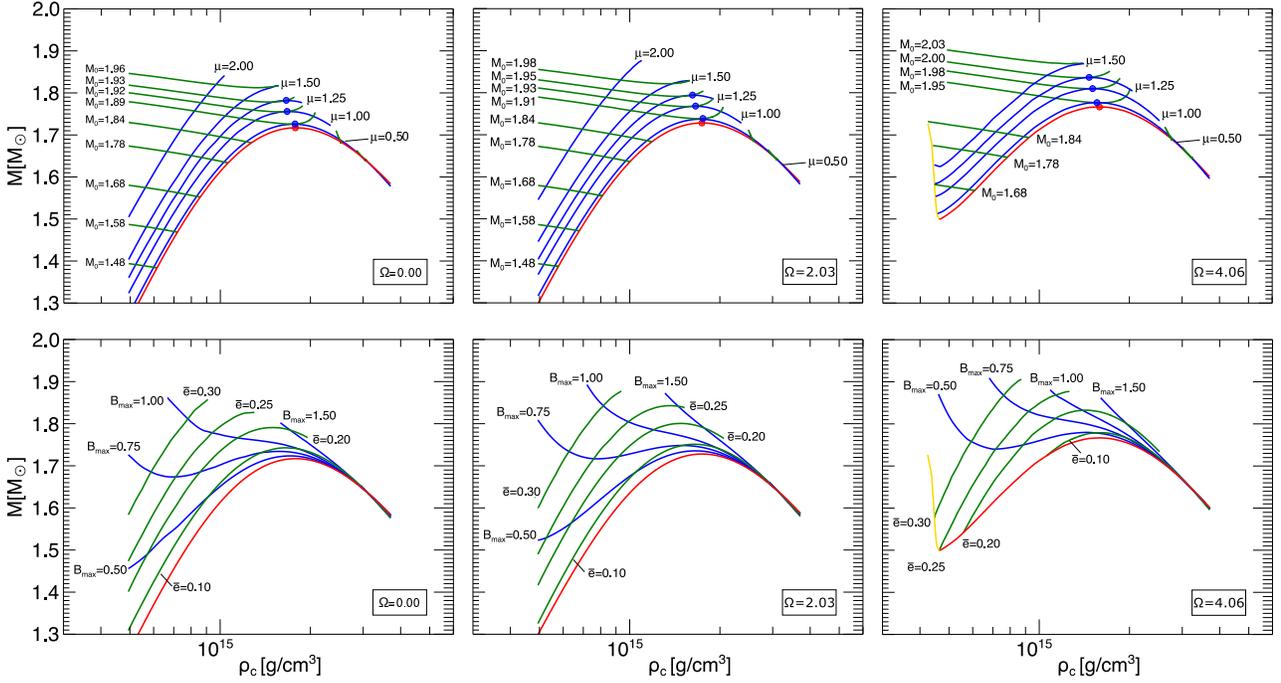}
	\caption{ Space of physical solutions for rigidly rotating NSs endowed with purely poloidal magnetic field,
	obtained assuming a current distribution with $\xi=0$ in Eq.~\eqref{eq:Mpol}.
	Top row: equilibrium sequences with fixed baryonic mass $M_0$ (green lines) 
	and fixed magnetic dipole moment $\mu$ (blue lines).
	Bottom row: equilibrium sequences with fixed matter deformation $\bar{e}$ (green lines) or with fixed
	maximum magnetic strength $B_{\rm max}$ (blue lines). The yellow lines trace the mass shedding limit,
	while the red lines represent unmagnetized equilibria. The baryonic mass is expressed in units 
	of $M_\odot$, $\mu$ is expressed in units of	 $10^{35} \mbox{ erg G}^{-1}$, $B_{\rm max}$ is expressed in
	 units of $10^{18}$~G and finally the rotational rate $\Omega$ is in units of $10^{3}\mbox{s}^{-1}$.
	}
	\label{fig:polparspace}
\end{figure*}
The parameter space is shown in Fig.~\ref{fig:polparspace}
in the $\rho_{\rm c}$-$M$ plane for slices with fixed $\Omega$.  In line with \citetalias{Pili_Bucciantini+14a}, in each plot 
we show equilibrium sequences at fixed baryonic mass $M_0$, magnetic dipole moment $\mu$,
maximum field strength $B_{\rm max}$, or deformation $\bar{e}$. As expected, 
in analogy with the toroidal case, both the gravitational and baryon mass rise with
the magnetization (or equivalently with the magnetic dipole moment $\mu$) 
and with the rotational frequency $\Omega$. Here, however, the magnetic field acts in the
the same way of the centrifugal force, flattening the star in the direction of the equatorial plane. 
As a consequence all the configurations are oblate and the surface ellipticity 
$e_{\rm s}$ is always positive.  Differences between the two effects are only evident in the low 
density region ($\rho_{\rm c}\lesssim 5\times 10^{15} \mbox{g}/\mbox{cm}^3$).
Indeed,  although the majority of the stellar surfaces can be well approximated with standard 
ellipsoids having  $n_{\rm s}\sim 2$, at low density the superellipsoid index ranges 
from  $\sim 1.6$, for the configuration at the mass-shedding limit, 
to $\eta \sim 2.8$  for the most magnetized NSs with $B_{\rm max}\gtrsim 7\times 10^{17} \mbox{G}$:
while rigid rotation acts preferentially on the external equatorial 
layers of the star, originating hypoellipsoidal with $n_{\rm s}<2$, the Lorentz force associated to the poloidal magnetic
field globally flattens the star, from the core to the external layers, reducing the polar radius and 
favouring the occurrence of hyperellipsoids with $n_{\rm s}>2$. 
Two representative equilibrium configurations are shown in Fig.~\ref{fig:confpol} and  Tab.~\ref{tab:confpol}. They both 
share the same value of the gravitational mass but, because of the different degree of magnetization and rotation rate, 
they have  $n_{\rm s}\sim2.5$ (left panel) and $n_{\rm s}\sim1.8$ (right panel). 
Notice that, even if the equatorial radius grows, the poloidal field does not inflate the outer layers of the 
star. As a consequence, the central density at mass shedding for a given $\Omega$ remains almost the same with respect to the 
unmagnetized case, as shown in the right panels of Fig.~\ref{fig:polparspace}. 
Our findings suggest that the poloidal field enhances the stability against the Keplerian limit:
the equatorial Lorentz force points outward in the inner region of the star causing its deformations,
but points inward in the outer layers playing a confining role.
\begin{figure*}
\centering
	\includegraphics[width=.40\textwidth]{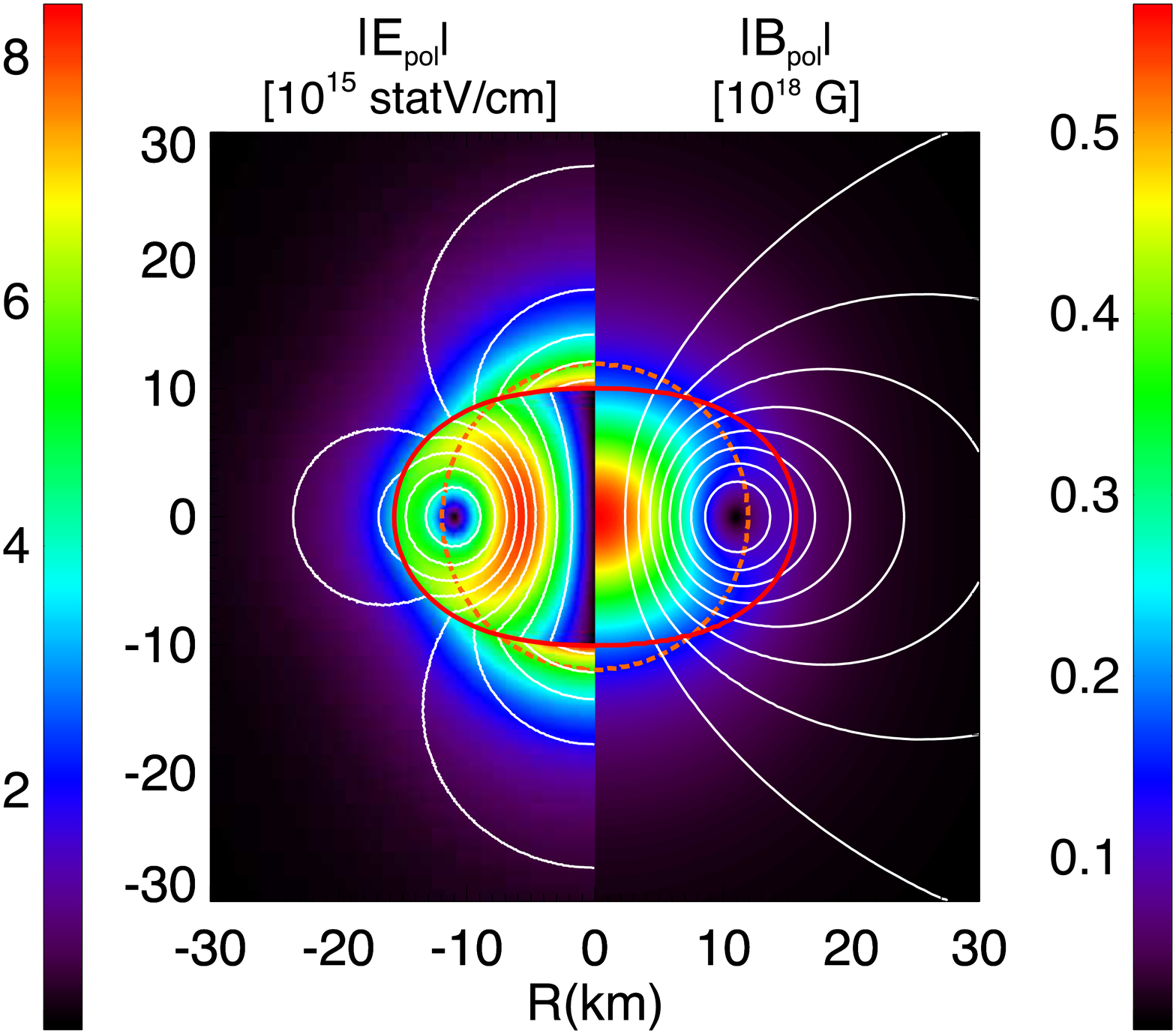}
	\includegraphics[width=.40\textwidth]{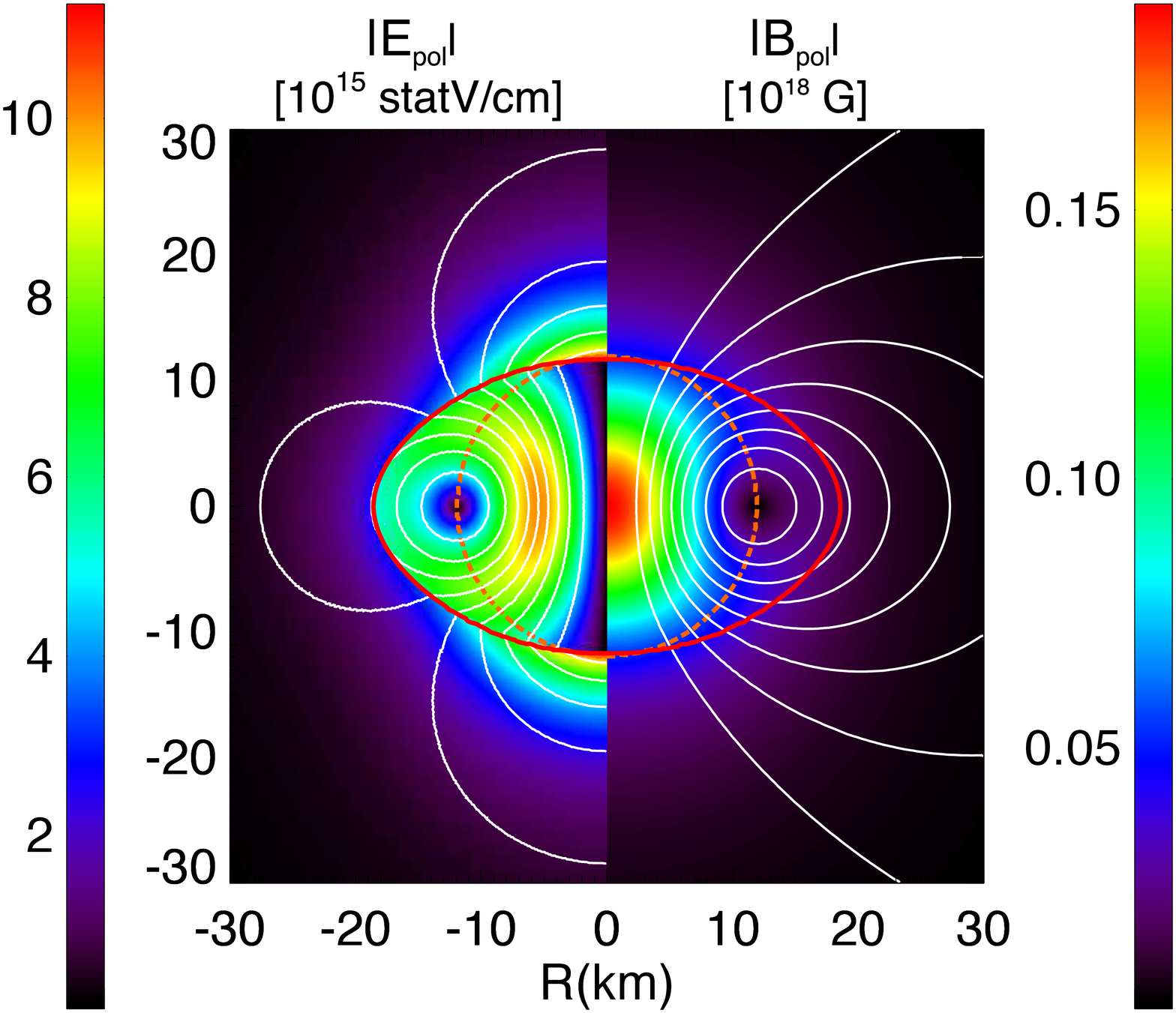}\\
	\includegraphics[width=.40\textwidth]{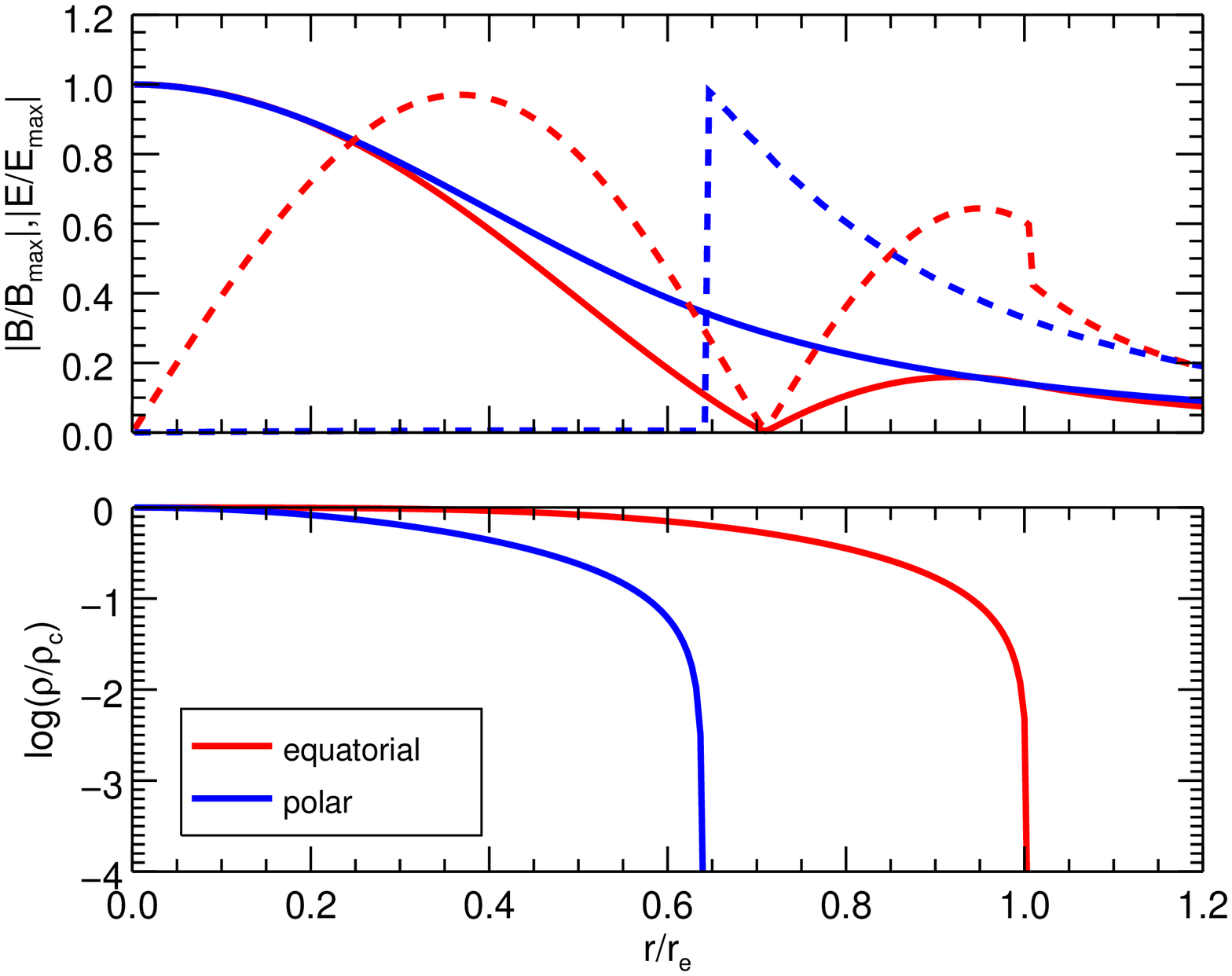}
	\includegraphics[width=.40\textwidth]{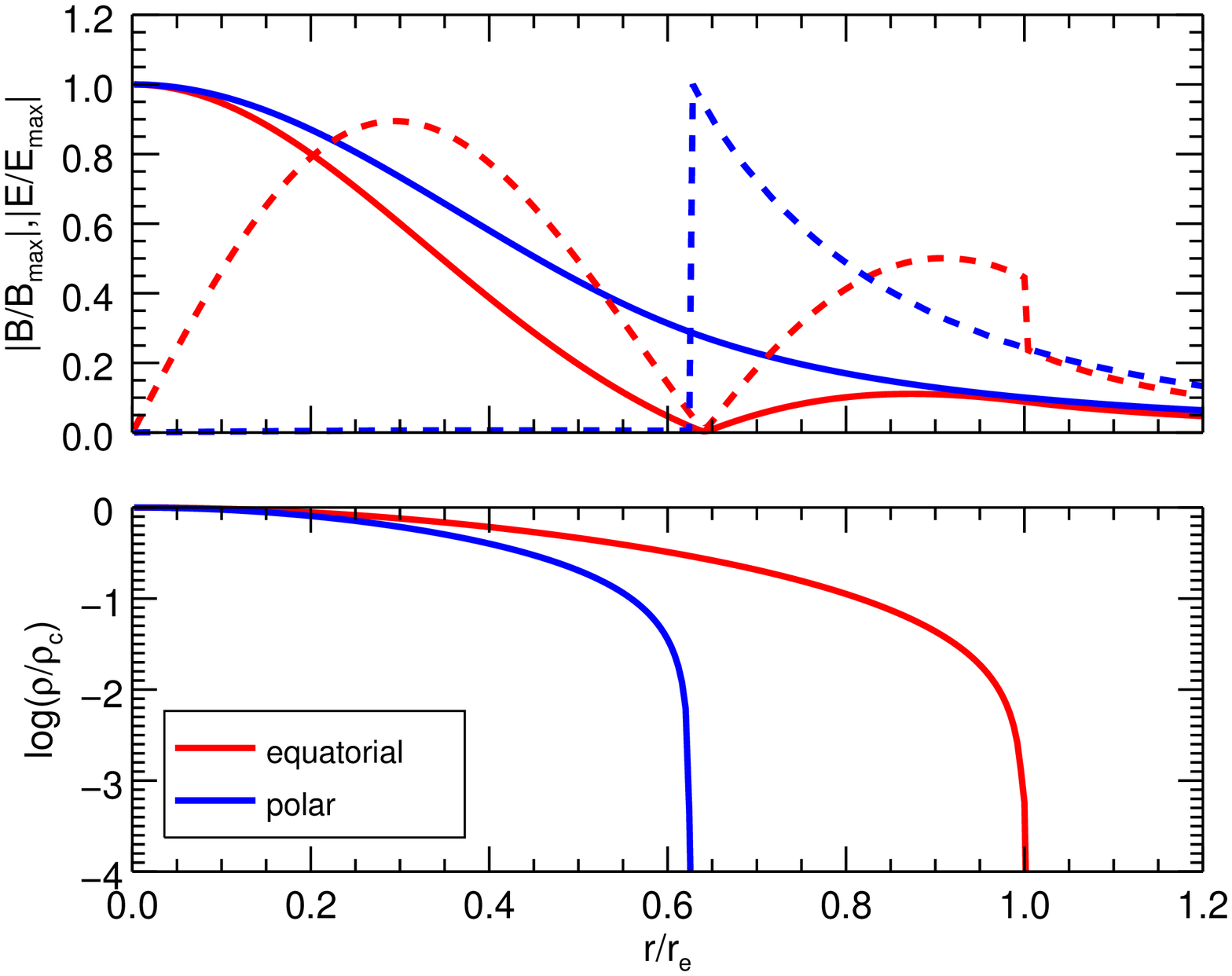}\\
	\caption{ Top row: electric (left half panel) and magnetic (right half panel) field distribution 
	together with the contours of the electric potential $\Phi$ (left half panel) and of the magnetic 
	potential $\Psi$ (right half panel) 	for two configurations sharing
	the same gravitational mass but with different $B_{\rm max}$ and $\Omega$ ($B_{\rm max}=5.72\times10^{17}$~G and 
	$\Omega=1.02\times10^3\mbox{s}^{-1}$ for the configuration on the left, $B_{\rm max}=1.88\times10^{17}$~G and 
	$\Omega=4.06\times10^3\mbox{s}^{-1}$ for the configuration on the right).
	Numerical details are shown in Tab.~\ref{tab:confpol}. Middle row: profile of the magnetic 
	field strength (solid lines) and of the electric field 
	strength (dashed lines) in the equatorial (red lines)
	and polar (blue lines) direction.  Bottom rows: polar and equatorial radial profiles of the baryon density $\rho_c$.  }
	\label{fig:confpol}
\end{figure*}
\begin{table*}
\centering
\caption{ \label{tab:confpol}
Global quantities for the configuration shown in Fig.~\ref{fig:confpol} with gravitational mass $M=1.55 \mbox{M}_\odot$.
}
\begin{tabular}{l*{11}c}
\toprule
\toprule
$B_{\rm max}$ & $\Omega$ & $\rho_{\rm c}$ & $M_0$ & $R_{\rm circ}$ & $r_{\rm p}/R_{\rm e}$ & $\bar{e}$ & $n_{\rm s}$ & $H/W$ & $T/W$ & $H/M$ & $T/M$  \\ 
$10^{17}$~G & $10^3\mbox{s}^{-1}$ & $10^{14} \mbox{g}/\mbox{cm}^3$ & $\mbox{M}_\odot$ & km &  & &  & $10^{-2}$ & $10^{-2}$ & $10^{-2}$ & $10^{-2}$  \\
\midrule
1.88 & 4.06 & 4.79 & 1.65 & 20.7 & 0.62  &  0.27 & 1.80 & 1.60 & 8.67 & 0.19 & 1.02 \\ 
5.72 & 1.02 & 4.30 & 1.64 & 17.7 & 0.64  &  0.32 & 2.20 & 14.0 & 0.49 & 1.79 & 0.06  \\ 
\bottomrule
\end{tabular} 
\end{table*}

At very high magnetization, the internal structure 
of the NS changes dramatically since the magnetic force can
evacuate the core so that the density reaches its maximum in a ring
located in the equatorial plane rather than at the centre (see Fig.~6 in \citetalias{Pili_Bucciantini+14a}). 
Because our numerical scheme is less accurate in this regime, in the present  analysis
we will exclude such kind of configurations. Moreover, as pointed out by  \citet{Cardall_Prakash+01a}, 
at even higher magnetization no stationary solution can be found because the magnetic
field pushes off-center a sufficient amount of mass that the gravitational force, near the center of the star, 
points outward.

In Fig.~\ref{fig:confpol} we show the morphology of  the magnetic  and electric field strength
together with the isocontours of the magnetic potential $\Psi$ and the electric potential $\Phi$.
Notice that inside the star, $\Phi$ traces the magnetic field lines $\Psi=\rm{cost}$ in agreement
with the MHD requirement.  
Outside the star the structure of the electric potential is mainly quadrupolar 
reflecting the fact that the monopolar component has been filtered out in order 
to achieve a globally uncharged configuration. The resulting
internal electric field strength reaches its maximum value between the rotational axis
and the neutral lines, where it vanishes together with the magnetic field. The exterior 
electric field instead reaches its maximum strength in correspondence of the pole of
the star. Obviously, different prescriptions on the total electric charge of the star
lead to different morphologies for the exterior electric field.  The interior one instead
remains unchanged. Hence the structure of the star is  only 
marginally affected by the choice for the electrosphere.
We postpone a  discussion about the structure of the electrosphere  to subsection~\ref{sec:electrosphere}.

\subsubsection{Results at fixed gravitational mass}

\begin{figure*}
\centering
	\includegraphics[width=.33\textwidth]{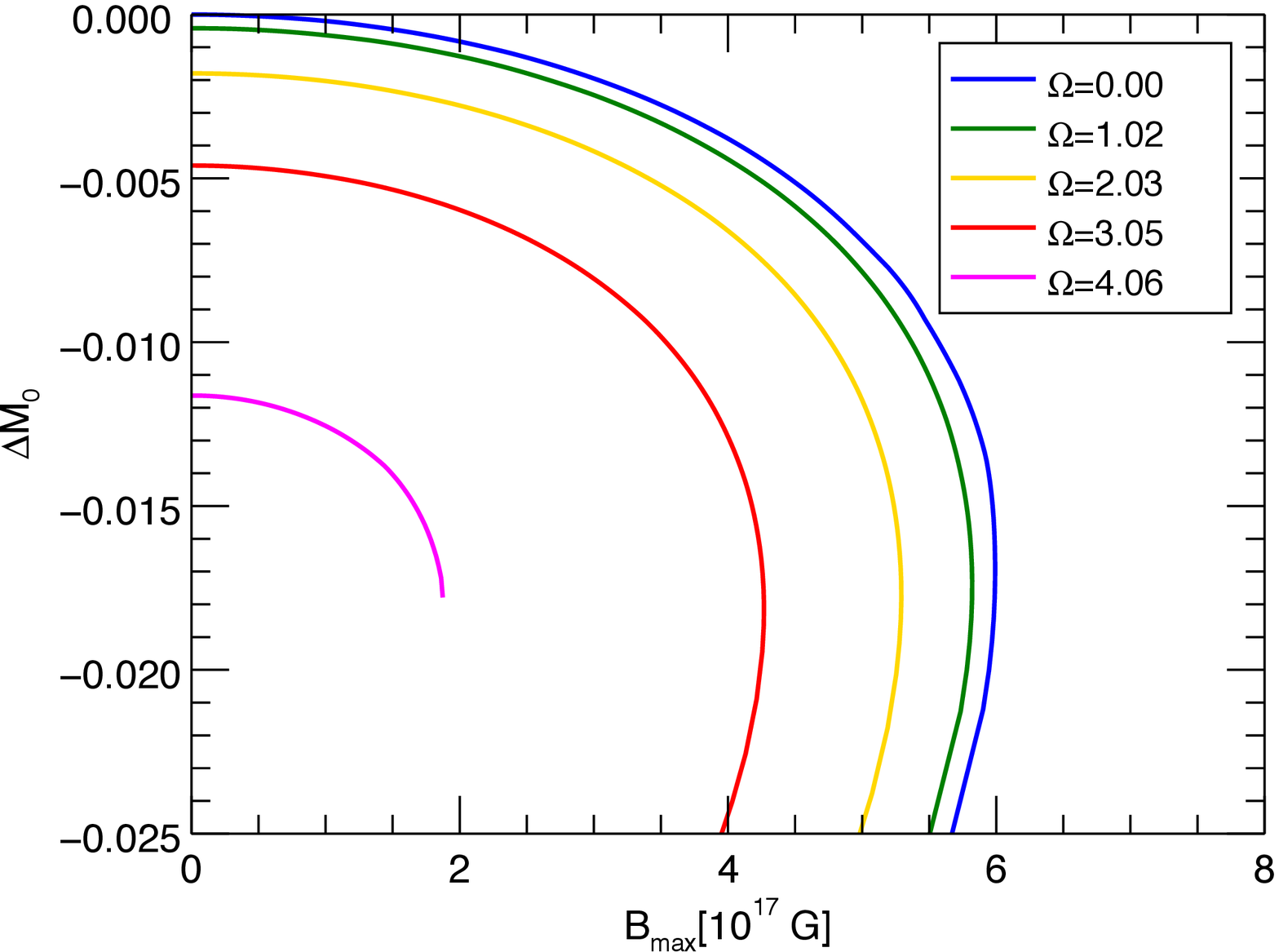}
	\includegraphics[width=.33\textwidth]{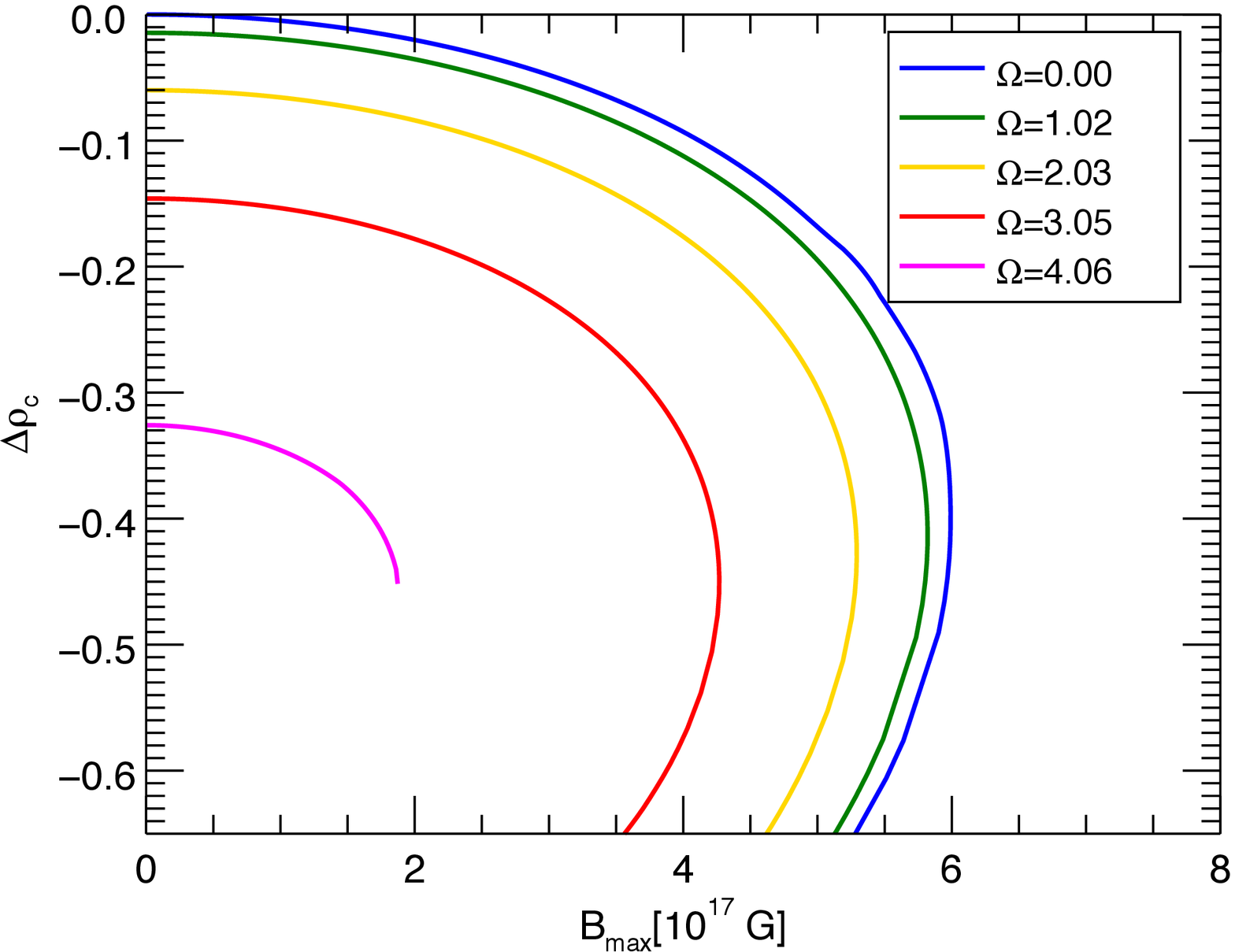}
	\includegraphics[width=.33\textwidth]{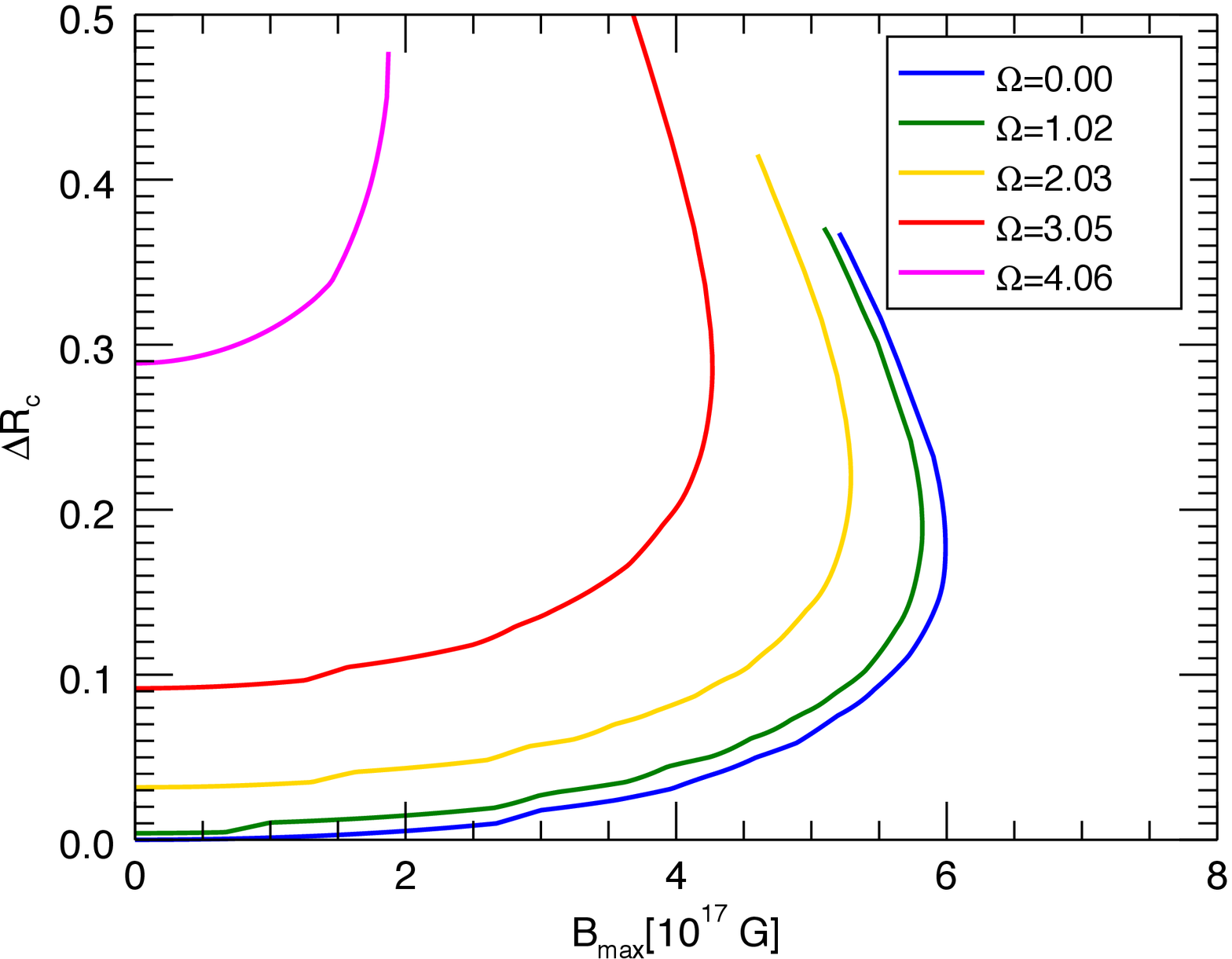}
	\caption{ Variation of the baryon mass $M_0$, central density $\rho_{\rm c}$ and circumferential radius $R_{\rm c}$
	with respect to the unmagnetized and static reference model, along sequences of constant gravitational mass 
	$M=1.55 \mbox{ M}_\odot$ but different $\Omega$, 
	as a function of  $B_{\rm max}$.  }
	\label{fig:M155pol}
\end{figure*}

Considering again a fixed gravitational mass $M=1.551 \mbox{M}_{\odot}$,
in Fig.~\ref{fig:M155pol} we show the variation of $\rho_c$, $M_0$ and $R_{\rm c}$,
with respect to the non rotating and unmagnetized equilibrium configuration,
as functions of $B_{\rm max}$ and $\Omega$.
Just as in the toroidal case, the qualitative effects of the poloidal magnetic field remain the same independently
from the rotational rate: both the baryonic mass $M_0$ and the  central density $\rho_{\rm c}$ decrease with
$B_{\rm max}$ while the circumferential radius $R_{\rm circ}$ and the deformation rate $\bar{e}$ grow. 
As it was  suggested in \citetalias{Pili_Bucciantini+14a}, the equilibrium sequences
are characterized by a turning point in  $B_{\rm max}$: 
the oblateness  initially grows  due to a rise of magnetic field, but, as soon as the maximum is reached, 
a further increase of the magnetization causes a rapid expansion of
the equatorial radius and a rapid drop in the central density corresponding 
also to a reduction of the polar radius. In the end this leads to configurations with an 
off-centered density distribution. Interestingly, independently of  $\Omega$, 
the configuration at the turning point shows a circumferential radius $R_{\rm circ}$ 
about $ 20\%$ larger than the unmagnetized model and $\bar{e} \sim 0.3$.

\begin{figure}
     \includegraphics[width=.45\textwidth]{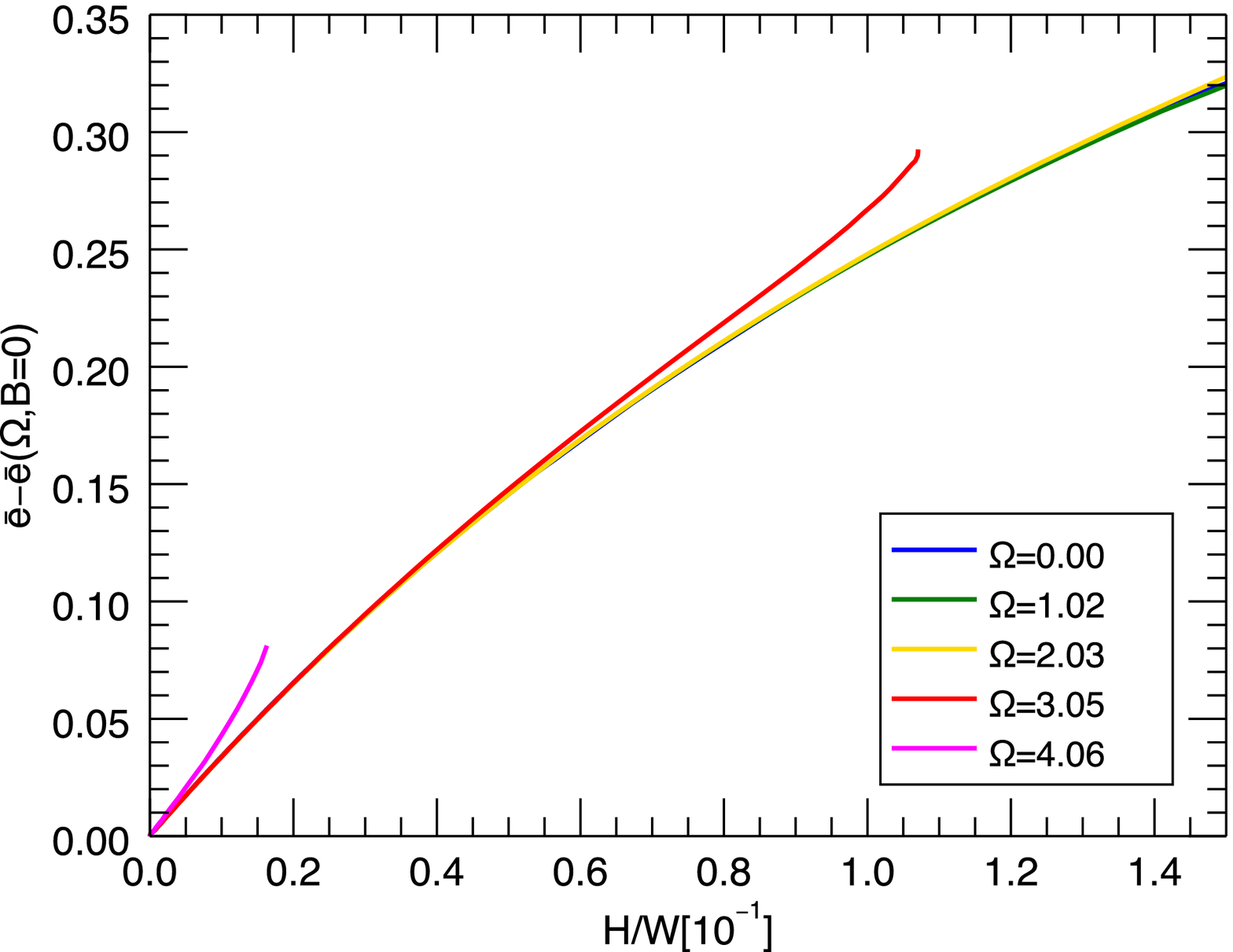}
     \caption{Variation of the deformation rate $\bar{e}$  with respect to the unmagnetized rotating reference 
     configuration as a function of the magnetic to binding energy ratio 
     along sequences with fixed mass $M=1.55 \mbox{M}_\odot$.}
     \label{fig:defHWpol}
\end{figure}

As for the  toroidal field case, the role of the rotation can be factored out as an offset, plus an enhancement 
of the effectiveness of the magnetic field and it is still possible to find  self-similarity scaling.
In particular Eqs.~\ref{eq:ss1} and~\ref{eq:ss2} are still valid with  $a_{\bar{e}}=-0.15$ and
$b_{\bar{e}}=0.27$ respectively, in the range  $\Omega \lesssim 3\times10^{3}\mbox{s}^{-1}$
and $T/M\lesssim5\times10^{-3}$ as shown in Fig.~\ref{fig:defHWpol}. However 
while the sign of $a_{\bar{e}}$  remains the same as for purely toroidal configuration,
the sign of $b_{\bar{e}}$ changes from negative to positive. This is because at a given
$H/W$ configurations with slower rotation have larger energy $H$ resulting in a more 
evident deformation.

We remark that now the electromagnetic energy is not uniquely confined inside the star: 
$\sim 25\%$ of the total energy is located outside the star. Moreover, the contribution of 
the electric field to the global energy is at most of the order of few percents, even for the
fastest rotators.

In the range $B_{\rm max}\lesssim 3\times 10^{17}\mbox{G}$  and 
$\Omega \lesssim 10^3 \mbox{s}^{-1}$  
(corresponding to $H/W\lesssim3.5\times10^{-2}$ and $T/W \lesssim 5.1\times10^{-2}$ or 
$H/M\lesssim 2.5 \times 10^{-3}$ and $T/M\lesssim 4.1 \times 10^{-3}$ ), which is comparable with 
the bilinear regime for the purely toroidal field case (see Eqs.~\ref{eq:param1} and~\ref{eq:parames1}),
the deformation rate $\bar{e}$ is approximated with an accuracy $\lesssim 10\%$ by the relation
\begin{equation}
\bar{e}=d_\Omega\, \Omega_{\rm ms}^2 + d_{\rm B}\, B_{17}^2,
\label{eq:eqBpol}
\end{equation}
where  $d_{\rm B}\simeq5.4\times10^{-3}$ and $d_\Omega\simeq0.31$.  Analogously, for the surface 
ellipticity we find
\begin{equation}
e_s=s_\Omega\,  \Omega_{\rm ms} +s_{\rm B} \, B_{17}^2,
\label{eq:espol}
\end{equation}
with $s_{\rm B}\simeq 4.5\times 10^{-3}$ and $s_\Omega\simeq0.38$.

\begin{figure*}
\centering
	\includegraphics[width=.33\textwidth]{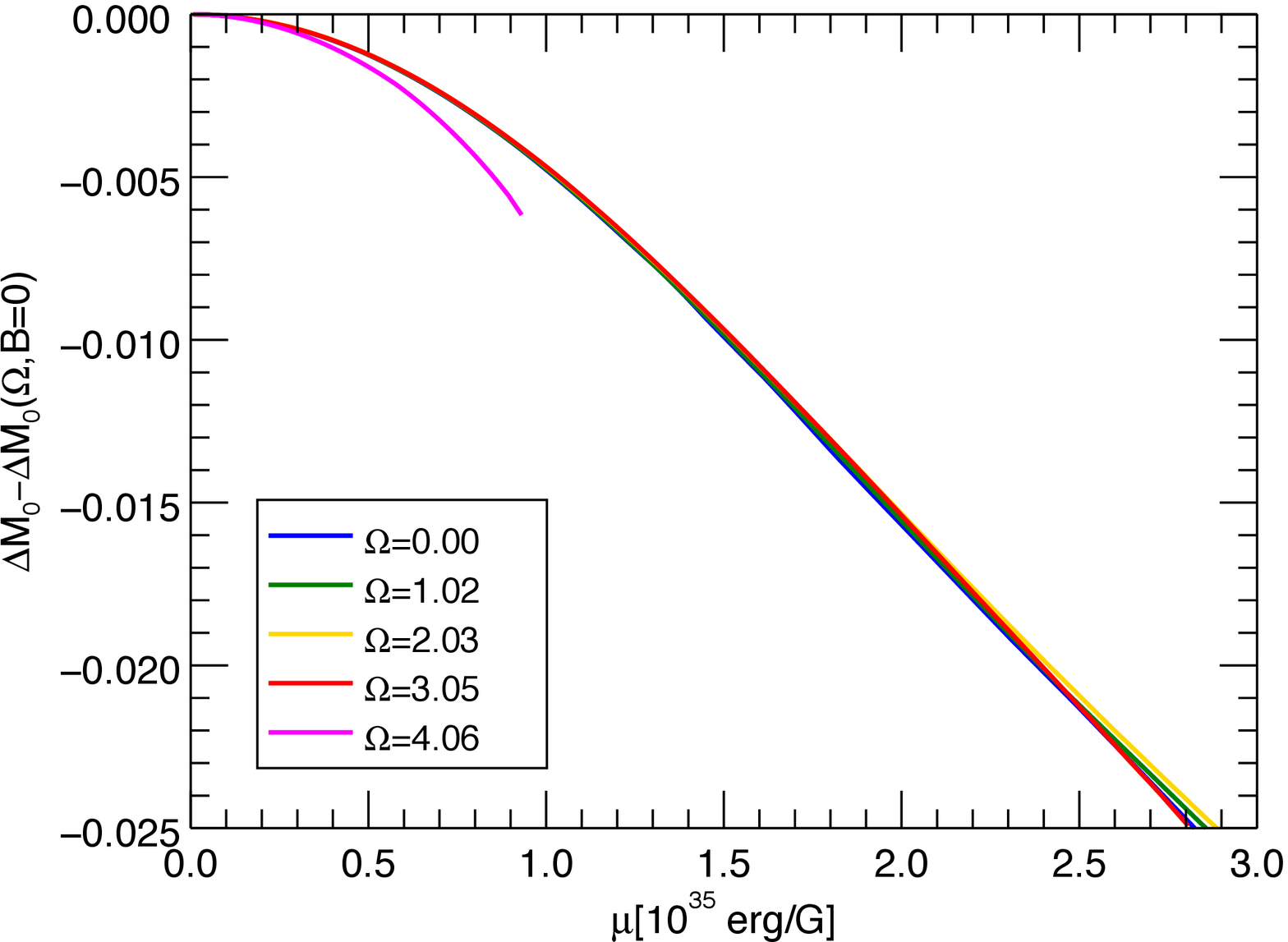}
	\includegraphics[width=.33\textwidth]{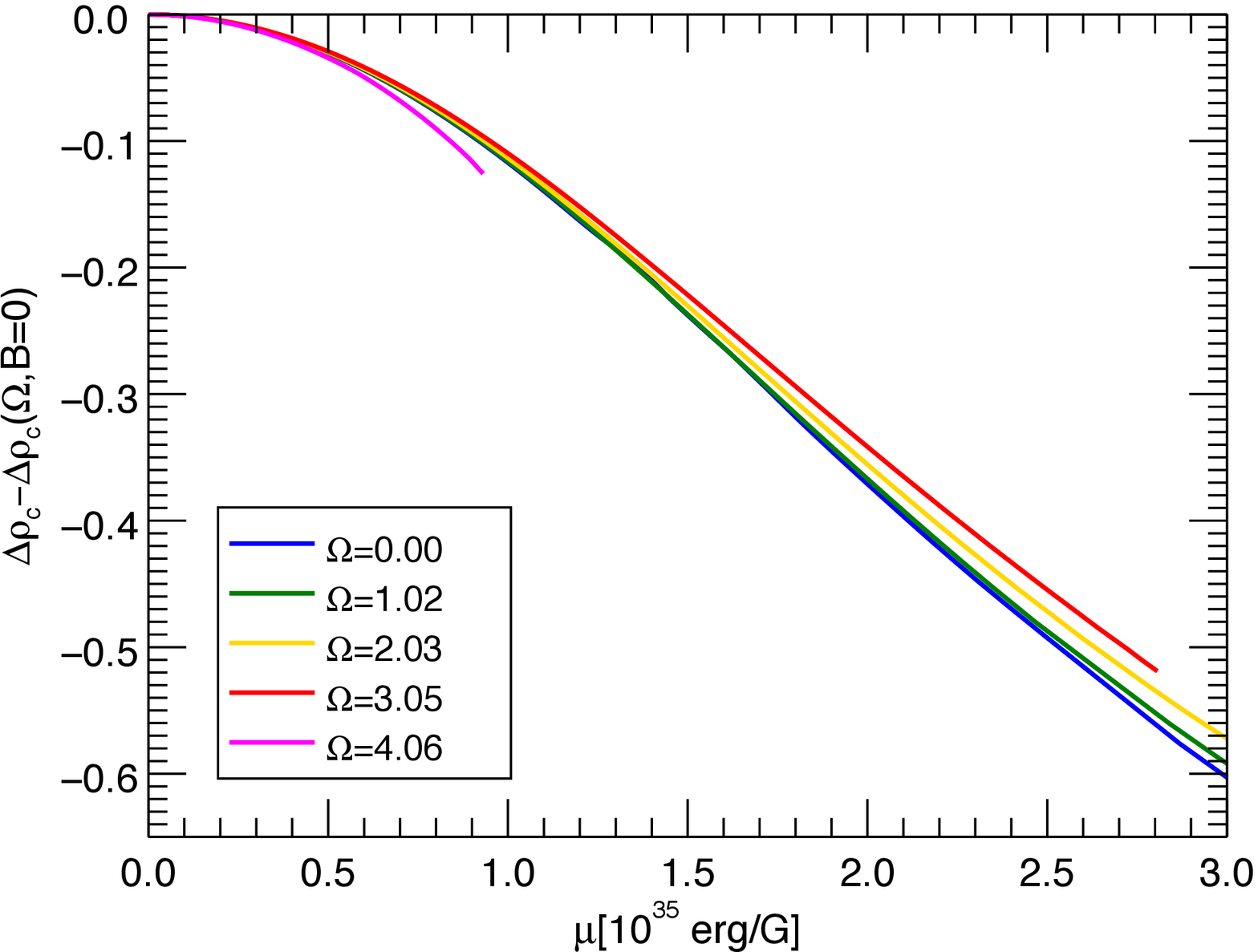}
    \includegraphics[width=.33\textwidth]{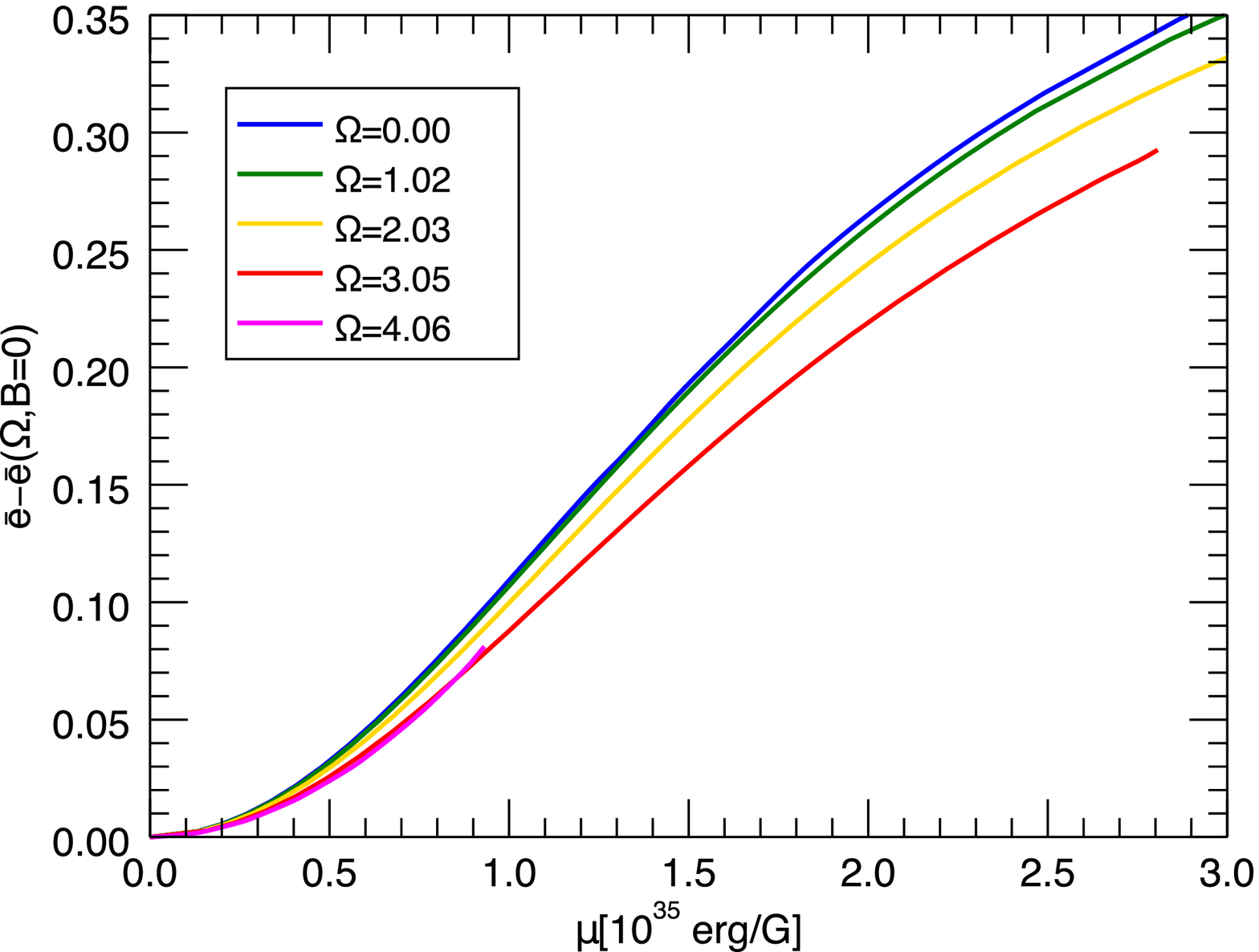}
 	\caption{ Variation of the baryonic mass $M_0$, central density  $\rho_{c}$ and deformation
 	rate $\bar{e}$ as a function of the magnetic dipole moment  $\mu$ along equilibrium configurations
 	with fixed gravitational mass $M=1.55 \mbox{M}_\odot$.}
 	\label{fig:M155mupol2}
\end{figure*}

These scalings can be given also in term of the magnetic dipole moment $\mu$
that, in contrast with $B_{\rm max}$, it is a  measurable quantity. In the same range as before we find
\begin{equation}
\bar{e}=d_\Omega\, \Omega_{\rm ms}^2 +d_\mu \,\mu_{35}^2,
\label{eq:eqmu}
\end{equation}
where $d_\mu=0.14$ and 
\begin{equation}
\bar{e}=s_\Omega\, \Omega_{\rm ms}^2 +s_\mu \,\mu_{35}^2,
\label{eq:eqmus}
\end{equation}
with $s_\mu\simeq 0.11$ ($\mu_{35}$ is the value of the magnetic dipole moment in 
unity of $10^{35}\mbox{erg G}^{-1}$).
In Fig.~\ref{fig:M155mupol2} we show the variation of different stellar quantities as functions of $\mu$.
Notice that, just as the magnetic energy, $\mu$ is a monotonic function of the magnetization $k_{\rm pol}$.
Moreover, in the case of $\Delta M_0$, a parametrization in terms of $\mu$ reduces the non-linear 
coupling between rotation and magnetic field.

\subsubsection{The effects of the current distribution}

The addition of non-linear current terms to the system can substantially modify the
structure of the magnetic field. In particular as discussed in \citetalias{Bucciantini_Pili+15a}
and \citetalias{Pili_Bucciantini+14a}  subtractive currents ($\xi<0$ in Eq.~\eqref{eq:Mpol})
tend to concentrate the magnetic field toward the magnetic axis causing a simultaneous demagnetization
of the outer layers. Additive current terms ($\xi>0$), instead, act to concentrate the 
magnetic field toward the stellar surface, causing also a global strengthening of the magnetic field.
As a result, at a given value of $B_{\rm max}$, the presence of additive current
gives a larger deformation $\bar{e}$; the opposite for subtractive ones. 

An extensive investigation of the parameter space in the case of 
rotating models, using different prescription for the current distribution, 
is however computationally expensive. The non-linear current term 
substantially slows down the convergence of the scheme 
even in the static case, where we have just to solve the Grad-Shafranov equation.
In order to get some handling on the effects of the current distribution on the 
deformation of the star, as we have done in the toroidal field case, we have just analyzed 
the simple static case with $\nu=1$, confident that in the bilinear regimes magnetic and rotation effects
can be separated. Away from the fully saturated regime discussed in \citetalias{Bucciantini_Pili+15a},
in the range $|\, \xi \, | \lesssim 30$, we have found that  the effects of the non-linear currents terms 
can be effectively reabsorbed with  a parametrization in terms of $H/W$ (or equivalently $H/M$) 
with an accuracy of $\sim 5\%$, using an effective energy ratio
 \begin{equation}
\bar{e}\simeq\mathcal{F}\left([1+ a_\xi \xi ]\frac{H}{W} \right)
\label{eq:poloidalxi}
 \end{equation}
where $a_\xi=-2.8\times 10^{-3}$. 
In the linear regime with $B_{\rm max}\lesssim 2. \times 10^{17} \mbox{ G}$,
the parametrisation in term of the magnetic field strength can be generalized as
\begin{equation}
 \bar{e}=[1+d_\xi \, \xi] \,d_{\rm B} B_{17}^2, 
\end{equation}
with $d_\xi=4.1\times10^{-3}$.

The difference between the signs of $a_\xi$ and $d_\xi$ may appear contradictory.
This discrepancy is however only apparent since, for a fixed value of  $H/W$, the configuration 
with $\xi<0$ has a larger value of $B_{\rm max}$ than the configuration with $\xi>0$.
This is because subtractive currents demagnetize to outer layer of the star and, in
order to achieve higher value of $H/W$, one has to increase the maximum strength of the 
magnetic field which in turn largely affects the core.
This holds also in the fully saturated regime discussed in \citetalias{Bucciantini_Pili+15a}.
Nevertheless, in this case, it is not possible to find a simple parametrization of 
the  effects of the current distribution  neither in term of energy ratios or other global quantities
such as the magnetic dipole moment. Contrarily to toroidal configurations, here the morphology 
of the current distribution may play a role in affecting the structure of the NS.
However,  if parametrized in terms of $H/W$, the deformation of the fully saturated regime 
ranges just within a factor 2 (see also Fig.~B1 of \citetalias{Bucciantini_Pili+15a}).

\subsubsection{Trends at different gravitational mass}

\begin{figure*}
\centering
	\includegraphics[width=.47\textwidth]{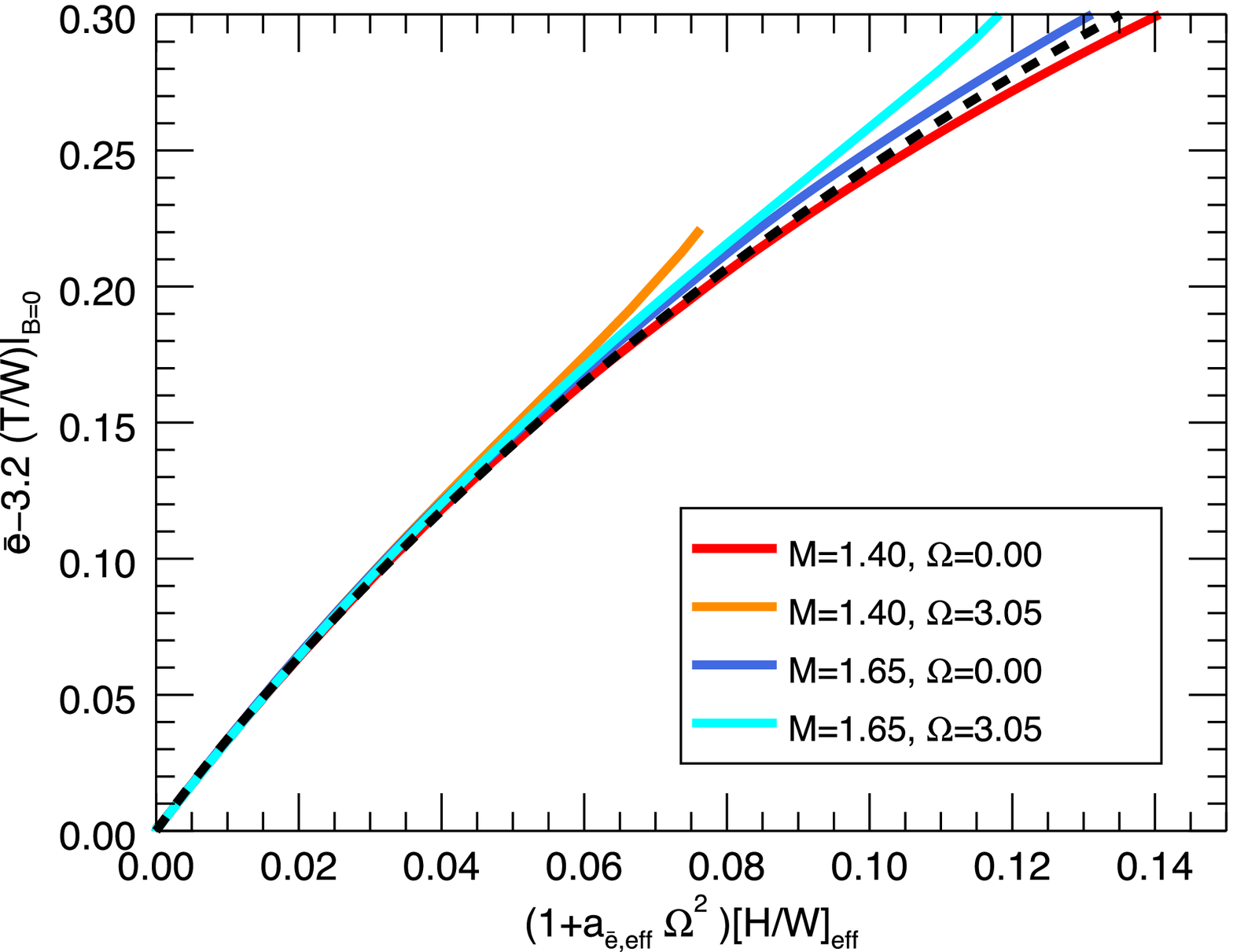}
	\includegraphics[width=.47\textwidth]{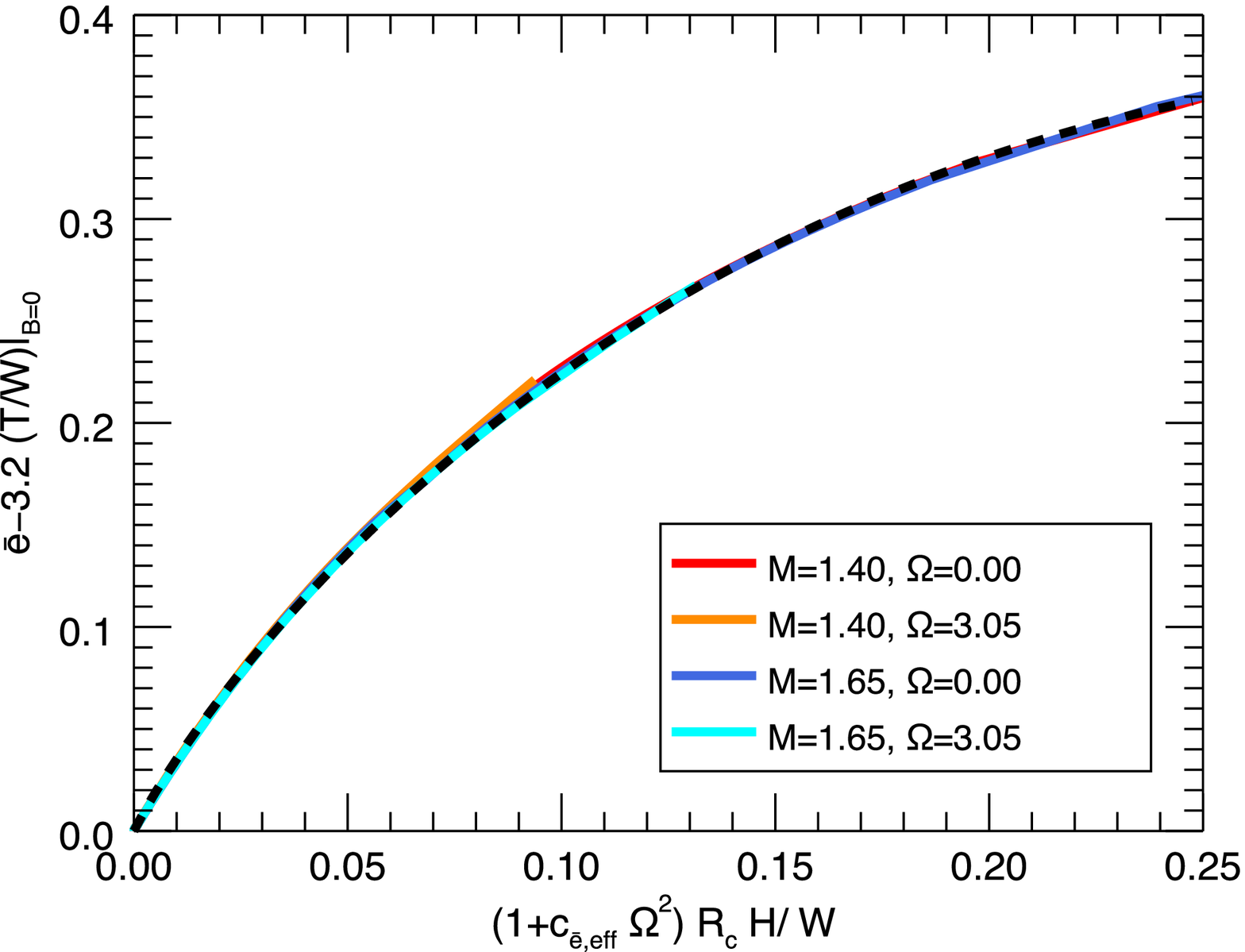}
	\caption{ Deformation $\bar{e}$ with respect to the unmagnetized model as a function
	of the effective mass energy ratio $H/W$ in the left panel or  $H R_c/W$
	(where $R_c$ is normalized to 14~km) in the right panel  for configurations with 
	mass between $1.40 \mbox{M}_{\odot}-1.65 \mbox{M}_{\odot}$ and 
	rotational frequency $\Omega=0.0-3.05\times10^{3} \mbox{s}^{-1}$.
	The dashed black lines show  Eq.~\eqref{eq:trendHWeffPol} (left panel) and Eq.~\eqref{eq:trendHWRcPol} (right panel).}
	\label{fig:HWeffpol}
\end{figure*}

As expected both the magnetic and rotational effects 
depend on the compactness of the star. To generalize the trends found in 
the previous sections we again make use of an effective energy ratio:
\begin{equation}
\left[ \frac{H}{W}\right]_{\rm eff}= \frac{1.55 \mbox{M}_{\odot}}{M}\frac{H}{W}.
\end{equation} 
By using this quantity the induced deformation can be  parametrized as
in the Eq.~\eqref{eq:trendM}, where the coupling term is now given by 
\begin{equation}
a_{\bar{e},{\rm eff}}=-\left(2.5 -2.4 \frac{M}{1.55 \mbox{M}_{\odot}} \right).
\label{eq:aeffpol}
\end{equation} 
while the functional form for $\mathcal{F}$ is provided by
\begin{equation}
\mathcal{F}(x)=3.8 x-4.3 x^{1.5}.
\label{eq:trendHWeffPol}
\end{equation}
As shown in the left panel of  Fig.~\ref{fig:HWeffpol}, this parametrization is 
able to describe the trends of the $\bar{e}$ up to $\sim0.15$ with an accuracy 
less than $5\%$. Notice that here, as in the toroidal field case, for a fixed value of
$H/W$ the coupling term $a_{\bar{e},{\rm eff}}$ reduces the absolute value of $\bar{e}$. 

Interestingly a more accurate parametrization of $\bar{e}$ can be 
obtained including 
the circumferential radius in place of the gravitational mass.
In particular we obtain that, by using the relation
\begin{equation}
\label{eq:trendpolR}
\bar{e}\simeq 3.2 \frac{T}{W}\bigg|_{B=0}+ \mathcal{G}\left( \left[ 1+ c_{\bar{e},{\rm eff}} \Omega^2_{\rm ms} \right] \frac{H}{W} R_{14} \right),
\end{equation}
where $R_{14}$ is the circumferential radius normalized to $14 \mbox{km}$,
the coupling term is given by 
\begin{equation}
c_{\bar{e},{\rm eff}}=-3.6+3.0\frac{M}{1.55\mbox{M}_{\odot}}
\end{equation} 
and the functional form of  $\mathcal{G}$ is
\begin{equation}
\mathcal{G}=4.8\,x - 5.1\,x^{1.3},
\label{eq:trendHWRcPol}
\end{equation}
we can fit with high accuracy the variation of $\bar{e}$ for all the value of 
$\Omega\lesssim3\times10^3\mbox{s}^{-1}$ and $M$ even in the strong magnetization regime as shown 
in Fig.~\ref{fig:HWeffpol}.

Limited to the bilinear regime, the coefficients appearing in the  Eqs.~\ref{eq:eqBpol}-\ref{eq:eqmus} 
are listed  in Tab.~\ref{tab:parampol} as a function of the gravitational mass. 
Interestingly the coefficients $d_\mu$ and $s_\mu$ are only weakly affected by the specific value
of the gravitational mass and they remain almost constant within $\sim 5\%$.

\begin{table}
\centering
\caption{ \label{tab:parampol}
Mass dependency for the $\bar{e}$ expansion coefficients 
in the case of purely poloidal magnetic fields.
}
\begin{tabular}{l*{7}c}
\toprule
\toprule
 $M$ & $ d_{ \Omega} $ & $d_{\rm B}$  & $d_{\mu} $ & $s_{\Omega}$ & $s_{\rm B} $ & $s_{\mu}$  \\ 
 $\mbox{M}_\odot$ & $10^{-1}$  & $10^{-3}$  & $10^{-1}$ & $10^{-1}$ & $10^{-3}$  & $10^{-1}$  \\
\midrule
1.40 &  4.5  &  9.8  & 1.4 & 6.0 & 9.0 & 1.1 \\
1.45 &  4.0  &  8.1  & 1.4 & 5.1 & 7.9 & 1.1 \\ 
1.50 &  3.6  &  6.7  & 1.4 & 4.6 & 5.9 & 1.1 \\ 
1.55 &  3.1  &  5.4  & 1.4 & 3.8 & 4.5 & 1.1 \\ 
1.60 &  2.8  &  4.4  & 1.4 & 3.5 & 3.9 & 1.1 \\
1.65 &  2.3  &  3.2  & 1.4 & 3.1 & 3.0 & 1.1 \\
\bottomrule
\end{tabular} 
\end{table}

In the perturbative regime of  $H,T\rightarrow 0$ the relation  in Eq.~\eqref{eq:trendM},
with Eq.~\eqref{eq:aeffpol} and Eq.~\eqref{eq:trendHWeffPol}, gives:
\begin{equation}
\bar{e}=\frac{C_{\bar{e}}}{W_0}\left[ T+ 1.8 \frac{H}{M/\mbox{M}_{\odot}}\right],
\label{eq:lineartrendBpol}
\end{equation}
which is the analogous of Eq.~\eqref{eq:selfsime} for the toroidal magnetic field.
The main difference between the two relations is that in the poloidal field
case the sign of the magnetic term is positive. This reflects the fact that a
poloidal field induces an oblate deformation. An analogous relation is found also for
the apparent ellipticity:
\begin{equation}
e_s=\frac{C_{e_{\rm s}}}{W_0}\left[ T+ 1.1 \frac{H}{M/\mbox{M}_{\odot}}\right].
\end{equation}

\subsubsection{The structure of the electrosphere}
\label{sec:electrosphere}

\begin{figure*}
\centering
	\includegraphics[width=.40\textwidth]{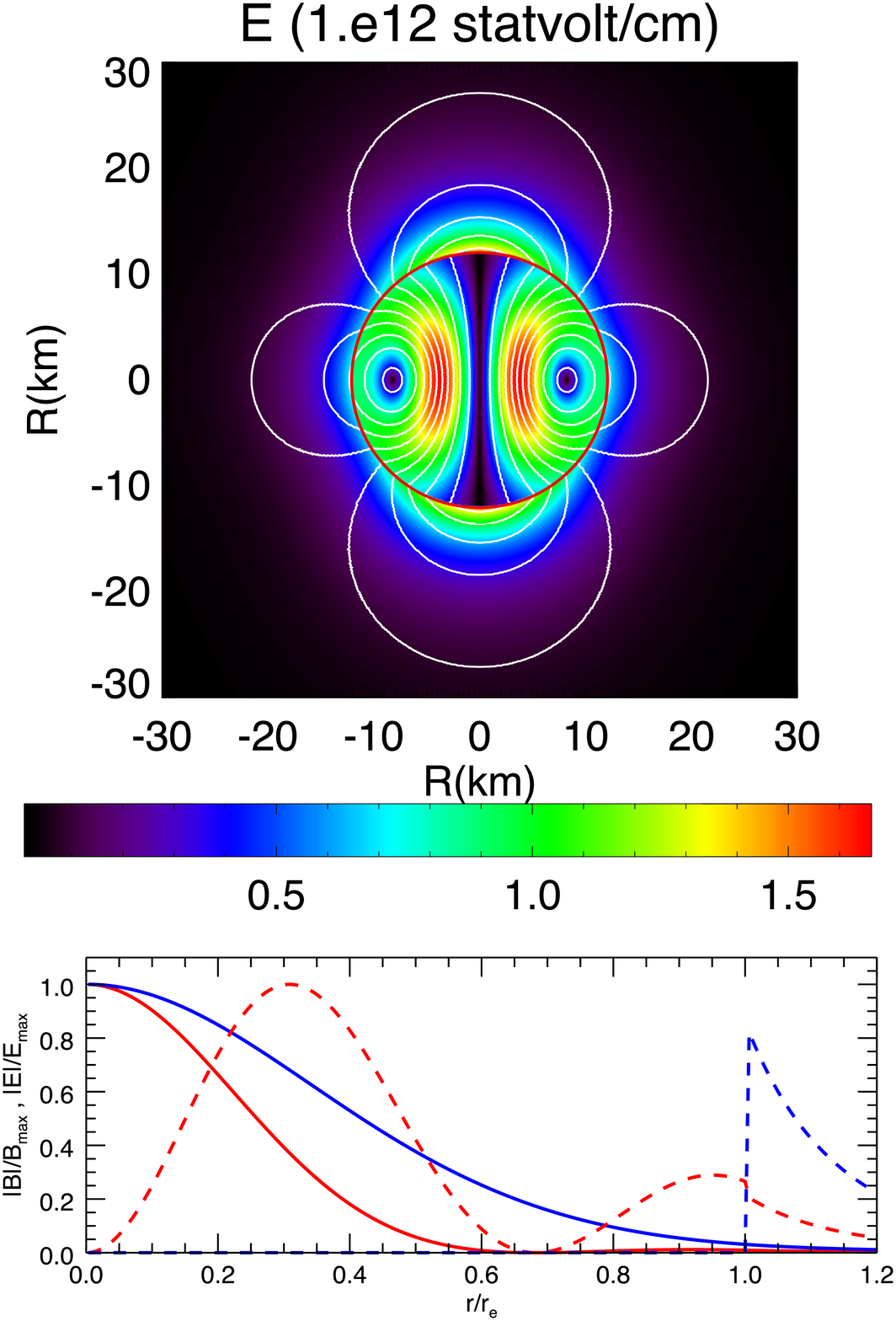}
	\includegraphics[width=.40\textwidth]{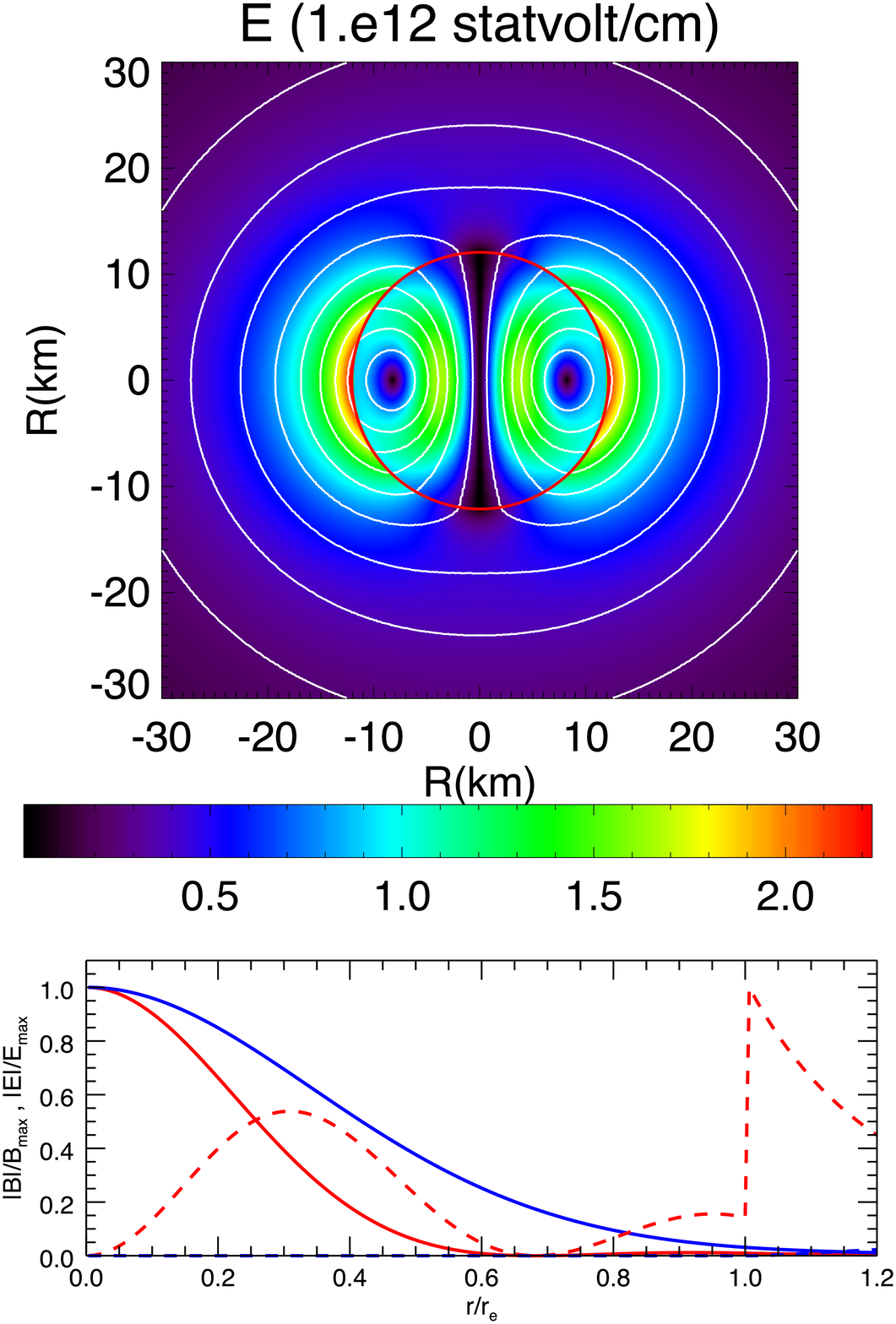}
 	\caption{ Top panels: distribution of the electric field strength $E=\sqrt{E^rE_r+E^\theta E_\theta}$ together
 	 with the isocontours of the potential $\Phi$ for configurations with vanishing net electric 
 	charge (left panels) and 	vanishing polar electric field (right panels) for a NS having $M=1.55 \mbox{M}_{\odot}$, 
 	$\Omega=200\,\mbox{s}^{-1}$, $B_{\rm max}=8.4\times 10^{14}\,\mbox{G}$ and $B_{\rm pole}=1.5\times10^{14}\,\mbox{G}$. 
 	Bottom panels: radial profiles of the magnetic (solid lines) and electric (dashed lines)
 	field strength along the equatorial (red lines) and the polar direction (blue lines).}
 	\label{fig:elec}
\end{figure*}
The electric field outside the star is not fully determined by the ideal MHD condition 
Eq.~\eqref{eq:MHDcond}, due to the arbitrary constant in the
definition of the harmonic function  $\Phi_a$. Such degree of freedom corresponds 
to an arbitrary charge that can be added to the system in order to obtain different 
physical outer structures.  As anticipated in Sec.~\ref{sec:nummax}, a typical choice is to consider a 
globally neutral NS \citep{Goldreich_Julian68a,Bocquet_Bonazzola+95a,Franzon_Schramm15a,Franzon_Dexheimer+16a},
but it is also possible to consider configurations where the charge distribution
reduces the Lorentz force in the polar region \citep{Michel74a} or minimizes the energy content 
of the electrosphere \citep{Ruffini_Treves73a}.

For simplicity in the following we limit our analysis to the weak magnetization and slow rotation limit, 
since in this regime the electromagnetic field does not alter substantially the structure
of the NS, and the solution can be rescaled with $B_{\rm max}$ up to $10^{16}$~G.
General trends, however, remain valid even at larger magnetic fields.

In the left panel of Fig.~\ref{fig:elec} we show the electric field strength for
a globally uncharged NS. In this case the external electric field peaks at the pole, in
the polar cap,  i.e. the region of magnetic field lines extending to infinity
beyond the Light Cylinder. In principle these electric fields are able to extract particle 
from the surface into the magnetosphere and beyond, charging the NS itself. 
In the right panel of the same figure we show a configuration for a NS  where 
the electric field vanishes at the poles. The NS is endowed with a net electric
charge corresponding to  $Q_{\rm e}\simeq  10^{24} \, B_{\rm pole, 14} \, \Omega_{100} \,  \mbox{statC}$
(here $ B_{\rm pole, 14}$ and $\Omega_{100}$ are the magnetic field
at the pole in units of $10^{14}$~G and the rotation rate in units of $100\,\mbox{s}^{-1}$, respectively).
Notice that $Q_{\rm e}$ is still far below the
critical value  $\sim 10^{29}\mbox{statC}$  (i.e. $~\sim 0.1\sqrt{G}M_0$ in cgs units) capable to 
induce substantial effects in the stellar structure \citep{Ghezzi05a}.  The electric field  now peaks at
the stellar equator, where it is a factor $\sim 2$ stronger than the internal one. Although this star is 
unable to extract particles from polar caps, it can in principle attract them from regions beyond 
the magnetosphere. Interestingly comparing configurations with different $Q_{\rm e}$ we have find that 
the configuration that minimizes the electromagnetic energy in the electrosphere
is the uncharged one. This is in contrast with the results obtained in \citet{Ruffini_Treves73a}, 
where the minimum energy configuration has a negative net charge. In that work however, the 
structure of the  magnetic field was chosen to depend on the specific value of the electric charge.

\begin{figure*}
\centering
    \includegraphics[width=.95\textwidth]{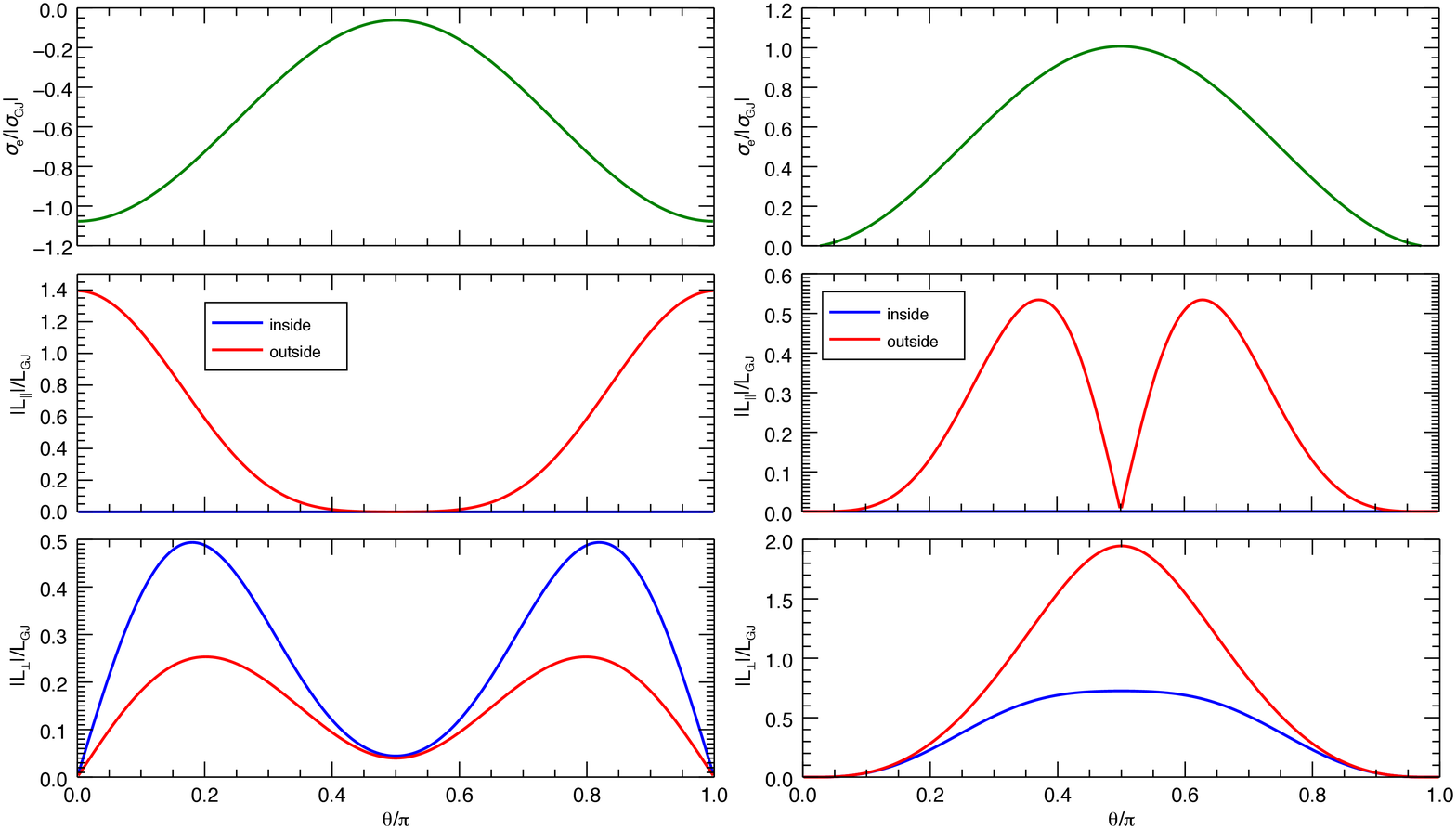}\\
    \includegraphics[width=.95\textwidth]{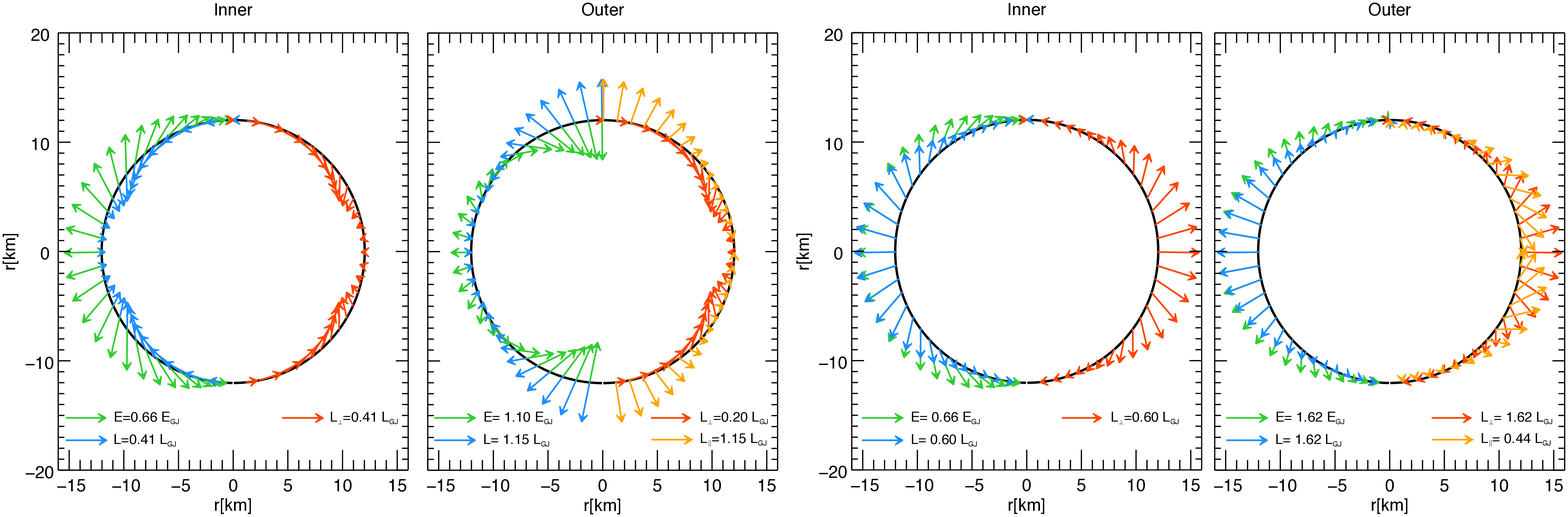}
 	\caption{  Top rows: trends of the surface charge density, of the parallel  $|L_\parallel |$ and
 	perpendicular component  $| L_\perp |$ of the Lorentz force along the surface of the star as a function 
 	of the colatitude (parallel and perpendicular refer to the direction of the magnetic field) for the same configurations
 	shown in Fig.~\ref{fig:elec}. Bottom row: vector plots of the surface electric field $E^{\hat{i}}$, of the Lorentz 
 	force $L^{\hat{i}}$ and its perpendicular and parallel component. Numerical values are normalized 
 	to $\sigma_{\rm GJ}=E_{\rm GJ}=\Omega B_{\rm pole} r_{\rm p} $ and
 	$L_{\rm GJ}= \rho_{\rm GJ} E_{\rm GJ}$ (corresponding respectively to
 	$\sigma_{\rm GJ}=9.5\times 10^{10} \mbox{statC cm}^{-2}$,
 	$E_{\rm GJ}=1.2\times 10^{12} \mbox{statVolt cm}^{-1}$ and $L_{\rm GJ}=1.1\times10^{23}\mbox{dyne}\,\mbox{cm}^{-2}$ ).
 	Panels on the left refers to the uncharged equilibrium configuration while panels on the right refers to the configuration 
 	with vanishing polar electric field.}
 	\label{fig:schargeLf}
\end{figure*}

In all cases the normal component of the electric field at the surface shows a discontinuity
corresponding to a surface charge density
\begin{equation}
\sigma_e=E^r_{\rm out}-E^r_{\rm in},
\end{equation}
where $E^{r}_{\rm out}$ and $E^r_{\rm in}$ are the radial components of the surface electric field
inside and outside the NS respectively.
The profile of  $\sigma_e$ for the configurations presented in Fig.~\ref{fig:elec} is shown in 
Fig.~\ref{fig:schargeLf} (numerical values are normalized to the  Goldreich-Julian density  
$\sigma_{\rm GJ} = \Omega B_{\rm pole} r_{\rm p}=9.5\times 10^{10} \mbox{statC cm}^{-2}$).
Notice that the sign of the surface charge, as well as the sign of the electric field, depends on
the relative orientation of the magnetic dipole moment and the angular momentum (our results are shown
in the aligned case). As a consequence the sign of the associated Lorentz force acting on the surface, $L^i=\sigma_e E^i$, does not change.
The latter is also shown in Fig.~\ref{fig:schargeLf}, where we plot both the orthogonal and the parallel component 
with respect to the magnetic field and where it is possible to see that, inside the star, the MHD condition guarantees 
that  the parallel component of the Lorentz force vanishes.

In the uncharged configurations the charge surface density is maximal at the pole. 
The  Lorentz force with respect to $E_{\rm in}$ vanishes on the rotation axis, 
reaches its maximum strength at latitude $\sim\pm 50\deg$ and  points
always toward the equator, remaining  mainly tangential to the stellar surface. 
The Lorentz force with respect to the external electric field instead is mainly parallel to 
the magnetic field in the polar region, and  becomes mainly orthogonal in the equatorial region
where it points inward. In the case of a negative surface charge (an
electron excess if angular momentum and dipole moment are aligned), the
Lorentz force is able to extract electrons.  In the case of a positive surface charge (an
ion excess if angular momentum and dipole moment are counter-aligned) the
Lorentz force will be unable to extract ions. Indeed, assuming a cohesive energy
$\sim 350 \mbox{keV}$ for an iron chain with spacing  $\sim 10^{-9} \mbox{cm}$ at
$B=10^{14} \mbox{G}$ \citep{Medin_Lai06a,Medin_Lai07a}, the critical electric field capable to
directly rip iron ions off the stellar crust is of the order of $\sim10^{13-14} \mbox{statV}/\mbox{cm}$ 
which is larger than the electric fields of our model ($\sim 10^{12-13} \mbox{statV}/\mbox{cm}$ up to
millisecond rotation).

The configuration with a vanishing polar Lorentz force is instead
characterized by a  surface charge density of opposite sign, with
respect to the uncharged case, with a maximum at the equator.
By consequence, the action of the surface Lorentz force is reversed: 
the maximum strength is reached in the equatorial region, and the force 
always points toward the exterior. Hence  if the surface charge is made of electrons (counter-aligned case),  
such force could pull them out from the surface and fill the closed magnetosphere within the Light Cylinder.

For a star described in terms of an ideal fluid, it is not possible to balance the surface Lorenz force, 
because no stress is present to counteract its tangential component.
In the presence of an elastic crust, on the other hand, a crustal deformation can give rise to stresses strong
enough to balance the Lorentz force.
In principle  the overall Lorentz force may act to stretch the crust 
of the NS  favoring a prolate deformation, in the uncharged case, or an oblate deformation in the case of
no polar Lorentz force. To estimate the crustal displacement required to balance the Lorenz force, 
we have solved the equations for the isostatic-like equilibrium of an elastic spherical thin shell
(see \citealt{Tanimoto97a} or \citealt{Tanimoto98a}) subject to a surface force, 
and taking into account also the role of buoyancy.
% describing how buoyancy reacts to an external surface force.
We assume a typical crust thickness of $\sim 1 \mbox{km}$, a density at the base of the crust  $\sim 10^{11}\mbox{g\,cm}^{-3}$
and Young modulus $\simeq 3\times10^{27}\mbox{erg\,cm}^{-3}$ \citep{Chamel_Haensel08a}. We find that
the vertical displacement is completely determined by buoyancy, and crustal elasticity plays a negligible effect. 
The amplitude of the vertical displacement is $|\Delta u_r| \simeq 6 \times 10^{-4} (B_{\rm pole,14} \Omega_{100})^2\, \mbox{cm}$
while for  the tangential one is $|\Delta u_\theta| \simeq 6 \times 10^{-2} (B_{\rm pole,14} \Omega_{100})^2\, \mbox{cm}$.
Such vertical displacement is negligible with respect to the one due to the global surface ellipticity derivable from
Eq.~\eqref{eq:espol}, $|\Delta u|\simeq (1.2 \,\Omega_{100}^2 + 0.46\, B_{\rm max,14}^2)\,\mbox{cm} $.
There is also a crustal quadrupole moment associated to such vertical displacement 
$|\bar{e}|\simeq 4 \times 10^{-13}  (B_{\rm pole,14}\, \Omega_{100})^2$,
which again can be shown to be negligible with respect to the global one as derivable from Eq.~\eqref{eq:eqBpol}.

Let us finally point here that once one allows for the presence of a singular surface 
charge  (and the related Lorentz force) in the Maxwell equations Eq.~\eqref{eq:MApot}, 
then there is no reason to exclude the presence  of any singular surface current. 
This is however usually done in building equilibrium models, because 
it guarantees the integrability for non rotating system and avoids the problem 
to consider arbitrary crustal currents that in general are not well constrained by physical arguments.
On the other hand the extra degree of freedom associated with a surface current,
can be used to modify the magnetic field structure, and the net Lorentz force at the surface. 
In particular if one chooses the following surface current:
\begin{equation}
J_{\rm surf}=\sigma_e\frac{\Omega-\omega}{\alpha}\delta(r-r_e),
\end{equation}
corresponding to the assumption that the surface charge corotates with
the star, then a discontinuity in the azimuthal component of the
magnetic field arises:
\begin{equation}
J_{\rm surf}=\frac{B^\theta_{\rm in}-B^\theta_{\rm out}}{\psi^2\sin{\theta}}.
\end{equation}
As we have numerically verified through the computation of a model including $J_{\rm surf}$,
both the radial and azimuthal component of the surface Lorentz force with respect to the 
the internal electromagnetic field now vanish. The surface charge and current are now in equilibrium 
with respect to the internal field. Notice however that, outside the star, no current can suppress 
the component of the Lorentz force parallel to the magnetic field. Finally, this current term induces only small 
deviation $\lesssim10^{-5}$  on the global structure of the electromagnetic field. The relevant effect is
on the Lorentz force itself.

\section{Discussion \& Conclusion}
\label{sec:conclusions}

\begin{table*}
\centering
\caption{ \label{tab:summary}
Summary of the global relations for the deformation ratio $\bar{e}$ in terms of the rotational energy ratio $T/W$ and the
magnetic energy ratio $H/W$.  $M_{1.5}$ and $R_{14}$ are respectively the 
gravitational mass and the circumferential radius in unity of $1.55 {\rm M}_{\odot}$ and $14$~km.
}
\begin{tabular}{lll}
\toprule
\toprule
Purely toroidal & $\quad \bar{e}\simeq 3.2 \frac{T}{W} \bigg|_{B=0}-2.7x-0.068(10x)^{3.2}$ & $ \mbox{with} \quad
                           x=\left(0.84+\frac{0.16}{m}\right)
                               \left[1-\left(3.94-2.98 M_{1.5} \right)\Omega_{\rm ms}^2\right]
                             M_{1.5}^{-1} \frac{H}{W}$ \\
\addlinespace
\midrule
Purely poloidal & $\quad \bar{e}\simeq 3.2 \frac{T}{W} \bigg|_{B=0} + 3.8x -4.3 x^{1.5} $ & $ \mbox{with} \quad 
                               x=\left[1-\left(2.5-2.4  M_{1.5}\right)\Omega_{\rm ms}^2\right]
                              M_{1.5}^{-1} \frac{H}{W}$ \\
\addlinespace
                        &  $\quad \bar{e}\simeq 3.2 \frac{T}{W} \bigg|_{B=0} + 4.8x -5.1 x^{1.3}$ & $\mbox{with} \quad
                             x=\left[1-\left(3.6-3.0 M_{1.5} \right)\Omega_{\rm ms}^2\right]
                               R_{14}  \frac{H}{W}  $\\
\bottomrule
\end{tabular} 
\end{table*}

In this work we present models of  rotating magnetized NSs
extending our previous results obtained in the static case 
(\citetalias{Pili_Bucciantini+14a}; \citetalias{Bucciantini_Pili+15a}; \citealt{Pili_Bucciantini+15a}). 
Our extensive investigation of the parameter space has allowed us to establish new
quantitative and qualitative relations among different quantities of interest
(such as the gravitational and baryonic mass, the energy content of the system, the 
current distribution) giving particular emphasis to the characterization and the 
parametrization of the stellar deformation $\bar{e}$. %($\bar{e}$ and $e_{\rm s}$).
This is indeed the relevant quantities in the context of  GWs astronomy, especially for 
newly born magnetars. Previous works (e.g. \citealt{Bocquet_Bonazzola+95a}, \citealt{Haskell_Samuelsson+08a} 
and \citetalias{Frieben_Rezzolla12a}) 
have already provided similar parametrizations, both in terms of the magnetic field strength and 
the rotational rate or in terms of the associated energetics. However, in the majority of cases, they were
limited to either the Newtonian or the perturbative regime, or to a limited set of reference masses. 
Here we have generalized such results up to the fully non-linear regime for a wide range of masses (excluding
supramassive configurations) and current distributions.

In the bilinear regime the coupling between magnetic and centrifugal effects can be safely neglected
and the stellar deformation can be accurately evaluated in terms of the magnetic field strength and the rotational 
rate through  Eqs.~\eqref{eq:param1m} and~\eqref{eq:eqBpol} or, equivalently, in terms of the energetic content
of the system using Eqs.~\eqref{eq:selfsime} and~\eqref{eq:lineartrendBpol}.
The deformability of the star mostly depends on the compactness of the star itself.
As shown also in Tab.~\ref{tab:tableparam1} and~\ref{tab:parampol} the deformation coefficients 
decrease with the mass or the gravitational binding energy of the star. This is indicative of the fact that 
the magnetic field is more efficient in less compact configurations with smaller central densities and larger radii.
Similar trends were also observed in  \citetalias{Frieben_Rezzolla12a} where, varying the EoS at 
fixed gravitational mass, the deformation coefficients grow with the stellar radius 
(notice that our results coincide with those of  \citetalias{Frieben_Rezzolla12a} in the case 
of $M=1.4\mbox{M}_\odot$). Interestingly, for purely poloidal field, we have find  that 
the deformation coefficient relative to the magnetic dipole moment $d_{\mu}$ weakly depends 
on the specific value of the mass, and it can be considered constant within $\sim 5\%$.
As a consequence the ratio between the magnetic and the gravitational spin-down timescales
(respectively $\tau_{\rm d}\propto \mu^2$ and $\tau_{\rm GW}\propto \bar{e}\sim d_{\mu}^2 \mu^2 $)
can be expressed as a function of the magnetic dipole moment and the rotational rate.
In particular in the case of orthogonal rotation one obtains  
$\tau_{\rm d}/\tau_{\rm GW}\sim 6\times 10^{-2}  \mu_{35}^2 \Omega_{\rm ms}$.

Comparing the purely toroidal/poloidal case, and in particular Eqs.~\ref{eq:selfsime} and~\ref{eq:lineartrendBpol}, 
it is apparent that, being the magnetic energy the same, a poloidal magnetic field is  more effective in 
deforming the star than the toroidal one. The situation is reversed if we look at the deformation in terms 
of the magnetic field strength, i.e. if we compare the absolute values of $d_{\rm B}$. This is due to 
the fact that, at a given value for the magnetic field strength, the toroidal configuration has typically a larger energy.

We can now establish the conditions favoring an efficient production of GWs. 
By considering the direct sum of the deformations induced by the two component
separately, we can roughly evaluate the total deformation induced by a mixed field.
According to \cite{DallOsso_Shore+09a}, we found that in the case of a
newly-born millisecond magnetar with a dipolar field of the order of $10^{14}$~G 
and an internal toroidal field of  $\sim 10^{16}$G, if the spin-flip mechanism occurs, 
the star gives rise to an observable GWs  emission.

Away from the bilinear regime the magnetic induced deformation can not be determined without considering
also the coupling effects with the rotation. This, reducing the compactness of the star,  actually enhances 
the effectiveness of the magnetic field. Moreover, the joint effects of a toroidal magnetic field
and the rotation can substantially reduces the frequency of mass-shedding, or determine characteristic solutions
with oblate surface deformation but globally prolate deformation (the transition $\bar{e}<0$ occurs near 
equipartition $H/T\sim 0.8 M/{\rm M}_{\odot}$ while the transition to $e_{\rm s}<0$ is at  $H/T\sim 4 M/{\rm M}_{\odot}$).
In the case of purely toroidal field the surface Lorentz force acts to stabilize the system against mass-shedding
(in line with the results by \citealt{Franzon_Schramm15a}).  In principle, in this case, one should also consider 
deformations due to the surface Lorentz stresses that, depending on the net charge of the NS, 
may counterbalance or enhance the global deformation. However our simple estimation suggests that
crustal deformation are generally negligible unless one considers millisecond rotators and high
magnetization with $B_{\rm pole}>10^{17}\,\mbox{G}$. Hence our general results, and in particular
the extrapolation to the bilinear regime, are not affected by the unbalanced surface stresses.

Interestingly we find that, in spite of the great complexity of the space of the solutions,
a parametrization of the induced deformation in terms of $H/W$ and $T/W$ 
allows us to reabsorb the dependency on the current distribution, the rotation and the compactness of 
the star in  unique scaling laws(summarized in Tab.~\ref{tab:summary} for the reader's convenience),
that hold up  to the fully non-linear regime with an accuracy $\lesssim 5$\%.  
This kind of self-similarity suggests that  apart from small residual effects, the stellar deformation 
is more sensible to the ratio of the magnetic and rotational energy to the gravitational binding energy
rather than the current distribution.  Unfortunately, this behavior fails if one considers fully saturated
or strongly concentrated currents as those presented in \citetalias{Bucciantini_Pili+15a}.
Being the energetics the same, concentrated currents located deeper inside the star
are indeed more effective than currents located in the outer layers. This has also important
consequences in the case of mixed field configuration. Reanalyzing the TT equilibrium sequences 
presented in \citetalias{Pili_Bucciantini+14a},  we have verified that the  deformation induced by a mixed 
field can not be trivially described as a direct sum of deformations induced by
the toroidal and poloidal component separately. As discussed in \citetalias{Bucciantini_Pili+15a}, 
in our TT models  the interplay between the two fields causes a rearrangement of the currents that 
tend to concentrate toward the surface.  As a result even a weakly energetic toroidal field 
(with $H_{\rm tor} \lesssim 10$\%$H_{\rm tot}$) is able to induce, at a given total energy $H_{\rm tot}$, 
a substantial reduction (of the order of 40\% for our models) of the oblateness of the star. 
This is in agreement with \citet{Mastrano_Suvorov+15a}. 
Computing TT models for non-barotropic NSs, they showed how it is possible to reduce  the effectiveness of the poloidal/toroidal 
component, augmenting the  weight of the quadrupolar component of the poloidal magnetic field 
or enlarging the volume occupied by the toroidal field. Hence, in  general, the condition $\bar{e}=0$ is not reached 
at equipartition $H_{\rm pol}=H_{\rm tor}$ but it strongly depends on the current distribution.
This may also suggests that the work done by the Lorentz force, evaluated as a volume integral of the Lorentz
force itself (e.g. as in \citealt{Fujisawa_Eriguchi13a} or \citealt{Fujisawa_Takahashi+13a}, including also the contribution
of possible unbalanced magnetic surface stresses), could be a more natural parameter 
to choose rather than the magnetic energy. Indeed it directly encodes a dependency on the currents distribution, 
distinguishing between the force-free and non force-free component of the magnetic field \citep{Fujisawa_Eriguchi15a}.

It is finally clear that, in order to provide  a more realistic estimation of 
the expected GWs emission from newly born millisecond magnetars, one needs to know in detail 
the current distribution  arising from the amplification of the magnetic field during core collapse. 
However if that currents are distributed rather than localized, we might expect that the 
direct addition of the  deformations produced independently  by the two component of the 
field gives a reasonable estimation of the global deformation.

\section*{Acknowledgements}

The authors acknowledge support from the INFN-TEONGRAV initiative and from the INFN-CIPE grant 
\emph{High performance data network: convergenza di metodologie e integrazione di infrastrutture per HPC e HTC}. 
LDZ also acknowledges support from the PRIN-MIUR project prot. 2015L5EE2Y 
\emph{Multi-scale simulations of high-energy astrophysical plasmas}.
This work has also been supported by an EU FP7-CIG grant issued to the NSMAG project (PI: NB).

%\bibliography{/Users/Antonio/Documents/PapersBD/MyBib.bib}{}
%\bibliographystyle{mn2e}

\end{document}